\documentclass[10pt,english]{article}

\usepackage[toc,page]{appendix}
\usepackage[]{graphicx}
\usepackage[]{color}
\usepackage{alltt}
\usepackage{caption}
\usepackage{subcaption}
\usepackage[T1]{fontenc}
\usepackage[utf8]{inputenc}
\usepackage{amsfonts}
\usepackage{amssymb}

\usepackage[table,xcdraw]{xcolor}
\usepackage{multirow}

\usepackage[top=100pt,bottom=100pt,left=68pt,right=66pt]{geometry}

\usepackage{lipsum}

\usepackage{setspace}
\usepackage[english]{babel}

\usepackage{fancyhdr}
\usepackage{titlesec}
\titleformat{\chapter}
   {\normalfont\LARGE\bfseries}{\thechapter.}{1em}{}
\titlespacing{\chapter}{0pt}{50pt}{2\baselineskip}

\usepackage{float}
\restylefloat{table}

\usepackage{multicol}

\usepackage{mathtools}
\usepackage{amsmath}
\numberwithin{equation}{section}

\usepackage[tableposition=bottom]{caption}

\usepackage{hyperref}
\hypersetup{hidelinks,linkcolor = black} 

\graphicspath{ {Figures/} }

\begin{document}

\selectlanguage{english}

\begin{titlepage}
	\clearpage\thispagestyle{empty}
	\centering

	\vspace{4cm}
	{\huge \textbf{Self-interaction of turbulent eddies in tokamaks with low magnetic shear}} \\
	\vspace{2cm}
	{\normalsize \textbf{Arnas Vol\v{c}okas, Justin Ball, Stephan Brunner} \par}
    \vspace{0.1cm}
	{Ecole Polytechnique F\'{e}d\'{e}rale de Lausanne (EPFL), Swiss Plasma Center (SPC), CH-1015 Lausanne, Switzerland \par}
    \vspace{0.1cm}
	{E-mail: \verb|Arnas.Volcokas@epfl.ch| \par}
    \vspace{1cm}
	{\normalsize \today \par}
    \vspace{1cm}
	\begin{abstract}
    Using local nonlinear gyrokinetic simulations, we demonstrate that turbulent eddies can extend along magnetic field lines for hundreds of poloidal turns in tokamaks with weak or zero magnetic shear $\hat{s}$. We observe that this parallel eddy length scales inversely with magnetic shear and at $\hat{s}=0$ is limited by the thermal speed of electrons $v_{th,e}$. We examine the consequences of these "ultra long" eddies on turbulent transport, in particular, how field line topology mediates strong parallel self-interaction. Our investigation reveals that, through this process, field line topology can strongly affect transport. It can cause transitions between different turbulent instabilities and in some cases triple the logarithmic gradient needed to drive a given amount of heat flux. We also identify a novel "eddy squeezing" effect, which reduces the perpendicular size of eddies and their ability to transport energy, thus representing a novel approach to improve confinement. Finally, we investigate the triggering mechanism of Internal Transport Barriers (ITBs) using low magnetic shear simulations, shedding light on why ITBs are often easier to trigger where the safety factor has a low-order rational value.
    \end{abstract}
	\vspace{1cm}
	
	\vspace{1cm}

	\vspace{2cm}
    
    
    \vspace{0.5cm}

	\pagebreak

\end{titlepage}


\newpage
\hspace*{0.3cm}

\newpage
\hspace*{0.3cm}

\section{Introduction}\label{section:Introduction}

An essential objective of contemporary fusion research is to develop techniques that effectively reduce particle and energy losses resulting from turbulent ("anomalous") transport in magnetically confined plasmas. Experimental observations have identified several improved confinement regimes for which transport is dramatically reduced. This can occur in different regions of the plasma and across some or all transport channels \cite{Wagner2018}. One significant class of these regimes, known as the Internal Transport Barrier (ITB), is characterized by a localized steepening of the pressure profile and an enhancement in confinement in the plasma core region \cite{Wolf2003, Connor2004, Ida2018}. Ion ITBs, which display transport reduction primarily in the ion channel, are of principle interest since ion temperature gradient (ITG) driven turbulence is expected to dominate in future high-performance tokamaks \cite{Transport_1999, Doyle2007, Garbet2010}, and ions are the particle species involved in fusion reactions. Even though there exists substantial experimental work on ITBs, their theoretical understanding is rather limited, which impedes their optimization and extrapolation to future devices. Hence, the aim of this paper is to help bridge this gap by numerically investigating turbulent self-interaction as a potential mechanism in ITB triggering and the reduction of turbulent transport.

It is non-trivial to precisely define ITBs or describe the conditions needed for their formation. They are observed under a broad range of experimental conditions and display varying properties \cite{Wolf2003,Connor2004,Ida2018}. Despite their variability, key features common to most ITBs have been identified through experimental observations across multiple devices \cite{Wolf2003, Garbet2010, Joffrin2002a, Tala2006}. In this work, we are interested in the distinct and critical role of the safety factor $q$ profile in the triggering of ITBs. 
It has been observed that ITBs are preferentially formed when a weak or slightly negative magnetic shear $\hat{s}$ region is present \cite{Lao1996}. Additionally, it is often beneficial if a magnetic surface within the weak magnetic shear region has a low order rational safety factor value or if the minimum safety factor $q_{min}$ in a reverse shear configuration is a low order rational (especially $q=2,3$) \cite{Joffrin2002}. Here the safety factor $q$ indicates the magnetic field line pitch on a given magnetic surface (i.e. how many toroidal turns the field line completes per poloidal turn), while the magnetic shear $\hat{s}$ is related to the radial gradient of $q$ \cite{wesson2004tokamaks}. Rational surfaces refer to magnetic surfaces with rational $q$ values. In Joint European Torus (JET) experiments, it has been shown that ITBs follow rational surfaces when the safety factor is varied throughout the shot after an initial single ITB forms \cite{Joffrin2002}. Zonal $\mathbf{E} \times \mathbf{B}$ flows, which have long been connected to turbulence stabilization \cite{BurrellExB1997, Hahm2002}, potentially play a secondary role compared to the safety factor profile in some cases. For example, it was observed that, while keeping the zonal flows fixed, ITBs form in plasmas with non-monotonic safety factor profiles, but not for standard monotonic profiles \cite{Eriksson2002}. Low magnetic shear is known to stabilize turbulence \cite{Antonsen1996}, but this effect alone is not enough to account for the observed reduction in transport and formation of ITBs. Hence, it is important to investigate how low (positive or negative) or zero magnetic shear combines with low order rational surfaces to affect plasma transport. Noteworthy in this respect is the fact that rational surfaces can facilitate strong parallel self-interaction of turbulent eddies, i.e. when individual eddies can "bite their own tails" by following a magnetic field line \cite{Ball2020}. Moreover, turbulent eddies get longer as magnetic shear is reduced \cite{Volcokas2023}, further accentuating self-interaction. Building on Ref. \cite{Volcokas2023}, this work explores self-interaction mechanisms and their role in ITB formation at low magnetic shear, a relatively unexplored area. There have only been a few limited prior studies into the role of self-interaction in ITB formation, but it has already been speculated that it can stabilize turbulence and lead to ITB triggering \cite{ChandrarajanJayalekshmi2020Thesis}.  \par 

Regarding prior numerical efforts to study ITBs, most past work has been done using global gyrokinetic codes \cite{Garbet2001, Candy2004, Waltz2006,  Sarazin2010, Kim2012}, which simulate a large fraction of the device. Due to their high computational cost, most of these studies were performed using adiabatic electrons with only some exceptions, namely reference \cite{Waltz2006}. Accordingly, reference \cite{Waltz2006} found stationary temperature, density, and flow profile corrugations at minimum $q$ (where $\hat{s}=0$), in agreement with experimental observations \cite{Austin2006} but with limited insight into the physics at play. Unfortunately, the global studies results were not conclusive and explored a very limited parameter space due to their high computational costs. More recent studies on stationary zonal flow structures around low order rational surfaces \cite{ChandrarajanJayalekshmi2020Thesis, Dominski2015, Dominski2017} have demonstrated that local simulations can correctly model the unique physics that occurs in the vicinity of these particular surfaces, namely the stationary temperature, density, and electric field corrugations observed in global simulations \cite{Waltz2006}. Given these results, we believe that the local flux tube approach, which is simpler, more reliable, and computationally cheaper than the global simulations used in the past, can capture the essential physics associated with the triggering and initial formation of ITBs. For example, the local approach is clearly well suited for studying an equilibrium like the JET pulse number 46050, which has $q \simeq 3$ and $\hat{s} \simeq 0$ over more than half of the plasma profile \cite{Joffrin2002}. Crucially, employing the local approximation allows us to explore a broad parameter space while including kinetic electron effects. According to studies on giant electron tails \cite{Hallatschek2005, Hardman2022} and turbulent self-interaction \cite{Ball2020, ChandrarajanJayalekshmi2020Thesis}, these effects are likely to play a significant role in the formation of ITBs.

This paper is organized in the following way. In sections \ref{section:gyrokinetics} and \ref{section:flux tube} we present a brief description of the considered gyrokinetic model as well as the flux tube simulation domain, including a discussion on the boundary conditions and the parallel boundary phase factor. The next four sections cover our numerical studies, split into linear and nonlinear parts: Section \ref{section:LinearFiniteShear} contains the linear study of low (but finite) magnetic shear, and section \ref{section:LinearZeroShear} contains a linear study of zero magnetic shear. Section \ref{section:NonlinearZeroShear} contains a nonlinear study at zero magnetic shear, and section \ref{section:NonlinearNonzeroShear} contains nonlinear simulations at low (but finite) magnetic shear. Lastly, section \ref{section:discussion} provides concluding remarks on our work. This work provides new findings on the physics of ultra-long turbulent eddies, their interplay with the magnetic field topology, and the effect of topology on turbulent transport.

\section{Theoretical model}

\subsection{Gyrokinetic model} \label{section:gyrokinetics}

Our study is based on the theoretical framework of gyrokinetics, a kinetic model that describes the self-consistent evolution of the distribution of the different plasma species in the presence of strong equilibrium magnetic fields. This model capitalizes on the concept of scale separation \cite{Brizard2007}. Key to this is the temporal scale separation between the slow timescale of fluctuations, denoted as $\tau_{fluc}$, and the rapid gyration timescale $\Omega$, where $\Omega$ stands for the cyclotron frequency of a given species. Given that $\tau_{fluc} \Omega \gg 1$, it allows one to average over the gyration of particles around the equilibrium magnetic field lines, effectively modeling them as charged rings attached to an averaged position called the gyrocenter. As a result, the phase space dimensions of the problem are reduce from six to five, removing the fastest timescale associated with particle gyromotion. Additionally, the spatial scale separation between turbulence $\sim \rho_{i}$ (the ion gyroradius scale) and the equilibrium $\sim a$ (the device minor radius), allows an expansion of the equations using the small parameter $\rho_{*} = \rho_{i}/a \ll 1$. These approximations make it possible to numerically solve the coupled system of integrodifferential equations for particle distributions and electromagnetic fields describing turbulent processes in magnetic fusion devices.

In the gyrokinetic formalism  \cite{Brizard2007} the gyrocenter distribution function $\Bar{F_{s}}( \Bar{\mathbf{X}}, \Bar{v}_{\parallel}, \Bar{\mu}, t )$  of species $s$ evolves according to the gyrokinetic Vlasov equation (collisions and plasma rotation are neglected here)
\begin{equation} \label{eq:fund_vlasovGK}
    \frac{\partial \Bar{F_{s}}}{\partial t} +\frac{d \Bar{\mathbf{X}}}{dt} \cdot \Bar{\nabla} \Bar{F}_{s} + \frac{d v_{\parallel}}{dt} \frac{\partial \Bar{F_{s}}}{\partial \Bar{v}_{\parallel}} + \frac{d \mu}{dt} \frac{\partial \Bar{F_{s}}}{\partial \mu} = 0,
\end{equation}
where $t$ stands for time, $\Bar{\mathbf{X}}$ is the gyrocenter position, $v_{\parallel}$ is the gyrocenter velocity along the magnetic field lines, $\mu$ is the gyrocenter magnetic moment, and an overbar indicates a gyroangle average at fixed guiding center position in a perturbed system. Thus, $d \Bar{\mathbf{X}}/dt$ denotes the velocity of the gyrocenters and $d v_{\parallel}/dt$ denotes the parallel acceleration. As $\mu$ is an adiabatic invariant, one finds that $d \mu / d t = 0$. \par

We further simplify the above gyrokinetic equation by assuming that the fluctuations are only electrostatic in nature, implying that the ratio $\beta$ of magnetic pressure over thermal pressure is small ($\beta \ll 1$). We also consider fluctuations that are much smaller than the equilibrium values, thereby separating the full distribution function $\Bar{F}_{s}=\Bar{f}_{0,s}+\delta \Bar{f}_{1,s}$ into an equilibrium $\Bar{f}_{0,s}$ and a fluctuating $\delta \Bar{f}_{1,s}$ parts ($\delta \Bar{f}_{1,s}/ \Bar{f}_{0,s} \ll 1$). The equilibrium distribution function is taken to be a local Maxwellian, assuming that the system is close to thermal equilibrium. Although collisions are neglected, a small numerical diffusion term is usually included when solving the equations numerically to avoid energy build-up at small scales (i.e. at the limit of the resolution) in the turbulence spectra. With these assumptions, the gyrokinetic equation can be expressed as \cite{Lapillonne2010Thesis}
\begin{align}
\label{eq:simple_GK_eq}
    \frac{\delta \Bar{f}_{1,s}}{\delta t} & +\mathbf{v}_{E \times B} \cdot \left( \nabla \Bar{f}_{0,s}+\Bar{f}_{0,s} \frac{\mu}{T_{0,s}} \nabla B_{0} \right) - \frac{\mu}{m_{s}} \hat{\mathbf{b}} \cdot \nabla B_{0} \frac{\partial \Bar{f}_{1,s}}{\partial v_{\parallel}}  \notag \\
    & + (v_{\parallel} \hat{\mathbf{b}}+ \mathbf{v}_{\nabla B}+\mathbf{v}_{c}+\mathbf{v}_{E \times B}) \cdot \left( \nabla \Bar{f}_{1,s} + \Bar{f}_{0,s} \frac{Z_{s}e}{T_{0,s}} \nabla \Bar{\phi} \right) = 0 
\end{align}
where $B_{0}$ is magnitude of the equilibrium magnetic field, $\hat{\mathbf{b}}$ is the unit vector along the magnetic field, $\mathbf{v}_{\nabla B}$ is the grad-B magnetic drift velocity, $\mathbf{v}_{c}$ is the curvature drift velocity, $\mathbf{v}_{E \times B}$ is the $\mathbf{E} \times \mathbf{B}$ drift, $m_{s}$ is the mass of species $s$, $e$ is the proton charge, $Z_{s}$ is the charge number of species $s$ and $\Bar{\phi}$ is the perturbed electrostatic potential. 

The Debye length is typically much smaller than the fluctuation scale and we have already assumed electrostatic fluctuations. Thus, instead of coupling Eq. \eqref{eq:simple_GK_eq} to the full set of Maxwell's equations, the gyrokinetic system is closed by imposing quasineutrality
\begin{equation}
    \label{eq:quaineutrality}
    \sum_{s} Z_{s} e n_{1,s} \simeq 0,
\end{equation}
where $n_{1,s}$ is the particle density of species $s$ calculated using the zeroth order moment of the gyrocenter distribution function $\Bar{f}_{1,s}$.

\subsection{Flux tube simulation domain} \label{section:flux tube}

Numerically solving the gyrokinetic model for a full torus is very expensive and in practice computationally feasible only for a limited number of simulations. It is much more common to solve the gyrokinetic model \cite{Abel2013} in a smaller flux tube --- a narrow computational domain that follows a set of neighbouring magnetic field lines. An in-depth discussion of the flux tube is given in the original paper \cite{Beer1995} and other works extending the model \cite{Ball2020, Ball2021, Ball2022}. Here we are only interested in discussing the key aspects of the flux tube formalism that are relevant to the present work --- namely the coordinate system, boundary conditions, the computational domain quantization condition, and the phase factor at the parallel boundary.

Flux tube considers a nonorthogonal, curvilinear, field-aligned Clebsch-type coordinate system defined by
\begin{equation}
\label{eq:coordinates_flux_tube}
    x=x(\psi), \quad y=C_{y}\left[q(\psi)\chi-\zeta \right], \quad   z=\chi,
\end{equation}
where $x,y,z$ are referred to as the radial, binormal, and parallel coordinates, and $\psi, \chi, \zeta$ are the poloidal flux, straight field line poloidal angle, and toroidal angle respectively. The radial profile of the safety factor is given by $q(\psi)$. $C_{y}$ is a normalization constant, which can be written as $C_{y}=r_{0}/q_{0}$ for circular flux surface in the large aspect ratio limit, where $r_{0}$ is the minor radius and $q_{0}$ is the safety factor of the reference magnetic surface on which the flux-tube lies. The radial coordinate can also be expressed in terms of the minor radius coordinate $r$ as $x=r-r_{0}$. Note that $x$ identifies a flux surface, $y$ identifies a magnetic field line on a given flux surface, and $z$ parameterizes the position along a given magnetic field line. Consequently, the equilibrium magnetic field $\mathbf{B}_{0}$ is parallel to $\nabla x \times \nabla y$. This is a standard definition for flux tube coordinates widely adopted in gyrokinetic codes, such as the GENE (Gyrokinetic Electromagnetic
Numerical Experiment) code \cite{Jenko2000, Gorler2011} used in this work. \par

The flux tube simulations take advantage of the anisotropy of the turbulence $k_{\bot} \gg k_{\parallel}$, where $k_{\bot}$ and $k_{\parallel}$ are the characteristic wavenumbers of the turbulence perpendicular and parallel to the magnetic field respectively. Hence the flux tube computational domain $\left( x,y,z \right) \in \left[ 0, L_{x} \right] \times \left[ 0, L_{y} \right] \times \left[ - \pi N_{pol}, \pi N_{pol} \right]$ can be relatively narrow in the perpendicular directions $x$ and $y$ compared to the full plasma cross-section but extended along the magnetic field lines. Here $L_{x}$ and $L_{y}$ are the radial and binormal computational domain widths respectively and $N_{pol}$ defines the parallel length of the flux tube by setting the corresponding integer number of poloidal turns the flux tube wraps around the torus. In the local approximation, equilibrium profiles (e.g. density, temperature and magnetic geometry) are Taylor expanded around the magnetic field lines of interest. This means that plasma equilibrium values and their gradients are constant across the flux tube and the geometric coefficients only vary along the magnetic field lines, i.e. are only functions of $z$. Note that in a tokamak system, the equilibrium is toroidally symmetric and thus independent of the toroidal angle $\zeta$. In a flux tube, equilibrium quantities are furthermore approximated to be independent of $x$. However, explicit radial coordinate $x$ dependence in field line geometry comes through the linearized safety factor profile 
\begin{equation}
    \label{eq:gene_q}
    q(x)=\left. q_{0}+\frac{dq}{dr} \right \vert_{r=r_{0}}(r-r_{0})=q_{0}+\hat{s}\frac{x}{C_{y}}
\end{equation}
where $\hat{s}=C_{y}(dq/dx|_{x=0})$ is the magnetic shear. \par

Given that the flux tube is sufficiently narrow such that the equilibrium quantities are independent of $x$ and $y$ but sufficiently broad with respect to the perpendicular correlation length of turbulent eddies, any fluctuating field $A$ will be statistically identical at the computational domain boundaries along these directions. Following this, the exact periodicity

\begin{equation}
\label{eq:radial_boundary}
    A(x+L_{x}, y, z)= A(x,y,z)
\end{equation}
\begin{equation}
\label{eq:binormal_boundary}
    A(x, y+L_{y}, z) = A(x,y,z),
\end{equation}
is applied for practical purposes. These boundary conditions straightforwardly guarantee statistical similarity across the flux tube. Generally in numerical simulations, it is crucial to make sure that the simulation domain is long enough in both the radial $x$ and binormal $y$ directions to avoid non-physical perpendicular turbulent self-interaction. Typically, this is accomplished by ensuring that the simulation domain is significantly larger than the turbulence correlation length, which is usually tens of gyroradii. However, it is important to note that, under specific circumstances, the perpendicular self-interaction of turbulent eddies can indeed be physical. For instance, if a turbulent eddy goes around the device multiple times and fully covers the full physical flux surface, it can then self-interact in the binormal direction. This particular scenario, while unusual, will be argued to be possible in this work. \par

The parallel boundary condition, called "twist-and-shift" \cite{Beer1995}, is more subtle due to magnetic shear and the choice of the parallel domain length. Due to axisymmetry, turbulence at the same poloidal locations (even if it is at a different toroidal location) has to be statistically identical and at the ends of the domain exactly identical turbulence is assumed
\begin{equation}
\label{eq:parallel_boundary}
    A(x, y(x, \chi+2\pi N_{pol}, \zeta), z(\chi+2 \pi N_{pol})) = A(x, y(x, \chi, \zeta), z(\chi)).
\end{equation}
Using Eqs. \eqref{eq:coordinates_flux_tube} and \eqref{eq:gene_q} gives the real-space parallel boundary condition
\begin{equation}
\label{eq:real_space_parallel_boundary}
    A(x, y(x, \chi, \zeta) + C_{y} q_{0} 2\pi N_{pol} + \hat{s} x 2\pi N_{pol}, z(\chi)+2 \pi N_{pol}) = A(x, y(x, \chi, \zeta), z(\chi)).
\end{equation}
The periodic boundary conditions in the perpendicular directions (Eqs. \eqref{eq:radial_boundary} and \eqref{eq:binormal_boundary}) allow us to represent fluctuations of any arbitrary fluctuating quantity with Fourier series in terms of radial $k_{x}=k (2\pi/L_{x})$, $k \in \mathbb{Z}$ and binormal $k_{y}=l (2 \pi / L_{y})$, $l \in \mathbb{Z}$ wavenumbers 
\begin{equation}
\label{eq:A_fourier_representation}
    A(x,y,z,t)=\sum^{\infty}_{k=-\infty} \sum^{\infty}_{l=-\infty} \hat{A}_{k_{x}, k_{y}}(z,t) e^{i(k_{x}x+k_{y}y)}.
\end{equation}
This Fourier representation implicitly implements the boundary conditions in Eqs. \eqref{eq:radial_boundary} and \eqref{eq:binormal_boundary}. Substituting Eq. \eqref{eq:A_fourier_representation} into the real-space parallel boundary condition of Eq. \eqref{eq:real_space_parallel_boundary} we obtain a coupling between modes with different $k_{x}$ values

\begin{equation}
\label{eq:finite_shear_mode_coupling}
   \hat{A}_{k_{x}, k_{y}}(z+2\pi N_{pol},t) =   \hat{A}_{k_{x}+k_{y}2 \pi N_{pol}\hat{s}, k_{y}}(z,t) e^{ik_{y}C_{y} q_{0} 2\pi N_{pol}}.
\end{equation}
As long as $|\hat{s}| \neq 0$, the modes will continuously couple to higher $k_{x}$ modes across the parallel boundary. 

When the magnetic shear is finite $\hat{s} \neq 0$, the phase factor
\begin{equation}
    \label{eq:phase_factor_full}
    C=e^{i 2\pi k_{y} C_{y} q_{0} N_{pol}}
\end{equation}
can be set to $C=1$ by slightly adjusting the radial position of the box to change $q_{0}$ \cite{Beer1995} such that $k_{y} C_{y} q_{0} N_{pol}$ is a large integer for all $k_{y}$ modes. This can be done since, based on Eq.  \eqref{eq:coordinates_flux_tube}, a given $k_{y}$ wavenumber is related to a toroidal mode number $n \in  \mathbb{Z}$ according to $n = - k_{y} C_{y}$ and this relation must hold for all wavenumbers including the smallest wavenumber $k_{y,min} = 2 \pi / L_{y}$ \cite{ChandrarajanJayalekshmi2020Thesis}. This allows us to rewrite the argument appearing in the phase factor as \par

\begin{equation}
    k_{y} C_{y} q(x=0) N_{pol} = - j n_{min} q_{0} N_{pol},
\label{eq:toroidal_wn}
\end{equation}
where $j=k_{y}/k_{y,min}$ and $j \in \mathbb{Z}$. Note that $|k_{y, min} C_{y}| = |n_{min}| \gg 1$, since $k_{y,min} \propto \rho_{i}^{-1}$ and $C_{y} \propto a$. In the case $\hat{s} \neq 0$, a small radial shift can infinitesimally adjust $q_{0}$ and it can then be arbitrarily well approximated by $q_{0} \simeq m/n_{min}$, $m \in \mathbb{N}$, $|m| \gg 1$. This results in
\begin{equation}
    k_{y} C_{y} q(x=0) N_{pol} = - j m N_{pol}  \in \mathbb{Z},
\label{eq:toroidal_wn_finalform}
\end{equation}
so that $C=1$, which leads to the final form of the standard flux tube parallel boundary condition for finite shear $\hat{s}$
\begin{equation}
\label{eq:finite_shear_mode_coupling_final}
   \hat{A}_{k_{x}, k_{y}}(z+2\pi N_{pol},t) =   \hat{A}_{k_{x}+k_{y}2 \pi N_{pol}\hat{s}, k_{y}}(z,t).
\end{equation}

One of the important consequences of the twist-and-shift boundary condition and the resulting mode coupling through finite magnetic shear is that it quantizes the computational domain size. If $L_{x}$ and $L_{y}$ are not chosen appropriately, imposing the twist-and-shift quasi-periodic parallel boundary condition can result in coupling to modes that do not exist on the computational grid. To avoid this, we require that all coupled Fourier modes $k_{x}+k_{y}2 \pi N_{pol}\hat{s}$ must be harmonics of $k_{x,min} = 2 \pi / L_{x}$ and that this condition must hold for the most limiting case of $k_{y, min}$. Thus, we require
\begin{equation}
    k_{y,min}2 \pi N_{pol} |\hat{s}| = k_{x,min} M
\end{equation}
where $M \in \mathbb{Z}^{+}$ is a positive integer. This gives the flux tube domain quantization condition
\begin{equation}
\label{eq:domain_quantization}
    L_{x} = \frac{M}{2 \pi N_{pol}|\hat{s}|} L_{y}=\frac{M}{k_{y,min}N_{pol}|\hat{s}|}.
\end{equation}
Since $M \geq 1$, as the magnetic shear is decreased at fixed $k_{y, min}$ and $N_{pol}$, the domain quantization condition necessitates larger radial domains, which increases the computational costs. 

However, the quantization condition vanishes completely when $\hat{s}=0$ as the shear no longer couples different $k_{x}$ modes, and one is free to choose both $L_{x}$ and $L_{y}$ independently.

Additionally, when the magnetic shear is finite the simulation domain is twisted into a parallelogram, which effectively introduces different order rational surfaces into the flux tube \cite{Ball2020}. This is because, at specific radial locations in the computational domain, the magnetic field lines pass through the parallel boundary an integer number of times and then exactly connect back onto themselves. 
For example, considering domain $x\in [-L_{x}/2, L_{x}/2]$, at $x=\pm L_{x}/2$ the parallel boundary condition \eqref{eq:real_space_parallel_boundary} in real space reads
\begin{equation}
    A(\pm \frac{L_{x}}{2}, y(x, \chi, \zeta) \pm \hat{s} \frac{L_{x}}{2} 2\pi N_{pol}, z(\chi)+2 \pi N_{pol}) = A(x, y(\pm \frac{L_{x}}{2}, \chi, \zeta), z(\chi)), 
\end{equation}
where we have already invoked $C=1$. Next, applying the domain quantization condition in Eq. \eqref{eq:domain_quantization} and setting $M=1$ we have
\begin{equation}
    A(\pm \frac{L_{x}}{2}, y \pm \frac{L_{y}}{2}, z+2 \pi N_{pol}) = A(\pm \frac{L_{x}}{2}, y, z).
\end{equation}
Hence, we can see that if we follow a given magnetic field line at $x=\pm L_{x}/2$ it will get shifted by $\pm L_{y}/2$ after crossing the parallel boundary once (i.e. after $N_{pol}$ poloidal turns). If we continue following the same magnetic field line until it passes the parallel boundary a second time, it will be shifted by $\pm L_{y}/2$ again --- ending up at the same coordinate. This results in an effective rational surface at $x= \pm L_{x}/2$ with order $n=2N_{pol}$.

Due to the standard parallel boundary condition, at least one lowest order rational surface always exists in the domain, corresponding to a location where a given magnetic field line connects with itself after one time through the parallel boundary, i.e. after $N_{pol}$ turns. The lowest order rational surface is usually at $x=0$ as a result of Eq.  \eqref{eq:toroidal_wn_finalform} leading to $C=1$, but the phase factor $C$ can also be set in such a way that some other radial location $x=x_{LRS}$ corresponds to the lowest order rational surface. As already mentioned, this corresponds to a simple radial translation of the simulation domain as it is periodic in the radial direction. Based on Eq.  \eqref{eq:domain_quantization}, when $M>1$, there will be $M$ such lowest order rational surfaces within the radial width of the flux tube. In finite $\hat{s}$ simulations, GENE sets $C=1$ so that the center of the domain corresponds to the lowest order rational surface. Away from the lowest order surface, other radial locations will correspond to higher order rational surfaces, up to a maximum order of $N_{max} =N_{pol} k_{y,max}/k_{y,min}$. Note that in the case that the flux tube only covers a fraction of the magnetic surface, i.e. $L_{y} < 2 \pi C_{y}$, these surfaces should be referred to as pseudo-rational. This is because they will not all necessarily correspond to a physical rational surface or if they do, they might not be of the correct order and are usually artifacts of the flux tube. The existence of these rational surfaces results in a unique behavior in conditions of low magnetic shear, in which even high order rational surfaces exhibit strong eddy self-interaction in the parallel direction. We will see that this can have a strong impact on turbulent transport. However, radial inhomogeneity due to rational surfaces does not arise in simulations with zero magnetic shear since the entire simulation domain is composed of surfaces with identical magnetic topology.

Returning to Eq. \eqref{eq:finite_shear_mode_coupling}, we now address the phase factor $C$ in the parallel boundary condition for the case when $\hat{s}=0$. If the magnetic shear is $\hat{s}=0$, a radial shift no longer changes the value of the phase factor and so we cannot set it to $C=1$ in all cases. Enforcing $C=1$ would implicitly assume that all surfaces within the simulation domain are lowest order (i.e. order $N_{pol}$) rational surfaces. However, a simulation should have this property only if one intends to model turbulence on surfaces where $q$ is a rational number with order $n=N_{pol}$. Otherwise setting $C=1$ imposes a nonphysical constraint. In fact, in simulations with zero magnetic shear $\hat{s}=0$, $C$ becomes a simulation parameter that allows one to model different order rational or irrational magnetic surfaces. In other words, artificially changing $C$ while keeping all other geometric parameters constant allows the study of different magnetic field topologies in isolation. 

This paper uses the phase factor $\eta=\Delta y / L_{y}$, appearing in the parallel boundary condition of the flux-tube in Eq. \eqref{eq:phase_factor_C}, as a free parameter to investigate the effects of magnetic field topology. Modifying this phase factor separately from the other coefficients that characterize the magnetic field geometry (which also depend on the safety factor) generates physically inconsistent magnetic surfaces. Note that in tokamaks, the topology of magnetic field lines is effectively determined by the safety factor. However, if linear modes are most sensitive to variations of the safety factor via the phase factor, which was the case for the linear simulations we investigated, it indicates that the primary effects originate from the actual field line topology rather than the values of the geometric coefficients through other safety factor dependencies. This suggests that topological safety factor effects can be examined by solely adjusting the phase factor. The parallel boundary phase factor has been previously studied at low and zero magnetic shear \cite{Volcokas2023, St-Onge2023}, and further investigation of its effects on linear modes and turbulence in zero magnetic shear simulations is one of the main focuses of this work. 

\begin{figure}[H]
\centering
\includegraphics[width=0.8\textwidth]{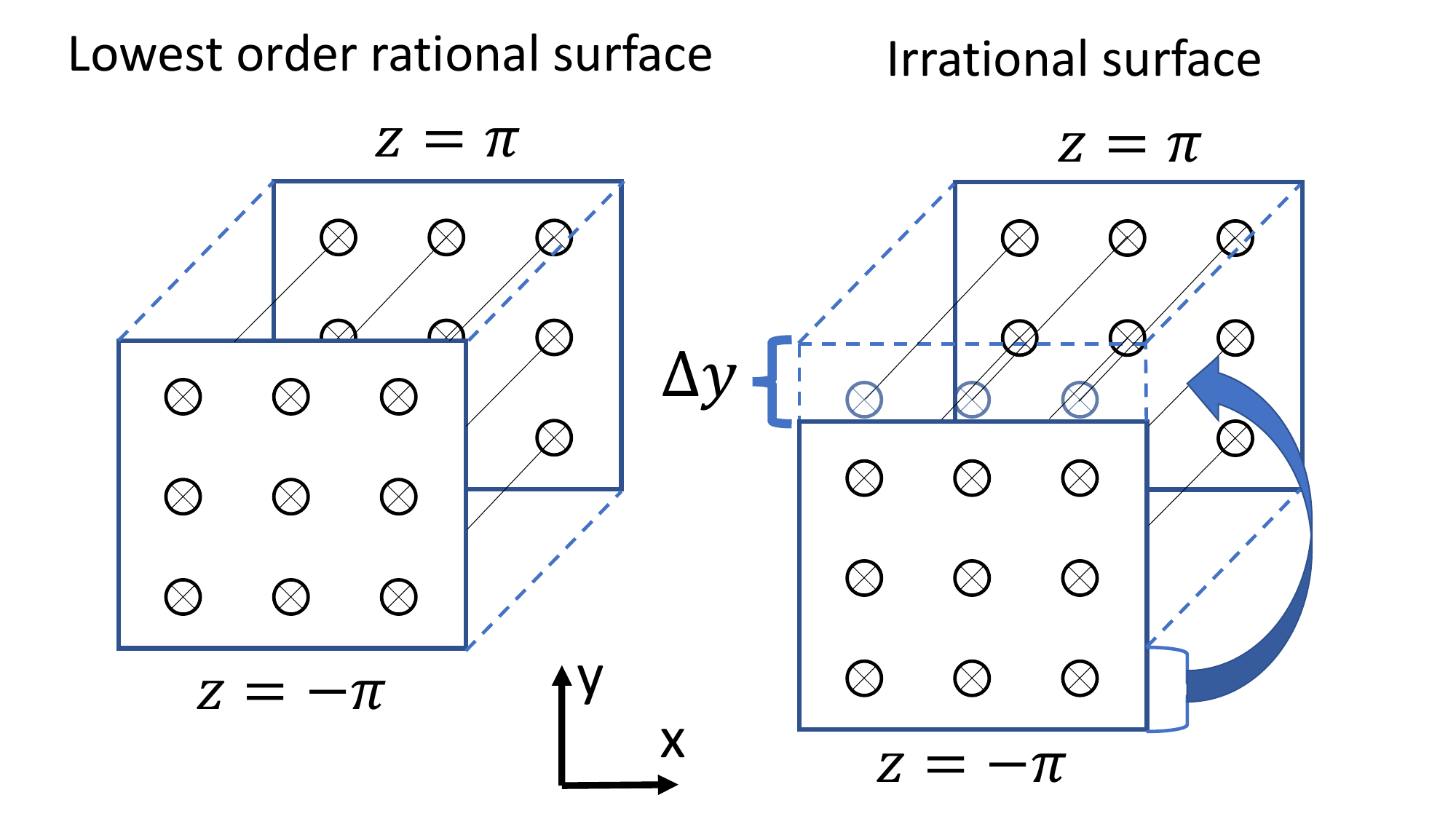}
\caption{An illustration of the phase shift at the parallel boundary for two cases: the lowest order rational surface (i.e. integer) using standard periodic boundary conditions with $\Delta y = 0$ (left) and an irrational surface featuring a phase shift of $\Delta y$ (right). In this illustration, the domain is assumed to span a single poloidal turn ($N_{pol}=1$). Crosses indicate the intersection of magnetic field lines (in black) with the inner mid-plane (blue) at $z= \pm \pi$.}
\label{fig:ParallelBoundary_170723}
\end{figure}

For the purposes of implementation, the phase factor in Eq. \eqref{eq:phase_factor_full} can be rewritten as
\begin{equation}
\label{eq:phase_factor_C}
     C = e^{i 2 \pi j \eta},
 \end{equation}
where we used $k_{y}= 2 \pi j / L_{y}, j \in Z$ and defined $\eta = \xi - NINT(\xi) \in (-0.5, 0.5]$ ($NINT$ is the nearest integer function) with $\xi =k_{y} C_{y} q_{0} N_{pol}$. Here $\eta$ controls the binormal shift of the field lines at the parallel boundary as a fraction of the box width. In this work we limit $\eta$ to $\eta \in [0, 0.5]$ as the mode frequency dependency on $\eta$ was found to be symmetric around $\eta=0$. Instead of $\eta$, one can use $\Delta y = \eta L_{y}$, the real-space shift of the field lines shift in the binormal coordinate $y$ (effectively in the toroidal direction when $z$ is fixed) after $N_{pol}$ poloidal turns as illustrated in Fig. \ref{fig:ParallelBoundary_170723}. 

With the phase factor given in Eq. \eqref{eq:phase_factor_C}, the parallel boundary condition in Eq. \eqref{eq:finite_shear_mode_coupling} for $\hat{s}=0$ becomes
\begin{equation}
\label{eq:zero_shear_parallel_boundry_phase}
       \hat{A}_{k_{x}, k_{y}}(z+2\pi N_{pol},t) =   \hat{A}_{k_{x}, k_{y}}(z,t) e^{i 2 \pi j \eta}.
\end{equation}
Implementing this phase shift for an arbitrary phase factor $\eta$ (equivalently, a $\Delta y$ shift) required a small modification to the parallel boundary conditions in GENE code and enables one to carry out both linear and nonlinear simulations with zero magnetic shear while keeping all other parameters constant. The numerical implementation of the phase factor was benchmarked against analytical results for two specific cases: (i) the parallel velocity gradient-driven instability and (ii) the ion temperature gradient-driven instability. These benchmarks were conducted in a slab geometry with adiabatic electrons and within the cold ion limit, utilizing calculations from \cite{Ball2019}. Further details and results of this benchmarking exercise can be found in Appendix \ref{appendix:eta_benchmark}.

Combining $\Delta y$ with $N_{pol}$ allows us to investigate a wider class of $\hat{s}=0$ surfaces, particularly those close to rational surfaces. Let us consider a couple of examples to illustrate how to appropriately set up $\hat{s}=0$ simulations for various physical situations. This could correspond to the minimum of a reversed shear safety factor profile or simply a region with a flat safety factor profile. For a low order rational surface $q=q_{0} \in m/n$, one wants to use $n = N_{pol}$ and $\Delta y =0$. The most straightforward situation is modeling an integer surface $q_{0}=1,2,3,$ etc. To simulate this, one should use a domain with a single poloidal turn, $N_{pol}=1$, and $\Delta y =0$ since magnetic field lines exactly close on themselves after one poloidal turn. On the other hand, a rational surface with a half-integer safety factor $q_{0}=5/2=2.5$ and $\hat{s}=0$ should use $N_{pol}=2$ and $\Delta y = 0$ since magnetic field lines exactly close on themselves after two poloidal turns (and five toroidal turns). If  $q_{0}=1251/500=2.502$ one would choose $N_{pol}=500$. However, this is only true if one is considering an infinitely large toroidal device. If the device is small enough that the magnetic field lines pass within a turbulent binormal correlation length of itself after two poloidal turns, then it is appropriate to set $N_{pol}=2$. This is because in the real device the eddy will bite its own tail (albeit with a small offset) even if the field line itself does not exactly do so. In this case, it is important to set $\Delta y$ to be non-zero and equal to the physical field line offset to capture the effects of the turbulent eddy closing on itself with an offset. Given the finite size of a physical tokamak, shearless surfaces with $q_{0}$ sufficiently close to a rational value should be simulated with a specific non-zero $\Delta y$. The details for correctly choosing $N_{pol}$ and $\Delta y$ for an arbitrary geometry is complicated and will be discussed in more detail in section \ref{section:combined_safety_factor}. \par

\subsection{Numerical setup}

For our numerical study, we use the local flux tube version of the GENE code. The numerical results presented in this work were obtained through collisionless electrostatic simulations using the local Miller equilibrium \cite{Miller1998}. We considered both kinetic and adiabatic electron models, and their comparison was made where appropriate. Our findings indicate that in most cases of interest, the adiabatic electron treatment is insufficient to accurately model self-interaction, necessitating a full kinetic treatment for both ions and electrons. As mentioned earlier, the focus of this work is on the study of standard ion temperature gradient (ITG) turbulence. To this end, we employed three different sets of driving gradients, as outlined in Table \ref{tab:compare_drive_cases}: the Cyclone Base Case (CBC) \cite{Dimits2000} that has finite density and temperature gradients for all particle species, a pure ITG (referred to as pITG) case with zero density gradients and electron temperature gradient, and an intermediate case (limited to linear simulations and referred to as 0.5CBC) where the density gradients and electron temperature gradient are halved with respect to the CBC case. We chose to study the CBC and pITG case extensively with nonlinear simulations as they were found to be linearly dominated by different ITG instability branches and present considerably different behavior.

\begin{table}[H]
\begin{tabular}{l|l|l|l}
& Cyclone base case & Pure ITG case & Intermediate case \\
\hline
Acronym & CBC & pITG & 0.5CBC \\
\hline
$\omega_{T,i}$ & 6.96 & 6.96 & 6.96 \\
\hline
$\omega_{T,e}$ & 6.96 & 0 & 3.48 \\
\hline
$\omega_{N}$ & 2.22 & 0 & 1.11
\end{tabular}
\caption{Driving gradients of different ITG turbulence cases. $\omega_{T,i}=-(R/T_{i})(dT_{i}/dx)$ is the normalized ion temperature gradient, $\omega_{T,e}=-(R/T_{e})(dT_{e}/dx)$ is the normalized electron temperature gradient and $\omega_{N}=-(L_{ref}/N)(dN/dx)$ is the normalized density gradient, where $R$ is the major radius.}
\label{tab:compare_drive_cases}
\end{table}

\section{Linear study}

\subsection{Linear modes at finite magnetic shear} \label{section:LinearFiniteShear}
It is expected that toroidal ITG modes (also referred to as curvature ITG modes) will play a dominant role in core plasmas of magnetic fusion devices and be responsible for the majority of energy transport in that region. This is indeed true for most large tokamak experiments operating today and will likely be the case in future devices, where ion species will ideally have higher temperatures than electrons. In this section, we will study how self-interaction affects such modes and demonstrate the different behavior of two toroidal ITG mode branches in linear gyrokinetics simulations with kinetic electrons and low magnetic shear $\hat{s} \lesssim 0.1$. 

We will primarily focus on cases with very low magnetic shear, where the most unstable mode transitions from a $k_{z} \neq 0$ toroidal ITG branch to a $k_{z} \simeq 0$ branch as magnetic shear approaches $\hat{s}=0$. Here $k_{z}$ denotes the dominant mode wavenumber along the magnetic field line and is defined through weighted average as
\begin{equation}
\label{eq:kz_definition}
    k_{z} = \left[ \frac{\sum_{k_{x}} \int |\partial_{z} \phi_{k_{x}}(z)|^{2} J_{xyz} dz}{\sum_{k_{x}} \int | \phi_{k_{x}}(z)|^{2} J_{xyz} dz} \right]^{1/2},
\end{equation}
where $\phi_{k_{x}}$ are the $k_{x}$-Fourier coefficients of the electrostatic potential $\phi$. Linear simulations with finite shear $\hat{s} \neq 0$ consider a mode with a single fixed $k_{y}$ and a set of coupled $k_{x}$, while in the case of zero magnetic shear $\hat{s}=0$ one considers only a single fixed $k_{x}$. As we will see, this investigation of the two different toroidal ITG modes at low magnetic shear will give useful insight into the nonlinear simulations presented in the second half of this paper. \par

\begin{figure}[H]
\centering
    \includegraphics[width=0.7\textwidth]{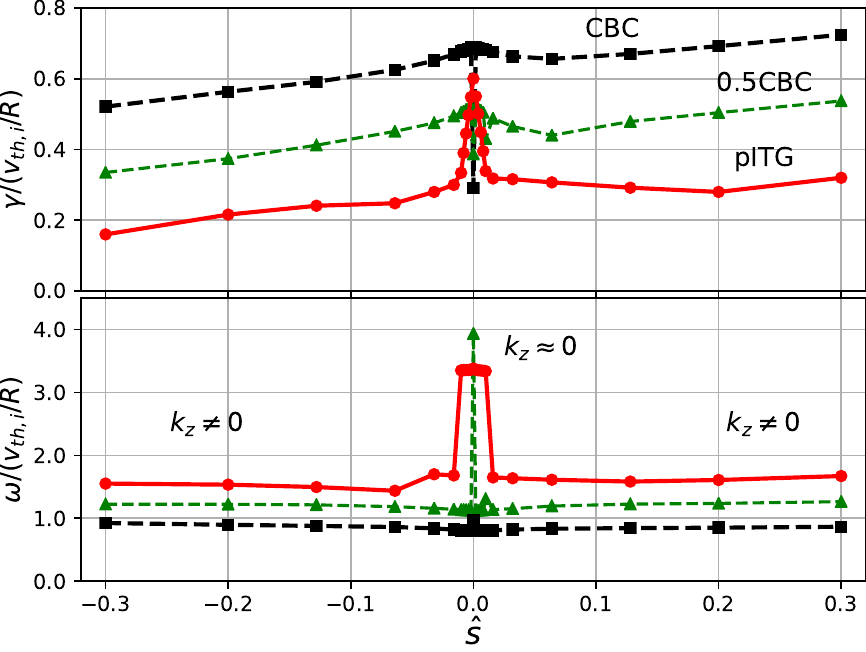}
\caption{The normalized linear mode growth rate $\gamma$ (top) and real frequency $\omega$ (bottom) with $\hat{s}$ for CBC (black squares), 0.5CBC (green triangles) and pITG (red circles) parameters. The binormal wavenumber is $k_{y} \rho_{i} = 0.45$, and the radial wavenumber is $k_{x} \rho_{i} = 0$. Here $v_{th,i}=\sqrt{T_{i}/m_{i}}$ is the ion thermal velocity. These simulations were performed using the physical and numerical parameters shown in Table \ref{tab:parameters_linear_miller} of Appendix \ref{appendix:simulation_parameters}, in particular $N_{pol}=1$ and $\eta = 0$.}
\label{fig:IoP_LinModeFreqVSshear}
\end{figure}

\begin{figure}[H]
\centering
    \includegraphics[width=0.7\textwidth]{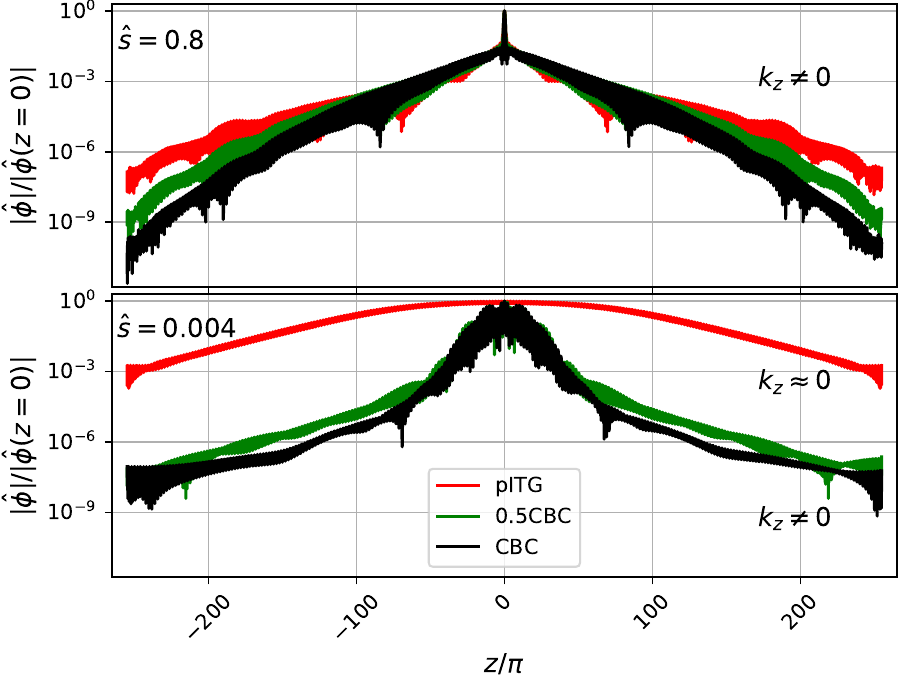}
\caption{The ballooning representation of the normalized electrostatic potential $\hat{\phi}$ for linear modes from simulations with $\hat{s}=0.8$ (top) and $\hat{s}=0.004$ (bottom) for CBC (black), 0.5CBC (green) and pITG (red) parameters. The binormal wavenumber is $k_{y} \rho_{i} = 0.45$, and the radial wavenumber is $k_{x} \rho_{i} = 0$. These simulations were performed using the physical and numerical parameters shown in Table \ref{tab:parameters_linear_miller} of Appendix \ref{appendix:simulation_parameters}.}
\label{fig:IoP_LinModeBallooningAngleVSShear_220223}
\end{figure}

A significant change in the linear mode structure is observed as the magnetic shear is decreased from a standard value of $\hat{s}=0.8$ to $\hat{s}=0$. Fig. \ref{fig:IoP_LinModeFreqVSshear} illustrates this transition, which is between different toroidal ITG mode branches. The most obvious indicator of this transition is the discontinuity in the real frequency $\omega$, as the $k_{z} \simeq 0$ mode exhibits a much higher value compared to the $k_{z} \neq 0$ branch. Additionally, as expected the $k_{z} \simeq 0$ mode is very extended in ballooning space compared with the $k_{z} \neq 0$ mode and has orders of magnitude larger amplitude in the tails as compared to the value at $z=0$ (as shown in Fig. \ref{fig:IoP_LinModeBallooningAngleVSShear_220223}). For $\hat{s}=0.8$ the most unstable mode lies on the $k_{z} \neq 0$ branch for all different sets of driving gradients. At $\hat{s}=0.004$ only the pITG case has undergone a transition from the $k_{z} \neq 0$ to $k_{z} \simeq 0$ toroidal ITG branch. For CBC and 0.5CBC the most unstable mode still has $k_{z} \neq 0$. Interestingly, the growth rate of the $k_{z} \simeq 0$ mode around $\hat{s}=0$ decreases when the density and electron temperature gradients are increased, indicating a stabilizing effect of these gradients on the $k_{z} \simeq 0$ mode. Special attention must be paid to the $\hat{s}=0$ point for the CBC-like case. If the flux tube length is set to $N_{pol}=1$ as is the case in Fig. \ref{fig:IoP_LinModeFreqVSshear}, there is an apparent discontinuity in the growth rate at zero shear. However, if $N_{pol}$ is increased until convergence, the growth becomes continuous at $\hat{s}=0$. This shows that the discontinuity at $\hat{s}=0$ for $N_{pol}=1$ is a consequence of linear self-interaction and demonstrates the importance of correctly treating the magnetic topology, which will be discussed in more detail in the next section. Moreover, it suggests that linear self-interaction has a stabilizing effect.

\begin{figure}[H]
\centering
    \includegraphics[width=0.7\textwidth]{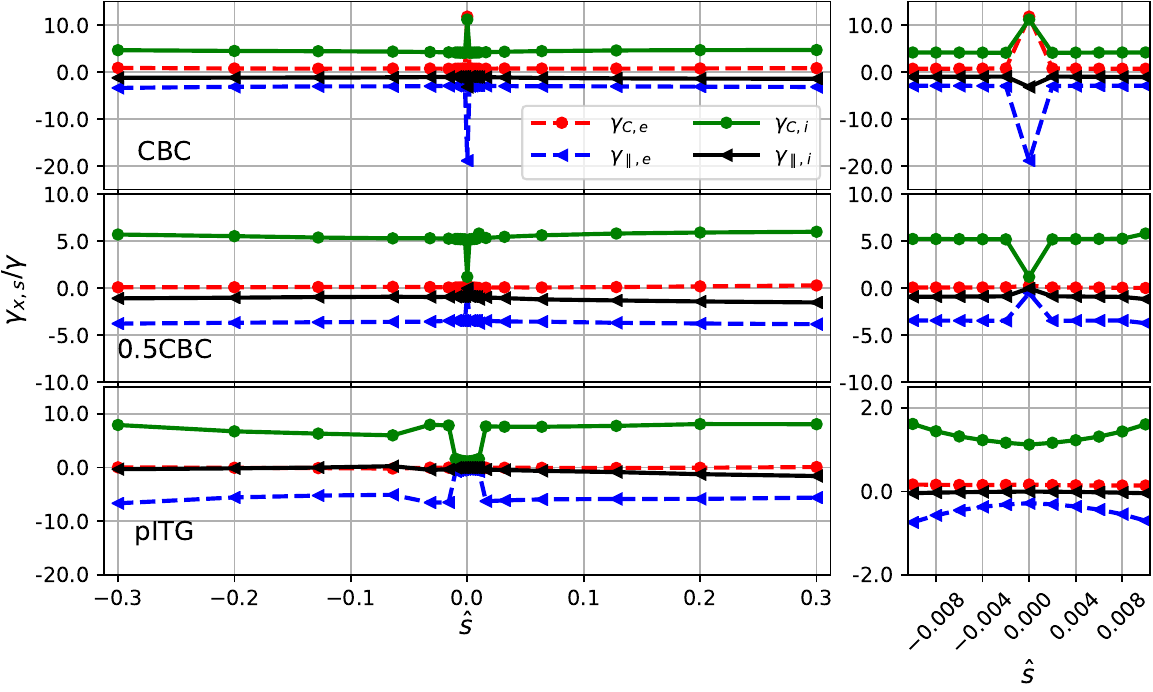}
\caption{Different contributions $\gamma_{x,s}$  to the growth rate normalized to the total growth rate $\gamma$ for CBC (top), 0.5CBC (middle) and pITG (bottom) parameters using $k_{y} \rho_{i} = 0.45$ and $k_{x} \rho_{i} = 0$. The subplots on the right show a zoom-in around $\hat{s}=0$. Curvature contributions from ions and electrons are plotted in green and red respectively, while contributions from parallel streaming are plotted in black and blue respectively. These simulations were performed using numerical and physical parameters shown in Table \ref{tab:parameters_linear_miller} of Appendix \ref{appendix:simulation_parameters}.}
\label{fig:IoP_LinModeEnergyVSShear}
\end{figure}

Additional insight into the different modes can be obtained by using the gyrokinetic free energy balance equation and evaluating the contributions to the growth rate $\gamma$ from the different linear terms. A detailed discussion of free energy balance in the gyrokinetic model is provided in reference \cite{BanonNavarro2011}. An evaluation of the growth rate $\gamma$ from the time evolution of the system's potential energy $E_{\phi}$ is shown in reference \cite{Manas2015} and is given by the following expression
\begin{equation}
    \gamma = \frac{1}{2}\frac{1}{E_{\phi}}\frac{\partial E_{\phi}}{\partial t} = \frac{1}{2} \frac{1}{\sum_{s} E_{\phi,s}}\sum_{s}\frac{\partial E_{\phi,s}}{\partial t},
\end{equation}
where $E_{\phi,s}$ is the contribution to the potential energy from species $s$. We can further separate the different contributions to the growth rate from the parallel streaming $\mathcal{L}_{\parallel,\phi}$, combined curvature and $\nabla \mathbf{B}$ drifts $\mathcal{L}_{C,\phi}$ and dissipation $\mathcal{D}_{\phi}$ terms in the following way
\begin{equation}
    \gamma = \frac{1}{2} \frac{1}{ E_{\phi}}\sum_{s} (\mathcal{L}_{\parallel, \phi,s}+\mathcal{L}_{C,\phi,s}-\mathcal{D}_{\phi,s}) = \sum_{s}(\gamma_{\parallel, s} + \gamma_{C,s} - \gamma_{D,s}),
\end{equation}
where the terms in the final expression correspond to the contributions of the different species to the overall growth rate. For species $s$, $\gamma_{C,s}$ is the part of the growth rate arising from the curvature term, and $\gamma_{\parallel, s}$ is the part from the parallel term. The dissipation term $\gamma_{D,s}$, resulting from collisions and numerical dissipation, is small compared to the other terms in the simulations we are considering. As shown in Fig. \ref{fig:IoP_LinModeEnergyVSShear}, for all values of magnetic shear and for all our parameter sets, the mode is destabilized due to the curvature term mainly from the ions. The parallel streaming terms, mainly from the electrons, provide only a stabilizing effect. Thus, we conclude that both modes are toroidal ITG. It is worth pointing out that in all cases the total growth rate $\gamma$ results from a near cancellation of $|\gamma_{C,i}|$ and $|\gamma_{\parallel,e}|$. 

In contrast, simulations using the adiabatic electron model do not exhibit a mode transition. The dominant mode has a large $k_{z} = 0$ component and its extent in ballooning space increases as magnetic shear is lowered, but nonetheless remains much less extended than the $k_{z} \simeq 0$ toroidal ITG mode when using kinetic electrons. The difference in behavior between the adiabatic and kinetic electron models results from the "giant tails" present in kinetic electron simulations \cite{Hallatschek2005}, which enables the mode to extend much further along the magnetic field lines. These observations align with previous studies \cite{Dominski2015, Ajay2020}, which have demonstrated the critical importance of kinetic electron physics around rational surfaces for standard shear values.  Hence, in accordance with the literature and our expectations, kinetic electron dynamics has a significant effect on mode structures and their stability at low magnetic shear and therefore must be accounted for when simulating low $\hat{s}$ conditions.

In summary, linear gyrokinetic simulations show a transition between different toroidal ITG modes as the magnetic shear approaches $\hat{s}=0$. Specifically, a toroidal ITG mode with $k_{z} \neq 0$ transitions to a toroidal ITG $k_{z} \simeq 0$ mode as the magnetic shear is lowered\footnote{In reference \cite{Volcokas2023}, this transition was inaccurately described as change from a toroidal ITG mode to a slab ITG modes. However, this terminological inaccuracy does not impact the overall conclusions derived from the study.}. A kinetic electron treatment is necessary to observe this. Additionally, a large number of connected $k_{x}$ modes (equivalently $N_{x} \geq 128$ of radial grid points) was needed to properly resolve the $k_{z} \simeq 0$ mode\footnote{In the light of our study, one should take care when performing resolution studies in radial grid spacing for linear simulations at low magnetic shear. The mode growth rate might seem to be saturated as the corresponding $k_{z} \neq 0$ mode is well resolved, but then increasing the radial resolution further is needed to properly resolve and destabilise the $k_{z} \simeq 0$ mode, which then becomes the most unstable mode. Otherwise, the $k_{z}=0$ mode can be artificially suppressed.}. 
The mode transition is characterized by a sharp mode broadening in ballooning space, a rapid but continuous increase in the growth rate and a discontinuous increase in the real frequency. We will later show that this transition between modes at low magnetic shear can have a substantial effect on the nature of the nonlinear turbulence that they drive. Additionally, this rapid change in the growth rate from the mode transition could be a plausible triggering mechanism for the formation of an ITB around low-order rational surfaces. \par

\subsection{Linear modes at zero magnetic shear} \label{section:LinearZeroShear}

The transition to the $k_{z} \simeq 0$ mode as $\hat{s} \rightarrow 0$, characterized by a significant lengthening in ballooning space, suggests that self-interaction will play a more significant role at low magnetic shear. In this section, we will focus on shearless simulations, i.e. with $\hat{s}=0$, allowing us to address the behavior of modes at the minimum safety factor in equilibrium with a reverse $q$ profile. More specifically, we will investigate how different magnetic field topology affects these modes by (i) changing the parallel domain length $N_{pol}$ and (ii) changing the field line offset through the parallel boundary using the binormal shift $\Delta y$.

\subsubsection{Effects of the parallel domain length on linear modes with $\hat{s}=0$}

In the case of $\hat{s}=0$ simulations, we recall that one uses a periodic parallel boundary condition (instead of twist-and-shift), implying that a given $(k_{x}, k_{y})$-Fourier mode couples to itself according to Eq. \eqref{eq:zero_shear_parallel_boundry_phase}. This means that a single radial Fourier mode $k_{x}$ is sufficient to represent an eigenmode in the system. Thus, increasing the number of radial modes in the simulation no longer extends an eigenmode as is the case for $\hat{s} \neq 0$. Instead, an eigenmode can be "given more space" only by increasing the length of the domain by considering a larger number of poloidal turns via $N_{pol}$. Hence, in some cases, to have a continuous growth rate across $\hat{s}=0$ it is necessary to have multiple poloidal turns at $\hat{s}=0$ to eliminate parallel self-interaction. This is the case for the CBC simulations with $\hat{s}=0$ in Fig. \ref{fig:IoP_LinModeFreqVSshear}. Physically interpreting the discontinuity seen in Fig. \ref{fig:IoP_LinModeFreqVSshear} is non-trivial since it is well justified to have a zero shear simulation with a single poloidal turn if one is simulating an integer surface, for example, a $q=2$ magnetic surface at the minimum of a non-monotonic safety factor profile. Therefore, we can use $N_{pol}$ parameter in two different ways. First, it can be used as a physical parameter to set the order of rational surfaces being simulated (e.g. $q=2/1$, $q=5/2$ or $q=7/3$ requires $N_{pol}=1$, $N_{pol}=2$ and $N_{pol}=3$ respectively). Second, $N_{pol}$ can be used as a purely numerical parameter to approximate low but finite shear $\hat{s} \simeq 0$ surfaces with exactly zero shear $\hat{s} = 0$, in which case it should be set sufficiently large to achieve convergence.

The periodic boundary condition in $\hat{s}=0$ simulations, combined with a single $(k_{x},k_{y})$-Fourier mode and $N_{pol} > 1$, aligns our study with concepts often seen in condensed matter physics, such as Floquet-Bloch's theorem and Brillouin zones. We outline the adaptation of Floquet-Bloch's theorem to linear gyrokinetics in Appendix \ref{appendix:Floquet_Bloch}. Utilizing these findings, the distribution function $f(z)$ is expressed as
\begin{equation}
    \label{eq:dist_func_s0}
    f(z) =F(z) e^{iK_{z}z},
\end{equation}
where $F(z)$ is a $2 \pi$ periodic function $F(z)=F(z+2 \pi)$ (periodic over single poloidal turn) and $K_{z}$ is a wave vector. The wave vector is defined as 
\begin{equation}
\label{eq:Kz_equation}
    K_{z}=\frac{p+\eta}{N_{pol}},
\end{equation}
where $p \in \{ -\lfloor (N_{pol}-1)/2 \rfloor, ..., \lfloor N_{pol}/2 \rfloor \}$ and $\lfloor ... \rfloor$ is the floor function. Here $p$ is a free integer parameter and the $p$ value that appears in simulations is the one that leads to the highest growth rate. It is important to make a distinction between $k_{z}$ and $K_{z}$: $k_{z}$ represents an average parallel wavenumber that takes into account both $2 \pi$ periodic fluctuations present in $F_{z}(z)$ and the longer wavelength modulations due to the wave vector $K_{z}$. Therefore, the $k_{z} \simeq 0$ toroidal ITG branch is dominated by the $K_{z}=0$ wave vector in Eq. \eqref{eq:dist_func_s0}.

\begin{figure}[H]
\centering
\begin{subfigure}{0.45\textwidth}
    \centering
    \includegraphics[width=1\textwidth]{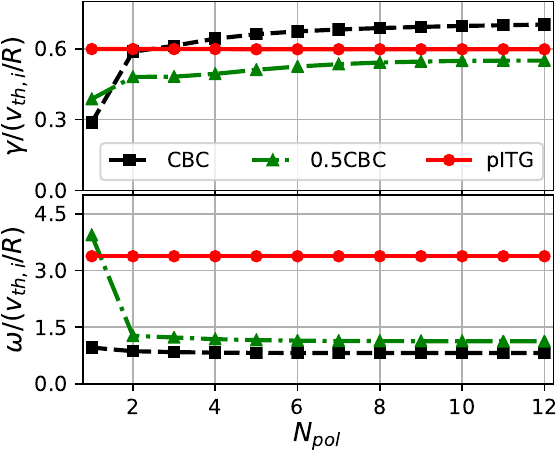}
    \caption{}
\end{subfigure}
\begin{subfigure}{0.465\textwidth}
    \centering
    \includegraphics[width=1\textwidth]{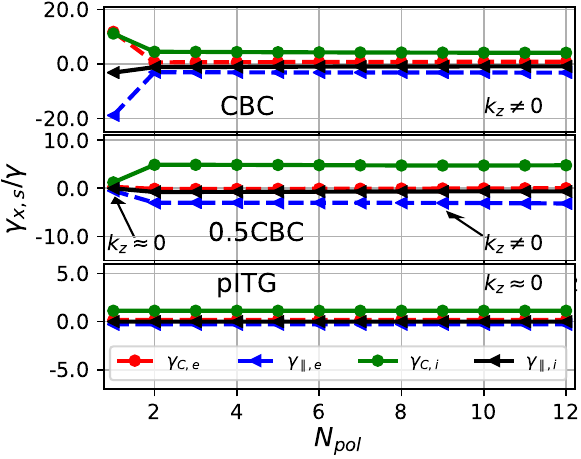}
    \caption{}
\end{subfigure}
\caption{The dependence on $N_{pol}$ of (a) the linear growth rate (top) and real frequency $\omega$ (bottom) and (b) different contributions to the total growth rate normalized to the total growth rate $\gamma$ for $\hat{s}=0$, $\Delta y = 0$ (i.e. $\eta = 0$), $k_{y} \rho_{i} = 0.45$, $k_{x} \rho_{i} = 0$ and various driving gradients. These simulations were performed using the physical and numerical parameters shown in Table \ref{tab:parameters_linear_npolScan_miller} of Appendix \ref{appendix:simulation_parameters}.}
\label{fig:IoP_LinModeEnergandFreqVSNpol}
\end{figure}

Considering the standard periodic boundary condition with $\eta = 0$, we see from Eq. \eqref{eq:Kz_equation} that if the most unstable mode for $N_{pol}>1$ has $p \neq 0$, it will not be allowed in a system with $N_{pol} = 1$. This suggests a strong dependence of the mode on $N_{pol}$. Fig. \ref{fig:IoP_LinModeEnergandFreqVSNpol} (a) confirms that linear mode stability can be significantly influenced by the length of the parallel domain though not for all driving gradients. For the pITG case, where the $k_{z} \simeq 0$ mode dominates, the growth rate is independent of the parallel domain length, as the mode with $p=0$ is the most unstable mode for all $N_{pol}$ and the limited mode poloidal variation is not changed by an increase in the parallel domain length. In contrast, for CBC, the dominant mode for long domains is $k_{z} \neq 0$ (i.e. $p \neq 0$) but this mode is stabilized as $N_{pol}$ decreases. At $N_{pol}=1$ the growth rate is a factor of two lower than the growth rate at larger $N_{pol}$. However, if at $N_{pol}=1$ the mode is constrained to a phase value that maximizes growth rate, i.e. if $\eta$ corresponds to optimal $p/N_{pol}$ in the large $N_{pol}$ limit, its stability is unaffected by the domain length. Otherwise, in an $N_{pol}=1$ simulation, the mode cannot achieve its optimal $K_{z}$ (equivalently $\eta$). It must adjust to the limited domain, which reduces its growth rate. This adjustment also alters the balance of linear terms contributing to it. These effects are illustrated in Fig. \ref{fig:IoP_LinModeEnergandFreqVSNpol} (b). Notably, even in CBC cases with $N_{pol}=1$, the growth rates of $k_{z} \simeq 0$ and $k_{z} \neq 0$ branches are comparable. Altering parameters like $k_{y}$, $k_{x}$, or the inverse aspect ratio $\epsilon = a/R$ can facilitate a transition between these branches.

In summary, our linear simulations with $\hat{s}=0$ and $\eta = 0$ show two distinct toroidal ITG mode branches: one with $k_{z} \simeq 0$ and the other with $k_{z} \neq 0$. The growth rate of the first is unaffected by the parallel domain length, while the $k_{z} \neq 0$ mode is stabilized when the domain is short. This is because the phase enforced by the parallel boundary conditions ($\eta= \Delta y / L_{y} = 0$) prevents the $k_{z} \neq 0$ mode from having its preferred $K_{z}$ value. The identification of these two toroidal ITG modes will help us to interpret nonlinear simulations results with $\hat{s}=0$.

\subsubsection{Effects of the parallel boundary phase factor on linear modes with $\hat{s}=0$}

In numerical simulations with $\hat{s} \simeq 0$, the presence of extremely extended modes in ballooning space underlines the strong dependence of the growth rate on $N_{pol}$. This observation highlights the importance of carefully considering the parallel boundary condition and appropriately choosing the parallel boundary shift $\Delta y$ (equivalently $\eta$) for $\hat{s} = 0$ simulations. As previously mentioned, incorporating a phase factor $C=\exp{(i k_{y} \Delta y)}$ (here in terms of $ \Delta y$) into the parallel boundary condition as in Eq. \eqref{eq:phase_factor_C} provides an additional degree of freedom in constructing the topology of the numerical simulations. More precisely, the phase factor is needed to accurately model all possible values of the safety factor. Moreover, the phase factor allows for a detailed exploration of self-interaction in both linear and nonlinear simulations. In linear simulations, adjusting the phase factor affects only how the mode connects with itself across the parallel boundary, effectively enabling Fourier modes along the $z$ direction that are non-integer harmonics of the parallel domain length. This is illustrated for the reduced analytical models in Appendix \ref{appendix:eta_benchmark} and in general in Appendix \ref{appendix:Floquet_Bloch}. Surprisingly, the parallel phase factor in zero magnetic shear flux tube simulations has received relatively little attention within the community, with only a couple of recent works studying its impact \cite{Volcokas2023, St-Onge2023}.

\begin{figure}[H]
\centering
\includegraphics[width=0.7\textwidth]{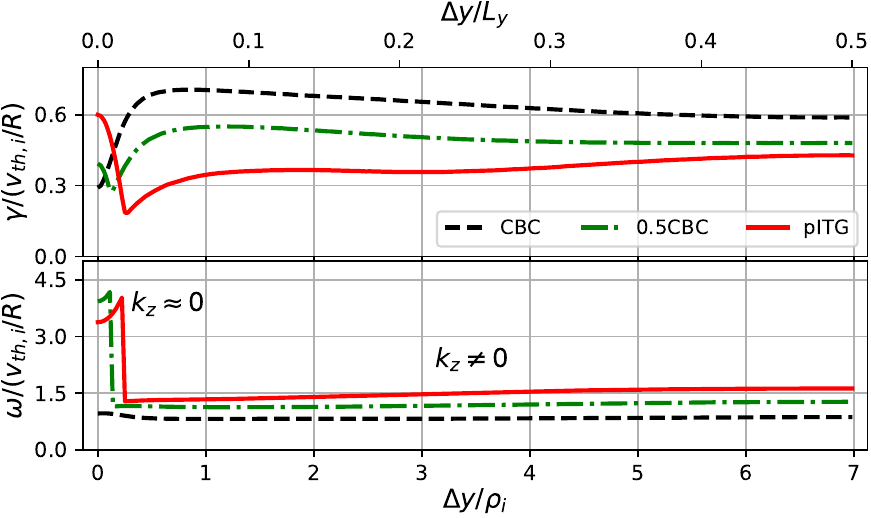}
\caption{The linear growth rate $\gamma$ (top) and real frequency $\omega$ (bottom) as a function of $\Delta y$ (or $\eta = \Delta y / L_{y}$) for $\hat{s}=0$, $N_{pol} = 1$, $k_{y} \rho_{i} = 0.45$, and $k_{x} \rho_{i} = 0$ and various driving gradients. These simulations were performed using the physical and numerical parameters shown in Table \ref{tab:parameters_linear_npolScan_miller} of Appendix \ref{appendix:simulation_parameters}.}
\label{fig:IoP_LinModeFreqVSEta}
\end{figure}

At zero magnetic shear, the growth rate of the mode strongly depends on the parallel boundary shift $\Delta y$ when the parallel domain length is short (i.e., a few poloidal turns or less). This is illustrated by Fig. \ref{fig:IoP_LinModeFreqVSEta} for the case of $N_{pol}=1$. Again, the behavior of the mode with $\Delta y$ depends on the driving gradients. With the pITG drive, the $k_{z} \simeq 0$ mode has a very sharp and narrow peak in the growth rate around $\Delta y=0$, but is stabilized at larger $\Delta y \simeq 0.3 \rho_{i}$ values. It is replaced by the $k_{z} \neq 0$ mode due to the phase factor enforcing a non-zero parallel wavenumber according to Eq. \eqref{eq:Kz_equation}. In the 0.5CBC case, i.e. with a drive between CBC and pITG, the dominant mode around $\Delta y =0$ is again the $k_{z} \simeq 0$ mode, but with a narrower and smaller peak in the growth rate compared to the pITG case. Finally, with the CBC drive, the $k_{z} \neq 0$ mode always dominates, even if it is strongly stabilized around $\Delta y=0$. For 0.5CBC and CBC, the $k_{z} \neq 0$ mode has a maximum growth rate at a non-zero $\Delta y$ value. This corresponds to the most unstable $k_{z}$ value in $N_{pol} \gg 1$ limit as changing the phase factor effectively scans all the possible $k_{z}$ values by changing $K_{z}$ according to Eq. \eqref{eq:Kz_equation}. In other words, when $\Delta y = 0$ the domain allows only modes with $k_{z}=2 \pi N/L_{z}$ where $N \in \mathbb{Z}$ and $L_{z}$ is the parallel domain length. However, as shown in appendix \ref{appendix:Floquet_Bloch}, finite $\Delta y$ shifts the allowed $k_{z}$ wavenumbers by changing $K_{z}$ and for CBC gradients such shifted wavenumbers are the most unstable modes. 

It is crucial to highlight that modifying certain physical parameters can alter the shape of the curves in Fig. \ref{fig:IoP_LinModeFreqVSEta}. For example, increasing the safety factor value, which leads to changes in the geometric coefficients, broadens the peak around $\Delta y=0$ in the growth rate for the pITG case, resulting in the stabilization of the $k_{z} \simeq 0$ mode only at larger $\Delta y$ values. This indicates that the shape of these growth rate curves depends on the flux tube geometry. Furthermore, it is also broadened by increasing the electron mass. Specifically, the width of the $k_{z} \simeq 0$ mode around $\Delta y =0$ is proportional to the square root of electron-to-ion mass ratio $\sqrt{m_{e}/m_{i}}$. When the mass ratio is increased by considering heavier electrons, the width of the growth rate peak around $\Delta y=0$ becomes wider for both the pITG and 0.5CBC cases. This dependence of the width of the growth rate peak hints at a relationship with the ability of the electrons to transfer information no faster than their thermal velocity $v_{th,e}=\sqrt{T_{e}/m_{e}}$. When the speed at which information travels is reduced (e.g. due to increased mass $m_{e}$, reduced temperature $T_{e}$ or $q_{0}$ through a single poloidal turn connection length $\propto q_{0} R$) at fixed domain length, the parallel boundary conditions matter less, and the effects of the phase factor diminish. This is analogous to increasing the parallel length $L_{z} = 2 \pi N_{pol}$ of the domain while keeping electron thermal velocity fixed. 

In fact, there is a direct relationship between the mode behaviour obtained when varying the parallel domain length $N_{pol}$ (as in Fig. \ref{fig:IoP_LinModeEnergandFreqVSNpol}) and the parallel boundary phase factor via $\Delta y$ (as in Fig. \ref{fig:IoP_LinModeFreqVSEta}). This is again based on the Floquet-Bloch theorem outlined in Appendix \ref{appendix:Floquet_Bloch}. It means that, given the dependence of the linear modes with $\Delta y$ for $N_{pol} = 1$, we can determine their dependence on $N_{pol}$ for arbitrary phase factor. Specifically, the growth rate $\gamma(\Delta y, N_{pol})$ of a linear mode in a flux tube with length $L_{z} = 2 \pi N_{pol}$ is concisely expressed as
\begin{equation}
\label{eq:gamma_deltay_relationship}
    \gamma(\Delta y, N_{pol}) = \max_{p \in \mathbb{Z}} \left[ \gamma(\frac{\Delta y + p L_{y}}{N_{pol}}, 1) \right],
\end{equation}
where $\Delta y$ is the shift at the parallel boundary after $N_{pol}$ poloidal turns, $p \in \{ -\lfloor (N_{pol}-1)/2 \rfloor, ..., \lfloor N_{pol}/2 \rfloor \}$ and the maximum growth rate for all available $p$ values is chosen. Notably, Fig. \ref{fig:IoP_LinModeEnergandFreqVSNpol} illustrates $\gamma(0, N_{pol})$ and Fig. \ref{fig:IoP_LinModeFreqVSEta} shows $\gamma(\Delta y, 1)$. This indicates that in a system with $N_{pol}$ poloidal turns and a constant $\Delta y$, the mode will choose from the possible values of $K_{z} =(\Delta y/L_{y} + p)/N_{pol}$ to maximize the growth rate. This maximized growth rate corresponds to the fastest growing mode in simulations with a single poloidal turn and a phase factor $\Delta y|_{N{pol}=1}= (\Delta y+pL_{y})/N_{pol}$.

\begin{figure}[H]
\centering
\includegraphics[width=0.7\textwidth]{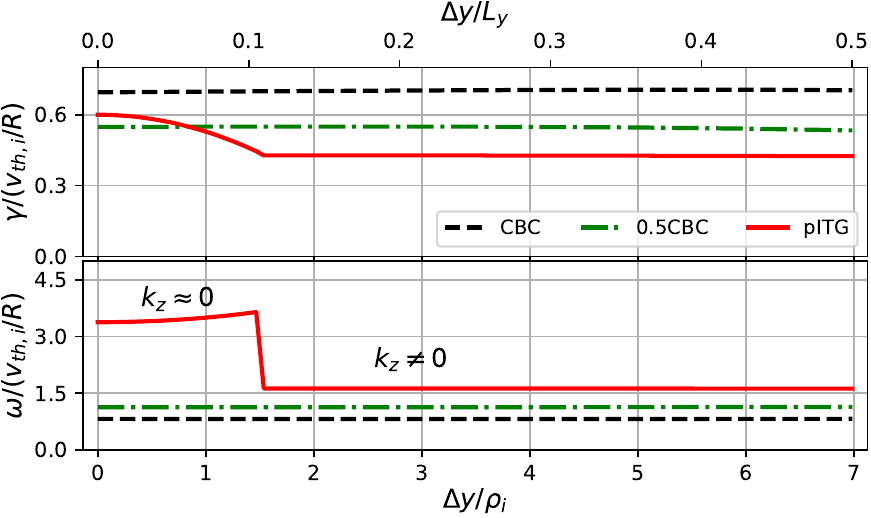}
\caption{The normalized linear growth rate $\gamma$ (top) and real frequency $\omega$ (bottom) as a function of $\Delta y$ for $\hat{s}=0$, $N_{pol} = 10$, $k_{y} \rho_{i} = 0.45$, and $k_{x} \rho_{i} = 0$ and various driving gradients. These simulations were performed using the physical and numerical parameters shown in Table \ref{tab:parameters_linear_npolScan_miller} of Appendix \ref{appendix:simulation_parameters}.}
\label{fig:IoP_LinModeFreqVSEta_Npol10_030323}
\end{figure}

To further illustrate the relationship described by Eq. \eqref{eq:gamma_deltay_relationship}, we present a $\Delta y$ scan in a long domain of $N_{pol}=10$ in Fig. \ref{fig:IoP_LinModeFreqVSEta_Npol10_030323}.  It is important to understand that each simulation with $N_{pol} = 10$ can be conceptualized as ten separate, but identical simulations, each with $N_{pol} = 1$ and $\Delta y|_{N{pol}=1}$. In this arrangement, the cumulative phase shift across all ten $N_{pol}=1$ simulations should equal the total phase shift $\Delta y$ required in the $N_{pol}=10$ simulation, adjusted modulo $L_{y}$. This requirement can also be seen from the relationship in Eq. \eqref{eq:gamma_deltay_relationship}. The freedom in selecting the integer $p$ allows, in this case, for 10 different $\Delta y|_{N{pol}=1}$ values. The mode that dominates in the $N_{pol}=10$ simulation, is the one with the largest growth rate among the $N_{pol}=1$ simulations with one of the 10 allowed $\Delta y|_{N{pol}=1}$ values. For the CBC and 0.5CBC simulations in Fig. \ref{fig:IoP_LinModeFreqVSEta_Npol10_030323}, the integer $p$ can always be chosen such that the corresponding $\Delta y|_{N{pol}=1}$ value is close to the value ($\Delta y \simeq \rho_{i}$) with the maximum growth rate in the $N_{pol}=1$ simulations regardless of $\Delta y$. However, for pITG drive, an even longer domain than $N_{pol} = 10$ would be required to eliminate the dependence on $\Delta y$ since the $k_{z} \simeq 0$ mode is highly sensitive to the imposed phase factor. Since the highest growth rate for pITG simulation with $N_{pol} = 1$ occurs in a very narrow region of width $\sim 0.1 \rho_{i}$ around $\Delta y = 0$ (as shown in Fig. \ref{fig:IoP_LinModeFreqVSEta}), $p$ cannot be easily adjusted to always achieve a growth rate close to the maximum, even when $N_{pol}=10$. Instead the $k_{z} \simeq 0$ peak around $\Delta y = 0$ broadens in the $N_{pol}=10$ simulations compared to $N_{pol}=1$ simulations. Hundreds of poloidal turns would be necessary for the system to be able to access a mode that is close to the peak at all values of $\Delta y$. These results demonstrate that as the simulation domain becomes longer, the mode will become increasingly independent of the phase factor imposed at the parallel boundary, showing once again that the phase factor only has a significant impact in short domains (i.e. on magnetic surfaces close to low order rational surfaces).

\subsubsection{Effects of the parallel boundary phase factor with the safety factor on linear modes with $\hat{s}=0$} \label{section:combined_safety_factor}

Reference \cite{Joffrin2002a} provides examples of experimental equilibria with flat or reversed safety factor profiles. These scenarios can be modeled using appropriate flux tube simulation domains. This can be achieved by correctly combining $N_{pol}$ and $\Delta y$ to faithfully represent the magnetic field line topology for a given value of the safety factor. We will now explicitly connect $N_{pol}$ and $\Delta y$ with the safety factor $q$ and show how the previous sections imply an extreme sensitivity of $\hat{s}=0$ simulations to the actual $q$ value.

\begin{figure}[H]
\centering
\begin{subfigure}{0.3\textwidth}
    \centering
    \includegraphics[width=1\textwidth]{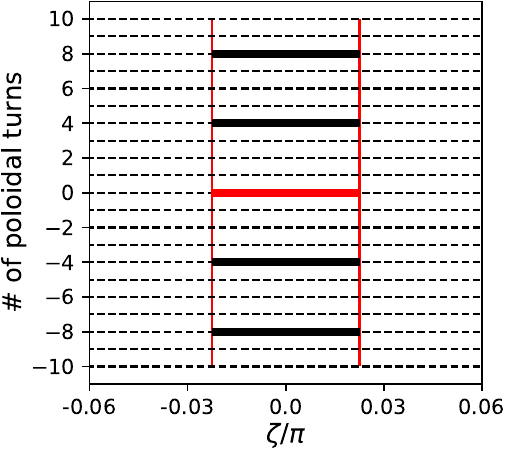}
    \caption{}
\end{subfigure}
\begin{subfigure}{0.3\textwidth}
    \centering
    \includegraphics[width=1\textwidth]{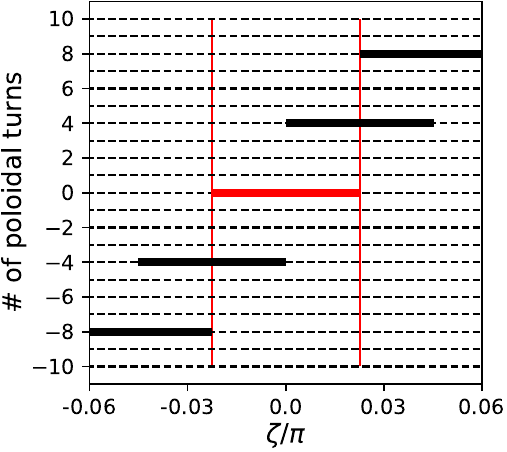}
    \caption{}
\end{subfigure}
\begin{subfigure}{0.3\textwidth}
    \centering
    \includegraphics[width=1\textwidth]{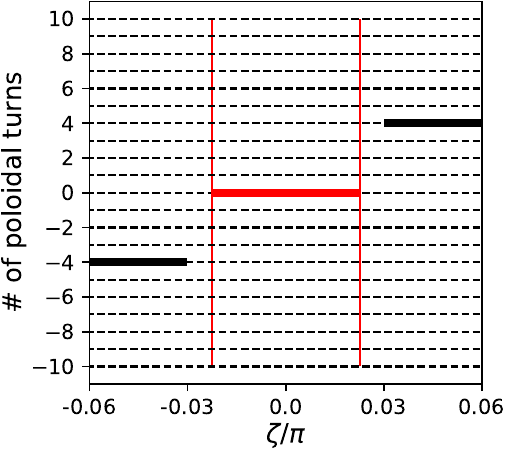}
    \caption{}
\end{subfigure}
\caption{A cartoon illustration of the toroidal location and extent of a single turbulent eddy (thick horizontal lines) at the outer midplane after different numbers of poloidal turns for (a) $q=2.25$, (b) $q=2.253$, and (c) $q=2.257$. An eddy overlaps with its central location (red horizontal line) when a black bar falls between the vertical red lines. Here the eddy's perpendicular size is $20 \rho_{i}$. The size of the flux surface is determined using $\rho_{*}=1/300$, a characteristic value for the DIII-D tokamak \cite{Luxon2002}.}
\label{fig:IoP_EddyOverlap_231023}
\end{figure}

We wish to stress the importance of selecting appropriate values for $N_{pol}$ and $\Delta y$ to accurately simulate a given safety factor $q$ when $\hat{s}=0$. For many cases of practical interest, it is quite clear what values of $N_{pol}$ and $\Delta y$ should be chosen. For a low order rational surface $q=q_{0} \in m/n$, one wants to use $N_{pol}=n$ and $\Delta y =0$. For example, the $q=2/1$, $\hat{s}=0$ surface in a tokamak should be simulated using $\Delta y=0$ and a computational domain that is a single poloidal turn long ($N_{pol}=1$) to reflect the physical periodicity of the field line. On the other hand, a $q=9/4$ ($q=2.25$) surface with $\hat{s}=0$ should also be simulated with $\Delta y=0$, but with a computational domain of $N_{pol}=4$ poloidal turns. This case is illustrated in Fig. \ref{fig:IoP_EddyOverlap_231023} (a). 

However, when dealing with surfaces very close to a rational value $q=q_{0}+\Delta q$, where $\Delta q \sim \rho_{*}$ and $q_{0} \in \mathbb{Q}$ the situation becomes less clear and a choice must be made regarding the appropriate values of $N_{pol}$ and $\Delta y$. Whether a magnetic surface is considered to be "close to rational" in our analysis will depend on the machine size through $\rho_{*}$ \footnote{Formally, we are including and considering a specific $\rho_{*}$ effect related to the finite turbulent eddy size while neglecting other $\rho_{*}$ effects. However, we specifically choose to consider equilibrium where this is an $O(1)$ effect due to strong self-interaction so it will dominate over other $\rho_{*}$ effects that we are neglecting.}. Specifically, due to the finite perpendicular size of turbulent eddies, they can exhibit significant self-interaction even when the field lines do not exactly close on themselves. We categorize such surfaces as being close to rational. For instance, consider a $q=2.253=2253/1000$ surface in a medium size device. As shown in Fig. \ref{fig:IoP_EddyOverlap_231023} (b), an eddy that is $20 \rho_{i}$ wide will partially overlap itself by $\simeq 10 \rho_{i}$ after 4 poloidal turns. Thus one should choose $N_{pol} = 4$ (as for the $q_{0}=2.25=9/4$ surface) and $\Delta y = 10 \rho_{i}$ to accurately account for the partial overlap of the eddy. This case is illustrated in Fig. \ref{fig:IoP_EddyOverlap_231023} (b). However, if we consider the $q=2.257=2257/1000$ surface, the turbulent eddies completely miss themselves after 4 poloidal turns. Thus one should model this surface with $N_{pol}=1000$ and $\Delta y = 0$ (in practice a smaller number of poloidal turns $N_{pol}$ would be sufficient as long as convergence is reached). This case is illustrated in Fig. \ref{fig:IoP_EddyOverlap_231023} (c). The choice of $\Delta y$ entirely depends on whether or not a turbulent eddy overlaps with itself after some finite number of poloidal turns around the device, which is related to the parallel and perpendicular correlation lengths of the eddy. Otherwise, if the appropriate $\Delta y$ and $N_{pol}$ cannot be determined, a full flux surface computation will always be accurate, albeit very expensive.

When considering both linear and nonlinear simulations, it is important to determine over what range of safety factor values $\Delta q_{c}$ around a rational surface $q_{0}=m/n$ does eddy overlap occur for a given device size. In other words, it is important to determine which surfaces are "close to rational". To estimate the relevant $\Delta q$ range leading to overlap after $n$ poloidal turns we consider the characteristic size of the turbulent eddy transverse to the magnetic field and within the flux surface, $l_{y, turb} \sim 2\pi/k_{y, turb}$, where $k_{y, turb}$ is the characteristic binormal turbulence wavenumber. It is important to mention that while turbulent eddies are absent in linear simulations, we will apply the same $\Delta q$ estimation since $k_{y}$ in linear simulations characterizes the scale of modes that lead to turbulence in nonlinear simulations. Thus considering the same $\Delta y$ shift in both the linear and the corresponding nonlinear simulation will ensure relevant linear growth rates. In other words, we will use $\Delta q$ to determine the safety factor range around rational surfaces where magnetic field topology plays an important role and needs to be correctly taken into account.

To obtain $\Delta q_{c}$, we begin by considering the change in the toroidal coordinate $\Delta \zeta_{c}$ of a magnetic field line at a constant poloidal angle on a $q=m/n+\Delta q_{c}$ surface, when the field line comes back past itself for the first time. The field line comes back closest to itself after $n = N_{pol}$ poloidal turns and, using the definition of the safety factor $q=d\zeta/d\chi$, we find that
\begin{equation}
\label{eq:delta_zeta_equation}
    \Delta \zeta_{c} = 2 \pi n \Delta q_{c}.
\end{equation}
Using Eq. \eqref{eq:coordinates_flux_tube}, we can relate this to a change in the binormal coordinate according to 
\begin{equation}
\label{eq:deltayturb}
    \Delta y_{c} = C_{y} \Delta \zeta_{c}.
\end{equation}
For a turbulent eddy to overlap with itself, the coordinate change $\Delta y_{c}$ cannot be larger than the turbulent eddy size $l_{y, turb}$. Hence, combining Eqs. \eqref{eq:delta_zeta_equation} and \eqref{eq:deltayturb}, the maximum change in the safety factor that still allows eddy overlap is
\begin{equation}
\label{eq:delta_q_physical_notnormalized}
    \Delta q_{c} = \frac{l_{y, turb}}{2 \pi C_{y} n}.
\end{equation}
Using the above equation, we can roughly estimate $\Delta q_{c}$ assuming ion scale turbulence $l_{y, turb}\sim \rho_{i}$ and a circular geometry (i.e. $C_{y} \sim a/q_{0}$) to find
\begin{equation}
\label{eq:delataq_estimate}
      \Delta q_{c} \sim \rho_{*} \frac{q_{0}}{n}.
\end{equation}
This implies that $\Delta q_{c}$ scales with $\rho_{*}$ and therefore the $q$ range over which eddy overlap plays a role gets narrower in larger devices. It also inversely scales with the order $n$ of the rational surface $q_{0}=m/n$. Thus, we can expect the effects related to eddy overlap due to small changes in the safety factor to be strongest around low order rational surfaces $q=2,2.5,3,...$. 

\begin{figure}[H]
    \centering
    \includegraphics[width=0.65\textwidth]{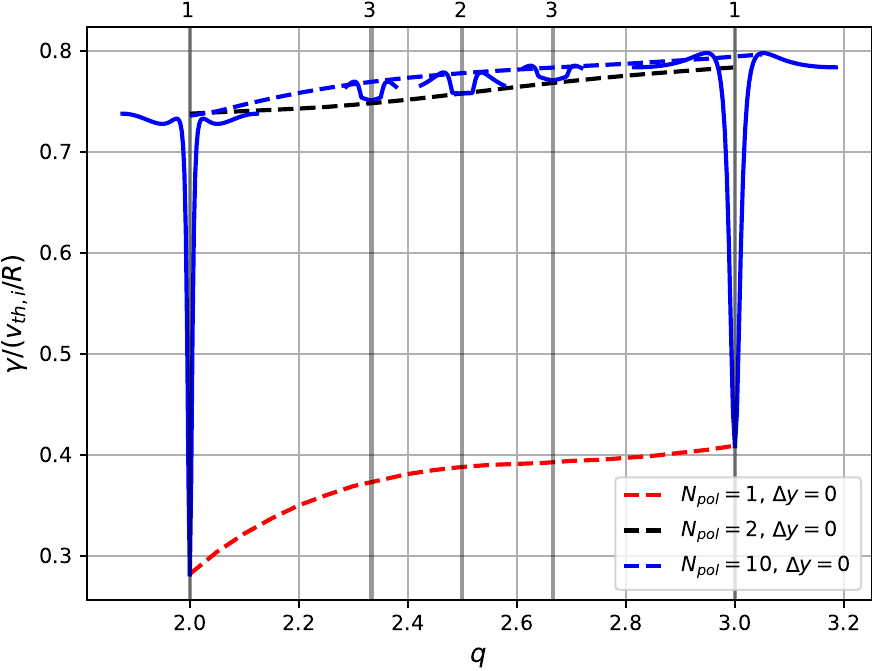}
\caption{The dependence of the linear growth rate on the safety factor $q$, with $N_{pol}$ and $\Delta y$ set to consistently reflect the actual field line topology around low order rational surfaces (solid blue). The dashed blue line indicates simulations with $N_{pol}=10$, the dashed black line indicates simulations with $N_{pol}=2$, and the dashed red line indicates simulations with $N_{pol}=1$ (all with $\Delta y = 0$). For determining the width of structures around rational surfaces, $\rho_{*}$ is set to $\rho_{*}=1/200$. These simulations were performed using CBC gradients; other physical and numerical parameters are shown in Table \ref{tab:parameters_linear_npolScan_miller} of Appendix \ref{appendix:simulation_parameters}.}
\label{fig:IoP_gamma_scanSafetyFactor_030323}
\end{figure}

To put the above discussion into perspective, we performed a $q$ scan using the parallel boundary phase factor together with the above expression for $\Delta q_{c}$ to construct an estimate of how the mode growth rate depends on the safety factor in the range $q\in [2,3]$ for magnetic surfaces with $\hat{s}=0$. Most importantly, around low order rational surfaces, $\Delta y$ scans were performed to emulate small deviations from the rational $q$ values. The linear simulations were performed at a fixed $\hat{s}=0$, $k_{x} \rho_{i} = 0$, $k_{y} \rho_{i} = 0.45$ using CBC gradients while only varying $q=m/n+\Delta q$ and changing $N_{pol}=n$ and $\Delta y$ self-consistently with the $q$ value. In Fig. \ref{fig:IoP_gamma_scanSafetyFactor_030323}, the solid lines show the result when the topology is treated consistently with the $q$ around low order rational surfaces. The figure also shows the growth rate if every $q$ value used the topology of an integer (red), half-integer (black), or irrational surface (blue). Irrational magnetic surfaces are represented using simulations with $N_{pol}=10$ since the linear growth rates are essentially converged for $N_{pol}=10$ (see Fig. \ref{fig:IoP_LinModeEnergandFreqVSNpol} (a)) and varying $\Delta y$ has a negligible effect (see Fig. \ref{fig:IoP_LinModeFreqVSEta_Npol10_030323}). It is noteworthy that for integer surfaces $q=2$ and $q=3$ the parallel boundary shift allows for a continuous mode growth rate change between $N_{pol}=1$ and the large $N_{pol}$ limit (i.e. growth rate value at $N_{pol}=10$) via the $\Delta y$ scan curve. 

According to Eq. \eqref{eq:delataq_estimate}, the width $\Delta q$ of the curves corresponding to the $\Delta y$ scans in Fig. \ref{fig:IoP_gamma_scanSafetyFactor_030323} depends on $\rho^{*}$ and in the limit of $\rho^{*} \rightarrow 0$ these structures will have infinitesimally small width. This will result in an effectively discontinuous growth rate change with $q$ at integer surfaces compared to all other higher-order surfaces. It is interesting to note that, if the $q$ scan is performed only with $N_{pol}=1$ and $\Delta y = 0$, the linear mode frequency is grossly underestimated for the majority of safety factor values. In tandem with previously discussed linear scans, these results indicate a candidate mechanism for the triggering of ITB --- the sensitivity of the linear growth rate to the magnetic topology. If one is considering a $q$ profile with $\hat{s}=0$, integer surfaces have a much lower growth rate than non-integer surfaces, which could result in a steepening of the plasma profile at the integer surfaces. As we will show shortly, these sharp changes in the growth rate with $q$ impact the nonlinear behavior and can lead to turbulence stabilization. 

In summary, linear flux tube simulations with $\hat{s}=0$ exhibit a strong dependence on the parallel domain length $N_{pol}$ and the binormal domain shift $\Delta y$ at the parallel boundary. Short parallel domains with $\Delta y = 0$ stabilize the $k_{z} \neq 0$ mode allowing only the $k_{z} \simeq 0$ mode. However, changing $\Delta y$ can substantially affect mode stability when the domain is short by effectively changing the allowed $k_{z}$ values. Based on these studies, \textit{linear self-interaction} can never increase the growth rate of the fastest growing mode. This is due to the additional constraints imposed on the modes by the geometry of the system. We have not found an instance where the growth rate at a single poloidal turn is larger than in the $N_{pol} \rightarrow \infty$ limit (i.e. all curves in Fig.  \ref{fig:IoP_LinModeEnergandFreqVSNpol} for $\gamma$ versus $N_{pol}$ are either constant or monotonically increasing). This result is proven analytically using Floquet-Bloch theory in Appendix \ref{appendix:Floquet_Bloch}. Finally, we discussed a method for identifying $N_{pol}$ and $\Delta y$ to correctly simulate an arbitrary physical magnetic field topology in a flux tube domain. Applying this to a scan in $q$ shows that the linear mode growth rate is very sensitive to the exact safety factor value due to the topological effects, especially around the lowest-order rational surfaces. Minor changes in $q$ can change the linear mode growth rate by a factor of 2.

\section{Nonlinear study}

\subsection{Nonlinear turbulence at zero magnetic shear} \label{section:NonlinearZeroShear}

Linear simulations provide an idealized understanding of plasma behavior. In this linear regime, each mode self-interacts through either destructive or constructive interference and is fully correlated with itself across the entire simulation domain. To capture the true physical dynamics nonlinear simulations are needed. The linear results discussed in the previous sections however serve as a valuable reference for interpreting the nonlinear results. In the upcoming sections, we aim to deepen our understanding of the effect of magnetic topology at $\hat{s}=0$ by carrying out corresponding nonlinear simulations.

\subsubsection{Effects of the parallel domain length on turbulence with $\hat{s}=0$}

We begin by examining the nonlinear effects of parallel self-interaction in simulations with $\hat{s}=0$, using adiabatic or kinetic electrons. Self-interaction has often been regarded as a numerical artifact arising from the constraints of flux tube computational domains, specifically resulting from the parallel boundary condition. To mitigate these artifacts, it is recommended to employ longer computational domains to ensure convergence \cite{Ball2020, Beer1995}. However, as we have argued in our linear mode study, short parallel domains are appropriate to use for simulating low order rational surfaces. The major motivation for using short simulation domains is to replicate the experimentally observed conditions for ITB formation --- ITBs are found to trigger preferentially near integer magnetic surfaces with low or zero magnetic shear. In this case, we expect parallel self-interaction to play a major role and one should numerically model such a situation by choosing a parallel domain length that corresponds to a single poloidal turn. As we will show next, in certain cases parallel self-interaction can have a dramatic effect on turbulence saturation. We will start by investigating how transport depends on the parallel domain length and quantifying the effects of parallel self-interaction when $\hat{s}=0$ and the parallel boundary condition is kept exactly periodic with $\Delta y = 0$. 

For quantitative analysis of the turbulent fluctuations, we make use of the two-point Eulerian correlation \cite{Wallace2014}
\begin{equation}
    C(\mathbf{x_{1}}, \mathbf{x_{2}}) = \frac{\langle \phi_{NZ}(\mathbf{x_{1}},t)\phi_{NZ}(\mathbf{x_{2}},t)\rangle_{t}}{\sqrt{\langle \phi^{2}_{NZ}(\mathbf{x_{1}},t)\rangle_{t}}\sqrt{\langle \phi^{2}_{NZ}(\mathbf{x_{2}},t)\rangle_{t}}},
\end{equation}
as a proxy for parallel self-interaction, where $\mathbf{x_{1}}$ and $\mathbf{x_{2}}$ are two real space locations between which the electrostatic potential correlation is measured, $\phi_{NZ}=\phi-\langle \phi \rangle_{y}$ is the non-zonal component of electrostatic potential, and $\langle . \rangle_{t}$ and $\langle . \rangle_{y}$ are temporal and binormal averages over the turbulence scale respectively. We assume that a higher spatial correlation of the fluctuations indicates a stronger turbulent self-interaction. Our primary interest lies in the parallel correlation $C_{\parallel}(x, z_{1}, z_{2}) = \langle C(\mathbf{x_{1}}(x, y, z_{1}), \mathbf{x_{2}}(x+\delta x, y+ \delta y, z_{2})) \rangle_{y}$, where $z_{1}=0$ will always be taken to be the \textit{central} outboard midplane of the domain (note that in a flux tube of length $N_{pol}$ there are $N_{pol}$ points along $z$ at the outboard midplane), and $\delta x$ and $\delta y$ are radial and binormal offsets respectively. In a tokamak, as a result of the axisymmetry of the underlying equilibrium, turbulence is statistically invariant in the binormal $y$ direction and we thus always average the correlation over $y$. When finite magnetic shear is considered, the presence of (pseudo-)rational surfaces causes the parallel correlation to depend on the radial position $x$. However, if there is no magnetic shear, the system is also homogeneous in $x$, so we perform an average in the radial direction as well, according to $C_{\parallel}(z_{1}, z_{2})= \langle \langle C(\mathbf{x_{1}}(x, y, z_{1}), \mathbf{x_{2}}(x+\delta x, y+ \delta y, z_{2})) \rangle_{y} \rangle_{x}$. Note that for computing $C_{\parallel}$, $\mathbf{x}_{1}(x, y, z_{1})$ and $\mathbf{x}_{2}(x+\delta x, y+ \delta y, z_{2})$ are in general not chosen to lie exactly on the same magnetic line ($\delta x = \delta y= 0$) but rather with $\delta x$ and $\delta y$ set to maximize $C_{\parallel}(\mathbf{x}_{1}, \mathbf{x}_{2})$. In other words, we adjust $\mathbf{x}_{2}$ to follow the turbulent eddy passing through $\mathbf{x_{1}}$ in case it drifts away from its initial magnetic field line.While we found such eddy drift to be unimportant on rational surfaces in simulations with kinetic electrons, it becomes significant in simulations with adiabatic electrons, where turbulent eddies closely follow ion drift trajectories (as calculated in the absence of turbulence). Neglecting to calculate the correlation along the eddy itself can lead to significant underestimation of the spatial correlation.

\begin{figure}[H]
\centering
\includegraphics[width=0.5\textwidth]{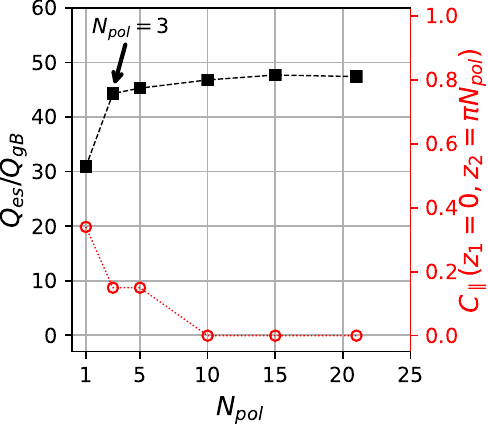}
\caption{The heat flux (black squares) and parallel correlation (red circles) as a function of the domain length in shearless toroidal simulations using adiabatic electrons. For each simulation, the correlation is measured from outboard midplane at $z_{1}=0$ to the furthest away inboard midplane at $z_{2}=-\pi N_{pol}$. $Q_{gB}= v_{th,i} p_{i} \rho_{*}^{2}$ is the gyro-Bohm heat flux unit, where $p_{i}$ is equilibrium ion pressure. These simulations were performed using the CBC-like parameters shown in Table \ref{tab:parameters_nonlinear_adiabatic_electron} of Appendix \ref{appendix:simulation_parameters} and $\Delta y = 0$.}
\label{fig:IoP_AE_s0_NpolScan}
\end{figure}

\begin{figure}[H]
    \centering
    \includegraphics[width=0.5\textwidth]{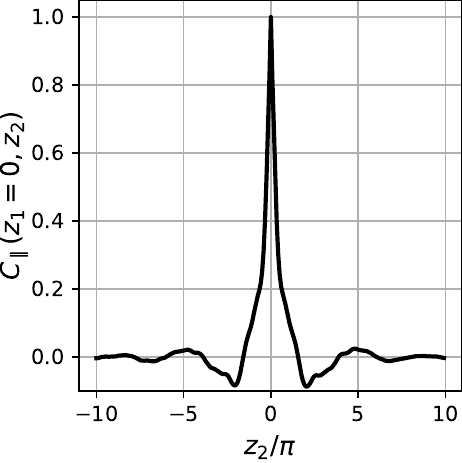}
    \caption{The parallel correlation along magnetic field lines for a $\hat{s}=0$ simulation with $N_{pol}=10$, CBC drive and adiabatic electrons. This simulations were performed using the parameters shown in Table \ref{tab:parameters_nonlinear_adiabatic_electron} of Appendix \ref{appendix:simulation_parameters} and $\Delta y = 0$.}
    \label{fig:IoP_AE_CorrEnvelopeNpol10_180324}
\end{figure}

In simulations with adiabatic electrons, we found that parallel self-interaction stabilizes turbulence and reduces the total heat flux. Specifically, at a single poloidal turn ($N_{pol} = 1$), the total electrostatic heat flux $Q_{es}$ is about $35\%$ lower compared to the large poloidal turn limit, as shown in Fig. \ref{fig:IoP_AE_s0_NpolScan}.
We benchmarked this result against the moment based gyrokinetic code \textsc{Gyacomo} \cite{Hoffmann2023} and found a good quantitative agreement (results not shown here). 
We believe the increased heat flux in longer domains is due to the decreased strength of parallel self-interaction. As shown in Fig. \ref{fig:IoP_AE_s0_NpolScan} the parallel length of an eddy $l_{\parallel, \text{turb}}$, as measured by the correlation function $C_{\parallel}$, is around one poloidal turn in these adiabatic simulations. Specifically, the correlation along the eddy falls to $C_{\parallel}\simeq 0.4$ at the first inboard midplane, half a poloidal turn away from the reference point at the outboard midplane. Therefore, simulations with $N_{pol} \simeq 10$ are sufficiently long to prevent any significant turbulent parallel self-interaction as illustrated in Fig. \ref{fig:IoP_AE_CorrEnvelopeNpol10_180324}.

We find that the parallel eddy length scale $l_{\parallel, \text{turb}}$ roughly corresponds to the distance a thermal ion travels with in the turbulence decorrelation time measured in the simulations. This is quantified by $l_{\parallel, \text{turb}} \simeq v_{th, i} t_{\text{turb}}$, $t_{\text{turb}}$ is the turbulence decorrelation time. We believe this is a consequence of critical balance \cite{Barnes2011, Terry2018}, implying that two spatial points along an eddy can only be causally connected if the information can propagate between them during the decorrelation timescale. 
Since the electrons are treated adiabatically, they only respond to the ions and cannot set the parallel length scale. We performed linear simulations with $\hat{s}=0$ and adiabatic electrons and found that the growth rate only weakly depends on $N_{pol}$ so it cannot be the main cause of the variation of $Q_{es}$ with $N_{pol}$ in Fig. \ref{fig:IoP_AE_s0_NpolScan}. Rather we find that zonal flows decrease in amplitude with increasing parallel domain length. This suggests that turbulence stabilization at short parallel domains partially results from the stronger zonal flows that effectively shear turbulent eddies. Finally, we will see that nonlinear simulations with adiabatic electrons do not provide an accurate picture near rational surfaces, where the adiabatic electron response $\omega / k_{z} \ll v_{th, e}$ breaks down as $k_{z} \rightarrow 0$ \cite{Dominski2015}. Thus kinetic electron effects are crucial. This is consistent with previous work on self-interaction \cite{Ball2020, ChandrarajanJayalekshmi2020Thesis} and the linear results discussed earlier. While the adiabatic electron model gives some insight into turbulence self-interaction, it fails to correctly account for physical processes at low magnetic shear.

\begin{figure}[H]
    \centering
    \includegraphics[width=0.5\textwidth]{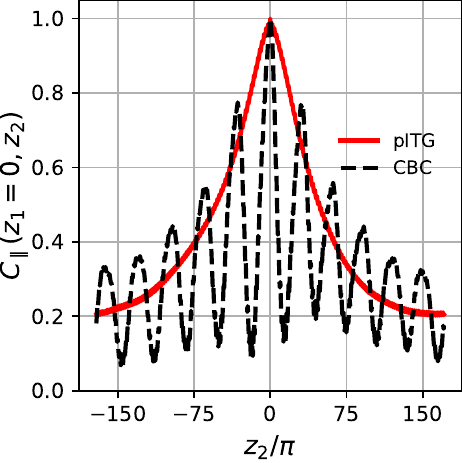}
    \caption{The parallel correlation along magnetic field lines for $\hat{s}=0$ simulations using an extremely long domain of $N_{pol}=171$, and CBC (black) and pITG (red) driving gradients. These simulations were performed using the parameters shown in Table \ref{tab:parameters_nonlinear_NpolScan_kinetic_electron} of Appendix \ref{appendix:simulation_parameters} and $\Delta y = 0$.}
    \label{fig:IoP_CorrEnvelopeNpol171_091122}
\end{figure}

Using kinetic electrons, we performed $\hat{s}=0$ simulations with up to $N_{pol}=171$ poloidal turns. This unusually high number of turns was necessary to allow the heat flux and parallel correlation to approach asymptotic values. The parallel correlation profile for CBC and pITG simulations with $N_{pol}=171$ poloidal turns is shown in Fig. \ref{fig:IoP_CorrEnvelopeNpol171_091122}. A key feature is the slow fall-off in correlation along the magnetic field lines, indicating the presence of ultra-long turbulent eddies spanning hundreds of poloidal turns. Importantly, both CBC and pITG simulations have similar correlation envelopes, CBC however has additional modulations with wavelength of $20$ poloidal turns that will be discussed shortly.

\begin{figure}[H]
    \centering
    \includegraphics[width=0.75\textwidth]{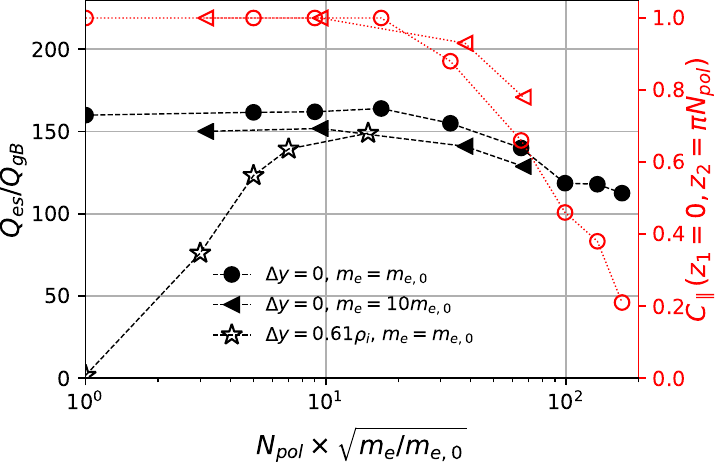}
    \caption{The total heat flux (black) and correlation along magnetic field lines (red) as a function of domain length for pITG drive with nominal electron mass and $\Delta y=0$ (circles), 10 times heavier electrons and $\Delta y =0$ (triangles), or nominal electron mass and $\Delta y=0.61\rho_{i}$ (stars). The horizontal axis is re-scaled by a factor $\sqrt{m_{e}/m_{e,0}}$ to show that parallel dynamics scale with the electron thermal velocity. These simulations were performed using the parameters shown in Table \ref{tab:parameters_nonlinear_NpolScan_kinetic_electron} of Appendix \ref{appendix:simulation_parameters}.
    }
    \label{fig:IoP_pITG_s0_NpolScan_wCollisions_011122}
\end{figure}

To establish how parallel correlation and heat flux change with the number of poloidal turns, we begin by analyzing the pITG drive simulations. The dependence of the total electrostatic heat flux and parallel correlation on the number of poloidal turns is shown in Fig. \ref{fig:IoP_pITG_s0_NpolScan_wCollisions_011122}. For $N_{pol} \lesssim 20$, neither quantity varies with domain length, which agrees with the linear trends seen in Fig. \ref{fig:IoP_LinModeEnergandFreqVSNpol}. For these low values of $N_{pol}$, the eddies are considerably longer than the domain length and the turbulence remains perfectly correlated across the entire parallel length. For $N_{pol} \gtrsim 20$, the volume-averaged heat flux and parallel correlation begin to decrease. We attribute this behavior to nonlinear effects as the growth rate does not behave this way in the linear simulations. Most importantly, the total heat flux curve follows the trend of the parallel correlation curve, indicating a direct influence of self-interaction on transport. Note that ion heat flux is an order of magnitude larger than electron heat flux and the ratio between the two remains approximately constant with changing $N_{pol}$. We see approximately a $25\%$ reduction in the total electrostatic heat flux when increasing the number of poloidal turns from $N_{pol}=1$ to the maximum considered value of $N_{pol}=171$. This implies that when $\hat{s}=0$, strong self-interaction destabilizes turbulence for these pITG parameters and $\Delta y = 0$. Due to the large computational cost, simulations with longer domains were not performed. However, extrapolating on the observed trends, we expect the heat flux to asymptote around $N_{pol} \simeq 250$ at a heat flux $ \sim 30 \%$ lower than the $N_{pol}=1$ simulation.

Following the critical balance postulate, the parallel correlation should be related to the velocity of information communication along the magnetic field lines when $\hat{s} \simeq 0$. In simulations with multiple kinetic species (in our case, kinetic ions and electrons), the species with the highest thermal velocity should dictate the parallel correlation $C_{\parallel}$. Hence we anticipated the eddy length to be associated with the electron motion, motivating a characteristic length scale of $l_{\parallel, \text{turb}} \simeq v_{th, e} t_{\text{turb}}$, where $v_{th, e}$ is the electron thermal velocity. In these simulations, we find that the turbulent decorrelation time $t_{\text{turb}}$ does not change significantly as the parallel domain length is varied and remains around $t_{\text{turb}} \simeq 2.5 R/v_{th,i}$ for all values of $N_{pol}$. The turbulent decorrelation time was estimated using temporal auto-correlation $C_{\tau}(t_{\text{turb}}) \sim 1/e$ with eddy drifts taken into account. This allows us to estimate the parallel correlation length to be $l_{\parallel, \text{turb}} \simeq 20 \, l_{conn}$, where $l_{conn} \simeq 2\pi q_{0} R$ is the connection length (i.e., the physical distance required for a magnetic field line to span one poloidal turn). This agrees with the observed trend in Fig. \ref{fig:IoP_pITG_s0_NpolScan_wCollisions_011122}, where the correlation decreases once $N_{pol} \gtrsim 20$, i.e. when the domain begins to exceed the length of the eddies. To further test the critical balance postulate, we conducted simulations with the electron mass increased by a factor 10 (while keeping all other parameters constant). Results from these heavy electron simulations are shown as black triangles in Fig. \ref{fig:IoP_pITG_s0_NpolScan_wCollisions_011122}. The resulting total electrostatic heat flux and parallel correlation match the case with lighter electrons, as long as the horizontal axis is rescaled by a factor of $\sqrt{10}$ (i.e. the decrease in parallel correlation actually begins at $N_{pol} \simeq 7$ instead of $N_{pol} \simeq 20$). This indicates that the characteristic parallel length scales as $\sqrt{m_{i}/m_{e}}$, providing further support that $l_{\parallel, \text{turb}}$ is proportional to the thermal electron velocity $v_{th,e} = \sqrt{T_{e} / m_{e}}$ as expected from critical balance.

\begin{figure}[H]
    \centering
    \includegraphics[width=0.75\textwidth]{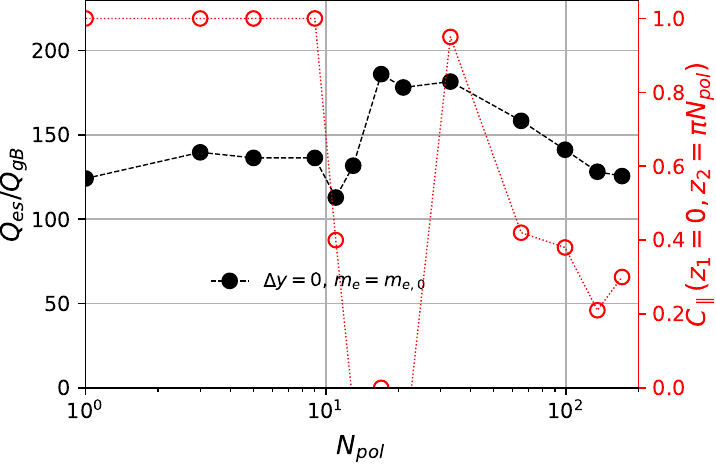}
    \caption{The total heat flux (black) and correlation along magnetic field lines (red) as a function of domain length for CBC drive. These simulations were performed using the parameters shown in Table \ref{tab:parameters_nonlinear_NpolScan_kinetic_electron} of Appendix \ref{appendix:simulation_parameters} and $\Delta y =0$. 
    }
    \label{fig:IoP_CBC_s0_NpolScan_wCollisions_011122}
\end{figure}

We further expand our study by including non-zero density and electron temperature gradients using CBC parameters but with $\hat{s}=0$. The dependence of the total heat flux and parallel correlation on domain length is shown in Fig. \ref{fig:IoP_CBC_s0_NpolScan_wCollisions_011122}. As the number of poloidal turns is increased from $N_{pol}=1$, the turbulent eddies show perfect correlation along the simulation domain until $N_{pol} \simeq 10$. For this range of $N_{pol}$ values, the heat flux does in fact slightly increase. This can be explained by the increase in linear growth rate shown in Fig. \ref{fig:IoP_LinModeEnergandFreqVSNpol}. For $N_{pol} \geq 10$, the correlation begins to vary dramatically. These abrupt changes in the correlation are due to the emergence of waves propagating parallel to the magnetic field with a wavelength of $\lambda \simeq 20 \, N_{pol}$. The modulations in the parallel correlation for $N_{pol}=171$ in the CBC case in Fig. \ref{fig:IoP_CorrEnvelopeNpol171_091122} are due to these developed waves. However, when the system is shorter than $N_{pol}=20$ but longer than $N_{pol}=10$, these waves can form, but their wavelength is discretized to be integer fractions of the length of the domain due to the parallel periodic boundary condition. The observed negative correlation for $N_{pol}$ values between $N_{pol}=10$ and $N_{pol}=20$ arises from the modulation of the correlation by the waves, where the furthest inboard midplane position from $z=0$ coincides with the wave trough.
Interestingly, \ref{fig:IoP_CorrEnvelopeNpol171_091122} shows that, despite these waves, the envelope of the correlation for the CBC simulation closely follows that of the pITG case. This suggests that the physical mechanism responsible for setting the envelope of the correlation function (i.e. the rate of parallel information transfer) is the same, despite the different driving gradients and presence/absence of the parallel waves. Importantly, the sharp increase in the heat flux in Fig. \ref{fig:IoP_CBC_s0_NpolScan_wCollisions_011122} around $N_{pol} \simeq 20$ is the result of the parallel waves being able to fully develop in the system. Note that electron heat flux is around $\sim 25 \%$ larger than ion heat flux and the ratio between the two remains approximately constant with changing $N_{pol}$. The combined effect of the initial increase in the heat flux and this later jump due to parallel waves almost completely counterbalance the gradual decrease in the heat flux past $N_{pol} \simeq 20$ due to weaker self-interaction. This leads to a small overall change in the heat flux for CBC when comparing $N_{pol}=1$ to $N_{pol}=171$ simulations, which is in contrast to the $\sim 20 \%$ heat flux decrease for the pITG case.

While the parallel waves were not a primary focus of this work, there are a few noteworthy observations. The occurrence of these waves requires a finite electron temperature gradient and a domain length exceeding $L_{z} \simeq v_{th,e} t_{turb} \simeq 200 \, R$. Importantly they disappear when modest collisionality is added, which also lowers the total heat flux indicating that the increase in the heat flux in Fig. \ref{fig:IoP_CBC_s0_NpolScan_wCollisions_011122} is due to these modes. Their wavelength is independent of the domain length once $L_{z} > \lambda$ is larger than their natural wavelength. The wavelength $\lambda$ scales as $\sqrt{m_{i}/m_{e}}$, implying that they are regulated by the finite inertia of electrons. Nevertheless, a comprehensive investigation into the physics underlying these long parallel waves falls outside the scope of the current paper, and remains a topic for future work.

We propose that the decrease of the heat flux with domain length beyond $N_{pol} = 20$ in Figs. \ref{fig:IoP_pITG_s0_NpolScan_wCollisions_011122} and \ref{fig:IoP_CBC_s0_NpolScan_wCollisions_011122}, can be at least partly attributed to a change in the turbulence from a 2D-like state to a 3D state. Until the critical length of $N_{pol}=20$ is reached, turbulent eddies are typically perfectly correlated in the parallel direction, effectively creating a 2D-like system. Note that the amplitude of the fluctuations may still vary between the inboard and the outboard planes so it is not strictly 2D turbulence. Nevertheless, because of the perfect correlation, short domains essentially only contain a single turbulent eddy in the parallel direction. However, once the critical length is exceeded, the domain becomes long enough to accommodate more than one eddy. As a result, fluctuations start to decouple in the parallel direction, and a 3D turbulent state emerges (this turbulence remains nonetheless anisotropic, with a much larger parallel than perpendicular correlation length). This has the effect of decreasing volume-averaged quantities. For instance, consider two eddies in the parallel direction that are perfectly correlated within themselves but not with each other. This situation would require a transition region between them. While the two eddies can organize themselves to maximize the radial transport, this will not be true for the transition region between them, leading to a smaller average radial transport per unit of length in the parallel direction. This general argument explains the similar decrease of the heat flux at sufficiently large $N_{pol}$ in both Figs. \ref{fig:IoP_pITG_s0_NpolScan_wCollisions_011122} and \ref{fig:IoP_CBC_s0_NpolScan_wCollisions_011122}, i.e. for different turbulent drives.

In this section, we presented a comprehensive nonlinear analysis of the impact of the parallel domain length on turbulence in tokamaks with $\hat{s}=0$. Our findings demonstrate the important role of parallel self-interaction at low order rational surfaces and highlight the need to include kinetic electron effects to accurately model surfaces with low magnetic shear. Kinetic electrons increase the length of turbulent eddies to more than a hundred poloidal turns, which means that the domain must be comparably long if one wishes to eliminate the effect of parallel self-interaction. Interestingly, this finding implies that in existing experiments, flux surfaces with low magnetic shear can be completely covered by a single turbulent eddy. Such an eddy would self-interact in both the perpendicular and parallel directions, a phenomenon we will explore in more detail in the subsequent section. When comparing pITG and CBC-like drives, we found that the reduction in both parallel correlation and the total heat flux follows a similar trend with parallel domain length. Furthermore, we have found no strong indication that zonal flows play a significant role in causing the change in the heat flux when the parallel domain length is varied in $\hat{s}=0$ simulations.
While the variation of the heat flux levels with domain length was found to be relatively modest, in the next section we will consider $\Delta y \neq 0$ and find a much stronger effect. Lastly, in our shearless simulations featuring CBC-like parameters, long parallel waves appeared in domains longer than a certain critical value --- roughly equal to the distance a thermal electron covers during the lifespan of a turbulent eddy. The value of $N_{pol}$ for which these waves emerge coincides with an increase in the heat flux.

\subsubsection{Effects of the parallel boundary phase factor on turbulence with $\hat{s}=0$}

One of the key findings from the previous section is that for zero magnetic shear, turbulent eddies can extend along field lines for hundreds of poloidal turns. On low order rational surfaces such long eddies will "bite their tails" and experience strong parallel self-interaction. On an irrational surface, an eddy will never exactly "bite its tail". However, given the typical size of current experiments ($\rho_{*} \sim 10^{-2}$), when eddies span hundreds of poloidal turns, they are likely to return to a position within at least a few ion gyroradii $\sim O(\rho_{i})$ of themselves, regardless of the $q$ value. Since eddies are approximately $\sim O(10\rho_{i})$ wide, they will experience some form of self-interaction. This leads us to an intriguing question: what happens on surfaces where eddies bite their own tail, but with some degree of misalignment (i.e. a binormal offset)? In this section, we will address this question by introducing a binormal shift $\Delta y$ at the parallel boundary of nonlinear simulations. 
This shift allows us to explore the more general form of self-interaction. In particular, we will be able to study aspects of perpendicular self-interaction, which occurs when an eddy's extent in the binormal direction causes it to run into itself. In experiments, surfaces with perpendicular self-interaction must be present near low order rational surfaces.


Before proceeding further, it's important to recognize that perpendicular self-interaction results from the finite $\rho_{*}$ of real machines. We can conceptualize this in two ways. First, we can consider the safety factor profile for ITB formation. Even in the limit of asymptotically small $\rho_{*}$, perpendicular self-interaction could always occur if one chooses the safety factor to be within $O(\rho_{*})$ of being rational, but not exactly rational. Given typical safety factor profiles, such values would only occupy an $O(\rho_{*})$ fraction of the radial profile and thus be negligible. However, in some ITB scenarios the safety factor profile is deliberately set to be close to rational across a large fraction of the profile (as a result of $\hat{s}$ being nearly zero) making the effect important. Second, we can consider the proportion of a flux tube surface occupied by an eddy. The area of a flux surface is $O(a^{2})$ and the area occupied by a single eddy is $O(\rho_{i} l_{\parallel})$, so an eddy will necessarily self-interact when $l_{\parallel} \sim a/\rho_{*}$. Hence, if an eddy extends hundreds of poloidal turns, it can entirely cover a flux surface, even given the $\rho_{*}$ of larger devices such as JET or ITER ($\rho_{*} \sim 10^{-3}$). In other words, eddies are so long that an effect that is formally small in $\rho_{*}$ can be numerically important for realistic machines. In this study, we assume that the finite $\rho_{*}$ effect due to the perpendicular eddy self-interaction is much larger than all other possible finite $\rho_{*}$ effects in the gyrokinetic formalism. We can justify this by considering an equilibrium where $q$ is very close to (but not exactly) rational. We will observe that modeling this effect for $\hat{s}=0$ conditions with a finite phase factor may lead to either complete turbulence stabilization, the emergence of intermittent turbulent states, or perpendicular eddy "squeezing".  

\begin{figure}[H]
    \centering
    \includegraphics[width=0.5\textwidth]{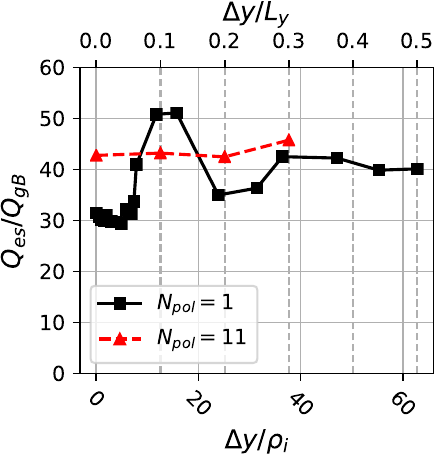}
    \caption{The dependence of the total electrostatic heat flux on the parallel boundary shift $\Delta y$ in shearless simulations using adiabatic electrons for domains with $N_{pol} = 1$ (black squares) and $N_{pol}=11$ (red triangles). These simulations were performed using the parameters shown in Table \ref{tab:parameters_nonlinear_adiabatic_electron_eta_scan} of Appendix \ref{appendix:simulation_parameters}.}
    \label{fig:IoP_AE_s0_etaScan_ky005_Npol1andNpol11}
\end{figure}

\begin{figure}[H]
\centering
\begin{subfigure}{0.45\textwidth}
    \centering
    \includegraphics[width=0.95\textwidth]{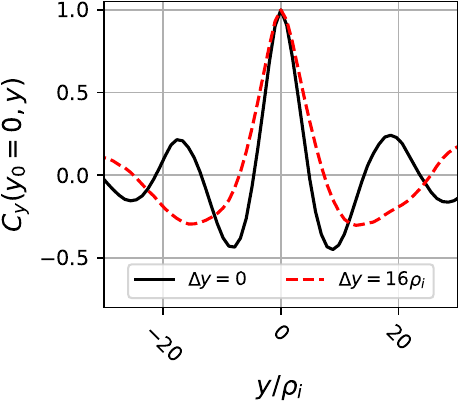}
    \caption{}
\end{subfigure}
\begin{subfigure}{0.45\textwidth}
    \centering
    \includegraphics[width=0.95\textwidth]{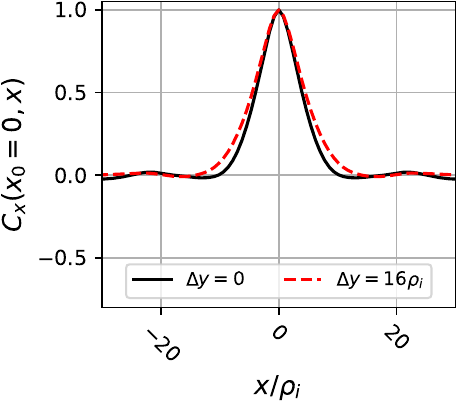}
    \caption{}
\end{subfigure}
\caption{A comparison of the (a) binormal correlation $C_{y}=\langle C(y_{0}=0, y)\rangle_{x,t}$ and (b) radial correlation $C_{x}=\langle C(x_{0}=0, x)\rangle_{y,t}$ at the outboard midplane ($z=0$) between $\Delta y =0$ (black line) and $\Delta y = 16 \rho_{i}$ (red broken line) simulations with $\hat{s}=0$ using adiabatic electrons. These simulations were performed using the parameters shown in Table \ref{tab:parameters_nonlinear_adiabatic_electron_eta_scan} of Appendix \ref{appendix:simulation_parameters}.}
\label{fig:IoP_AE_s0_etaScan_ky005_strechingIlustration}
\end{figure}

In this section, as we did in the previous one, we commence with adiabatic electron simulations. This allows us to establish a preliminary understanding of finite $\Delta y$ effects and provides a point of comparison for subsequent kinetic cases. For simulations involving adiabatic electrons, we observe a significant dependence of the heat flux on the parallel boundary shift $\Delta y$ when the domain is one poloidal turn in length, as depicted in Fig. \ref{fig:IoP_AE_s0_etaScan_ky005_Npol1andNpol11}. This is anticipated since turbulent eddies are approximately one or two poloidal turns long in the parallel direction (as seen in Fig. \ref{fig:IoP_AE_s0_NpolScan}), so they can still weakly self-interact after a single poloidal turn. For $0 \leq \Delta y \lesssim 5 \rho_{i}$, the total heat flux does not change significantly. This can be understood from the linear growth rate trends (not shown), since the linear growth rate does not vary significantly with $\Delta y$ for these specific simulation parameters. However, compared to the $\Delta y =0$ case, self-interaction can result in a $70\%$ increase in the total heat flux when $\Delta y \simeq 10 \rho_{i}$. To understand why the heat flux increases when $\Delta y \simeq 10 \rho_{i}$, we compare the perpendicular correlation function for $\Delta y \simeq 16 \rho_{i}$ against $\Delta y = 0$ in Fig. \ref{fig:IoP_AE_s0_etaScan_ky005_strechingIlustration}. From the case $\Delta y =0$ we see that an eddy has a region of anti-correlation at a binormal distance $\simeq 10 \rho_{i}$ and a second region of positive correlation at $\simeq 20 \rho_{i}$ away from its center. Thus, for a shift of $\Delta y \simeq 10 \rho_{i} $ after a single poloidal turn, the turbulent eddy comes back and "bites its anti-tail". This interplay of parallel and perpendicular self-interaction changes the binormal correlation shape, in fact making the eddy broader along $y$. This moves the peak in the binormal heat flux spectra towards lower binormal wavenumbers. In agreement with the quasilinear estimate $Q \propto \gamma / k^{2}_{\perp}$ \cite{Kadomtsev1965}, this results in higher heat flux. We label this effect "eddy stretching" as it results in eddies that are larger in the poloidal direction. Finally, when $\Delta y \gtrsim 20 \rho_{i}$ the eddy no longer encounters itself after one poloidal turn and we observe a heat flux intermediate between that of $\Delta y = 0$ and $\Delta y \simeq 10 \rho_{i}$. This suggests that the strong parallel self-interaction present when $\Delta y = 0$ is stabilizing in the adiabatic electron case. Note that the simulation with $N_{pol} = 1$ and $\Delta y = L_{y}/2=62 \rho_{i}$ in Fig. \ref{fig:IoP_AE_s0_etaScan_ky005_Npol1andNpol11} has a heat flux $Q_{es}/Q_{gB} \simeq 40$ that is consistent with the $N_{pol}=2$ simulation with $\Delta y = 0$ shown in Fig. \ref{fig:IoP_AE_s0_NpolScan}. Specifically, as long as the binormal box size is sufficiently large for perpendicular self-interaction to be weak, the heat flux should be the same for the $N_{pol}=2$, $\Delta y = 0$ and $N_{pol}=1$, $\Delta y = L_{y}/2$ cases. This is because, with a $\Delta y = L_{y}/2$ shift, the eddy effectively makes two poloidal turns before meeting itself. Finally, to verify that $\Delta y$ has a negligible effect on the heat flux when the domain is long enough to eliminate parallel self-interaction, we performed a set of simulations with $N_{pol} = 11$, shown by the red dashed line in Fig. \ref{fig:IoP_AE_s0_etaScan_ky005_Npol1andNpol11}. For this large value of $N_{pol}$, the heat flux is indeed essentially independent of $\Delta y$.

We do not expect that the exact changes in the heat flux with $\Delta y$ as seen in Fig. \ref{fig:IoP_AE_s0_etaScan_ky005_Npol1andNpol11} are universal. The heat flux dependence on $\Delta y$ will generally also be a function of the driving gradients, geometry, and other parameters in the problem. However, based on our observation we can expect some general trends to hold. Namely, around $\Delta y = 0$, the change in the heat flux will reflect changes in the linear growth rate and that $\Delta y$ will have a strong effect on the turbulence so long as the domain is short enough to allow self-interaction.

\begin{figure}[H]
    \centering
    \includegraphics[width=0.8\textwidth]{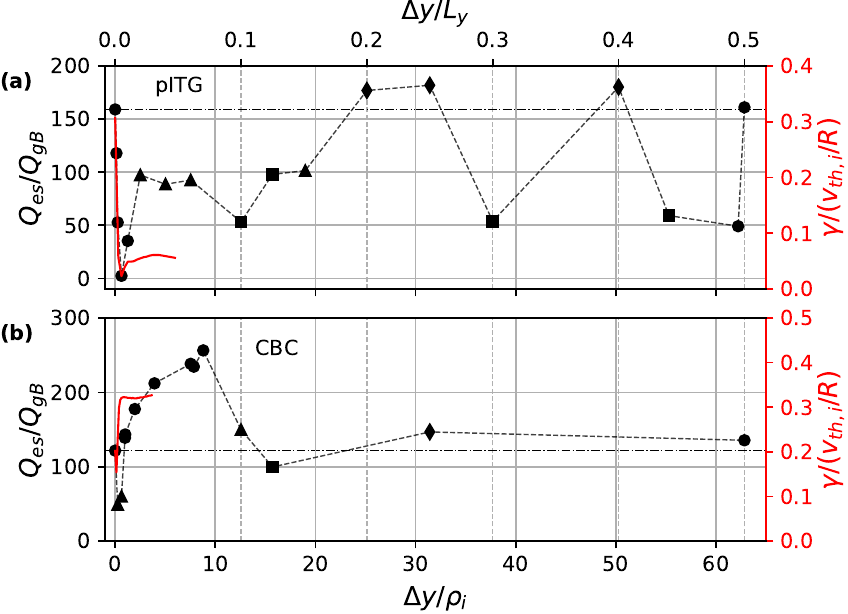}
    \caption{The dependence of the total electrostatic heat flux on the parallel boundary shift $\Delta y$ in shearless simulations using kinetic electrons for domains with $N_{pol}=1$ and either (a) pITG or (b) CBC parameters. The different markers indicate different simulation groups: "linear" (circles), "bursty" (triangles), "eddy squeezing" (squares), and "eddy stretching" (diamonds), which are explained in the text. The red solid line indicates the linear growth rate $\gamma$ for the binormal wavenumber $k_{y} \rho_{i} = 0.2$ at which the nonlinear heat flux spectra peaks. These simulations were performed using the parameters shown in Table \ref{tab:parameters_nonlinear_etaScan_kinetic_electron} of Appendix \ref{appendix:simulation_parameters}.
    }
    \label{fig:IoP_CBCpITGetaScan_full_221122}
\end{figure}

When using kinetic electrons, the picture becomes much more complex. Compared to the adiabatic electron simulations, our kinetic electron simulations display an even stronger dependence of the total heat flux on $\Delta y$ in both pure ITG and CBC-like turbulence regimes, as shown in Fig. \ref{fig:IoP_CBCpITGetaScan_full_221122}. Since turbulent eddies in this case can stretch across hundreds of poloidal turns instead of just a few, it leads to even stronger self-interaction effects. These simulations were performed with $N_{pol}=1$ to study the radial region around integer values of $q$ and to maximize the strength of parallel self-interaction. Investigating the results in Fig. \ref{fig:IoP_CBCpITGetaScan_full_221122}, we find that depending on the value of $\Delta y$ these simulations can be separated into four different groups based on their qualitative behavior. In the rest of this section, we will discuss these different groups, explaining their unique turbulence self-organization properties and comparing the pITG and CBC turbulent regimes.

To start with, we test the self-consistency of our simulations by comparing results for $\Delta y = 0$ and $\Delta y = L_{y}/2$. Given that the eddies in this case have very long parallel correlation lengths and the fact that the box width $L_{y}$ in these simulations is large enough for the same eddy to fit twice in the binormal direction without significant perpendicular interaction, the heat flux in the simulations with $\Delta y = 0$ should closely match those with $\Delta y = 0.5 L_{y}$. In this scenario, simulations with $N_{pol}=1$ and $\Delta y = 0.5 L_{y}$ can essentially be viewed as equivalent to $N_{pol}=2$ and $\Delta y = 0$ simulations. In the pITG scenario, the heat flux of the $N_{pol}=1$, $\Delta y = 0.5 L_{y}$ simulation as expected precisely matches that of the $N_{pol}=1$, $\Delta y = 0$ simulation. In the CBC-like scenario, the heat flux of the $N_{pol}=1$, $\Delta y = 0.5 L_{y}$ simulation only slightly exceeds that of the $N_{pol}=1$, $\Delta y = 0$ case. These observations are perfectly in line with the corresponding $N_{pol}$ scans with $\Delta y=0$ shown in Figs. \ref{fig:IoP_pITG_s0_NpolScan_wCollisions_011122} and \ref{fig:IoP_CBC_s0_NpolScan_wCollisions_011122} for pITG and CBC cases respectively, which display no measurable change in the heat flux between $N_{pol}=1$ and $N_{pol}=3$ for the pITG drive and a small increase in heat flux for the CBC-like drive (from linear interpolation we expect an increase for $N_{pol}=2$ as well). This comparison would not hold if the binormal size of the eddy was comparable to $L_{y}$, as the eddy would interact with itself in the perpendicular direction when $\Delta y = 0.5 L_{y}$.

Although we have established that our selected simulation domain is wide enough in the binormal direction to accommodate the same eddy twice without self-interaction in the perpendicular direction, it is essential to note that the binormal domain size can significantly influence simulations when $\Delta y \neq 0$. This is true even when using values of $L_{y}$ that would be considered converged for a standard simulation with $\Delta y = 0$. This is because ultra long turbulent eddies can maintain a strong correlation in the parallel direction even after passing through the parallel boundary many times. Effectively, an eddy can tile the entire binormal extent by passing through the parallel boundary, shifting in the binormal direction due to the $\Delta y$ offset, and experiencing strong perpendicular self-interaction. 
    
\begin{figure}[H]
    \centering
    \includegraphics[width=0.5\textwidth]{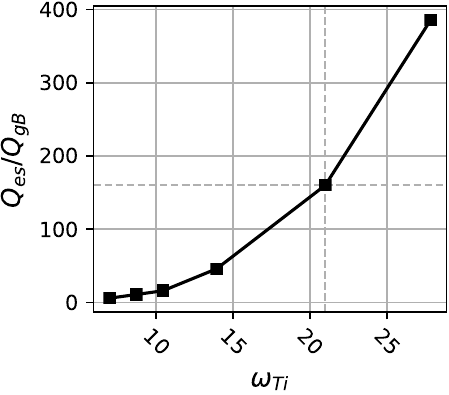}
    \caption{The dependence of the total electrostatic heat flux on the normalized logarithmic ion temperature gradient $\omega_{Ti}$ in shearless simulations with $\Delta y = 0.628 \rho_{i}$ and pITG parameters. The vertical dashed line indicates the $\omega_{Ti}$ value at which the heat flux from the $\Delta y = 0$ case is recovered. The heat flux closely follows a second order polynomial $\propto (\omega_{Ti, crit}-\omega_{Ti})^{2}$, where $\omega_{Ti, crit}$ is the critical normalized logarithmic ion temperature gradient. These simulations were performed using the parameters shown in Table \ref{tab:parameters_nonlinear_etaScan_kinetic_electron} of Appendix \ref{appendix:simulation_parameters}.}
    \label{NL_pITG_eta0005_omtScan_110424}
\end{figure}

Let us now consider the nonlinear heat flux when $\Delta y$ is very close to zero. In this group of points (indicated by circles in Fig. \ref{fig:IoP_CBCpITGetaScan_full_221122}) the heat flux closely follows the trends of the linear growth rate $\gamma$ for the mode with $k_{y} \rho_{i} = 0.2$. This mode was chosen as it corresponds to the peak in the nonlinear heat flux spectra. A close match is observed for the pITG parameters, as depicted in Fig. \ref{fig:IoP_CBCpITGetaScan_full_221122} (a), where the trend of the linear growth aligns closely with the nonlinear heat flux. We also note that in this case, near-complete turbulence stabilization is achieved at $\Delta y = 0.628 \rho_{i}$. This is remarkable, given how small of a shift this is. This strong stabilization due to self-interaction at $\Delta y = 0.628 \rho_{i}$ can be eliminated by increasing the domain length. This can be seen by repeating the $N_{pol}$ scan in Fig. \ref{fig:IoP_pITG_s0_NpolScan_wCollisions_011122} using $\Delta y = 0.628 \rho_{i}$ instead of $\Delta y =0$, which shows dramatic increase in the heat flux with $N_{pol}$ as expected. Note that for $N_{pol} \geq 10$, the heat flux approaches the same value as for the simulations with $\Delta y =0$. 
Additionally we investigated the impact of the stabilization on experimental profiles by setting $N_{pol}=1$ and increasing the ion temperature gradient until heat fluxes for the $\Delta y =0.628 \rho_{i}$ case matched the $\Delta y =0 $ case. We found that a \textit{three} times larger value of $R/L_{Ti}$ was needed to match the heat fluxes as shown in Fig. \ref{NL_pITG_eta0005_omtScan_110424}. These findings suggest that considerable turbulence stabilization could occur in a region where $\hat{s}=0$ and $q$ is close to, but not exactly, a low order rational. A similar conclusion was reached in reference \cite{Waltz2006}, where the triggering of an ITB was seen just before the minimum $q$ became an integer value. Finally, we note that similar agreement is also found between the variation of $\gamma$ and $Q_{es}$ with $\Delta y$ for CBC-like parameters, though not as good as for the pITG case. The CBC-like simulations at $\Delta y = 0.628 \rho_{i}$ and $\Delta y = 1.25 \rho_{i}$ are close to marginal stability and exhibit intermittent turbulence. 

In the $\Delta y$ scan of the linear growth rate $\gamma$ shown in Fig. \ref{fig:IoP_LinModeFreqVSEta}, we saw that the peak in $\gamma$ broadens in $\Delta y$ with decreasing the electron to ion mass ratio. Accordingly, we predict that the ion mass can have a substantial effect on the shape of the features seen in the nonlinear $\Delta y$ scan around $\Delta y = 0$. However, such an isotope effect has not been directly addressed in this study and is left for future work.
    
\begin{figure}[H]
    \centering
\begin{subfigure}{0.45\textwidth}
    \centering
    \includegraphics[width=0.95\textwidth]{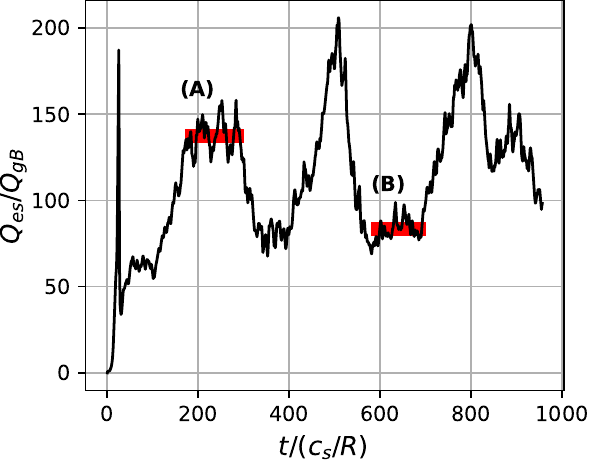}
    \caption{}
\end{subfigure}
\begin{subfigure}{0.45\textwidth}
    \centering
    \includegraphics[width=0.95\textwidth]{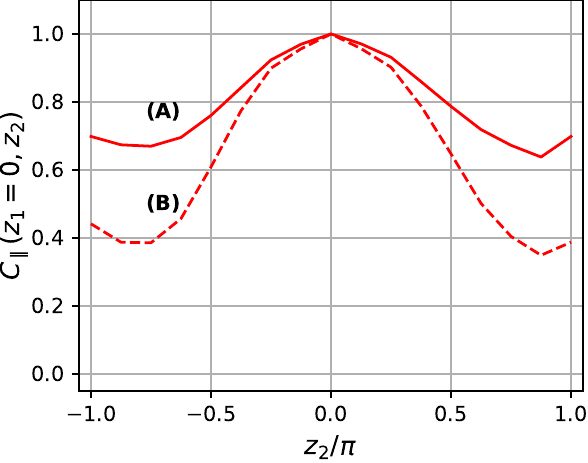}
    \caption{}
\end{subfigure}
    \caption{(a) The total electrostatic heat flux time trace for pITG drive with $\Delta y =7.5\rho_{i}$ where a "bursty" heat flux pattern is observed. The solid red lines next to markers (A) and (B) indicate the time windows over which the parallel correlation was averaged. (b) The parallel correlation with poloidal angle $z_{2}$ for time windows (A) (full line) and (B) (dashed line). These simulations were performed using the parameters shown in Table \ref{tab:parameters_nonlinear_etaScan_kinetic_electron} of Appendix \ref{appendix:simulation_parameters}.
    }
    \label{fig:s0_pITG_eta006_nrg_Q_es_ions_190522}
\end{figure}

When $q$ is such that the $\Delta y$ shift is comparable to the binormal size of the eddies, the dynamics become more complex. The eddies can come back close and partially "bite their own tails". As part of this, they can intermittently connect and disconnect with themselves due to the chaotic nature of the nonlinear system. Simulations presenting this feature are indicated by triangles in Fig. \ref{fig:IoP_CBCpITGetaScan_full_221122}. Under these conditions, the turbulence displays a "bursty" behavior, as illustrated in Fig. \ref{fig:s0_pITG_eta006_nrg_Q_es_ions_190522}(a) for the pITG case with $\Delta y = 7.5 \rho_{i}$. It is important to note that, for all the pITG intermittent cases, throughout the peak of a turbulent burst the parallel correlation is $C_{\parallel}(z_{1}=0, z_{2}=\pi) \simeq 0.8$, which is significantly larger than $C_{\parallel}(z_{1}=0, z_{2}=\pi) \simeq 0.4$, the typical value during the period between the bursts. This trend is shown in Fig. \ref{fig:s0_pITG_eta006_nrg_Q_es_ions_190522}(b). This further indicates that high parallel correlation destabilizes pITG-driven turbulence, aligning with the nonlinear parallel length study results discussed alongside Fig. \ref{fig:IoP_pITG_s0_NpolScan_wCollisions_011122} and also consistent with the linear results shown in Fig. \ref{fig:IoP_LinModeFreqVSEta}. 

"Bursty" turbulence cases are also observed in CBC-like simulations (triangles in \ref{fig:IoP_CBCpITGetaScan_full_221122} (b)). However, compared to the pITG case, the timescales of the bursts are considerably longer. An analogous but inverse trend is observed concerning the degree of parallel correlation and the amplitude of the heat flux. Namely, when the correlation measured along the magnetic field is high, the heat flux is low, and when the correlation is low, the heat flux is high. This pattern again aligns with both the nonlinear $N_{pol}$ study results, shown in Fig. \ref{fig:IoP_CBC_s0_NpolScan_wCollisions_011122}, and the linear results for CBC-like simulations, as presented in Fig. \ref{fig:IoP_LinModeFreqVSEta}. These findings suggest that a $\Delta y$ scan of the linear growth rate may be utilized to determine whether, for a given physical equilibrium, the effect of turbulent self-interaction could be expected to stabilize or destabilize the system around rational surfaces.

\begin{figure}[H]
\centering
\begin{subfigure}{0.45\textwidth}
    \centering
    \includegraphics[width=1\textwidth]{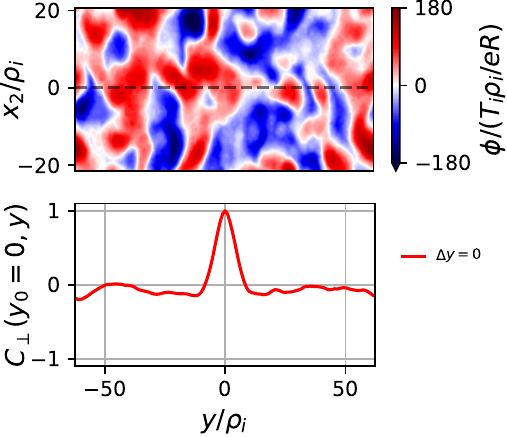}
    \caption{}
\end{subfigure}
\begin{subfigure}{0.45\textwidth}
    \centering
    \includegraphics[width=1\textwidth]{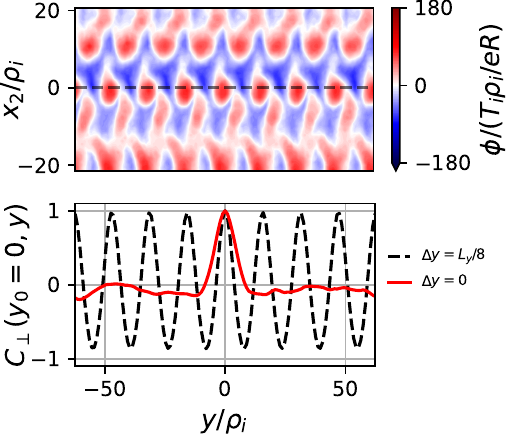}
    \caption{}
\end{subfigure}
\begin{subfigure}{0.45\textwidth}
    \centering
    \includegraphics[width=1\textwidth]{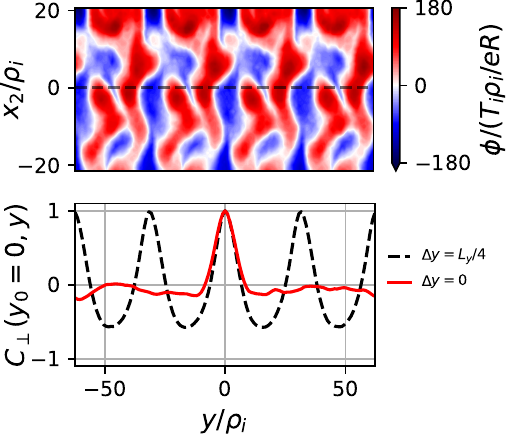}
    \caption{}
\end{subfigure}
\caption{Snapshots of electrostatic potential at $t \simeq 700 R/c_{s}$ (top subplot) and binormal correlation $C_{\perp}(y_{0}=0, y) = \langle C(y_{0}=0, y)\rangle_{x, t}$ at $z=0$ (bottom subplot) for pITG simulations with (a) $\Delta y / L_{y}=0$, (b) $\Delta y / L_{y}=0.125$, and (c) $\Delta y / L_{y}=0.25$. Note that the snapshots of electrostatic potential are shown only for the fraction $x \in [-20,20] \rho_{i}$ of the radial simulation domain ($L_{x} \simeq 150 \rho_{i}$), while the full binormal domain ($L_{y} \simeq 126 \rho_{i}$) is shown. These simulations were performed using the parameters shown in Table \ref{tab:parameters_nonlinear_etaScan_kinetic_electron} of Appendix \ref{appendix:simulation_parameters}.}
\label{fig:s0_pITG_etaScan_corrcontour_test}
\end{figure}

Finally, we discuss two distinct but related groups of simulations in Fig. \ref{fig:IoP_CBCpITGetaScan_full_221122} that arise due to the unique balance between parallel and perpendicular self-interaction --- turbulence "stretching" and "squeezing". In these cases the $\Delta y$ shift at the end of the domain is such that the different poloidal turns of an eddy either don't quite leave enough space to be their natural size (eddy squeezing) or have a bit too much space (eddy stretching). The cases of eddy squeezing are indicated by squares in Fig. \ref{fig:IoP_CBCpITGetaScan_full_221122}. In these cases, the eddies are forced by the geometry to become narrower in the binormal direction, which can be calculated using the binormal correlation diagnostic. Accordingly, the dominant $k_{y}$ moves towards higher values. An example of this is shown in the binormal correlation profile in Fig. \ref{fig:s0_pITG_etaScan_corrcontour_test}(b). This phenomenon triggers a significant decrease in transport, which, as illustrated in Fig. \ref{fig:IoP_CBCpITGetaScan_full_221122}, can be as large as a factor of three. In contrast to eddy squeezing, turbulence stretching (indicated by diamonds in Fig. \ref{fig:IoP_CBCpITGetaScan_full_221122}) was most noticeable in simulations with adiabatic electrons, as already depicted in Fig. \ref{fig:IoP_AE_s0_etaScan_ky005_strechingIlustration}. In that scenario, the eddies widened in the binormal direction, and the dominant $k_{y}$ value shifted towards lower values leading to increased heat transport. In contrast, in kinetic electron simulations, such pronounced "stretching" is not observed. Instead, eddies remain a similar size in the binormal direction. An example of this can be seen in the binormal correlation profile in Fig. \ref{fig:s0_pITG_etaScan_corrcontour_test}(c). Nevertheless, even though it does not change the eddy size substantially, it still modestly increases the heat flux compared to the base case of $\Delta y =0$ (as can be seen in Fig. \ref{fig:IoP_CBCpITGetaScan_full_221122}).

It is not clear if there is any link between eddy squeezing, which stabilizes turbulence, and the formation ITBs. Given the typical size of current tokamaks, the squeezing effect occurs on higher order rational surfaces where $\Delta y / L_{y} \in \{0.1,0.125,0.3\}$ (corresponding to 8th and 10th order surfaces), which doesn't explain the observation that ITBs often form near low order rational surfaces. However, "eddy squeezing" could offer an alternative method for reducing turbulent transport. This could be accomplished by choosing a safety factor profile such that perpendicular self-interaction restricts the size of turbulent eddies. In experiments, this could be achieved by creating a wide region with zero magnetic shear and scanning $q$ until a value that induces eddy squeezing is found. The precise $q$ value would depend on the perpendicular size of the eddies and the size of the machine.

\begin{figure}[H]
\centering
\begin{subfigure}{0.65\textwidth}
    \centering
    \includegraphics[width=1\textwidth]{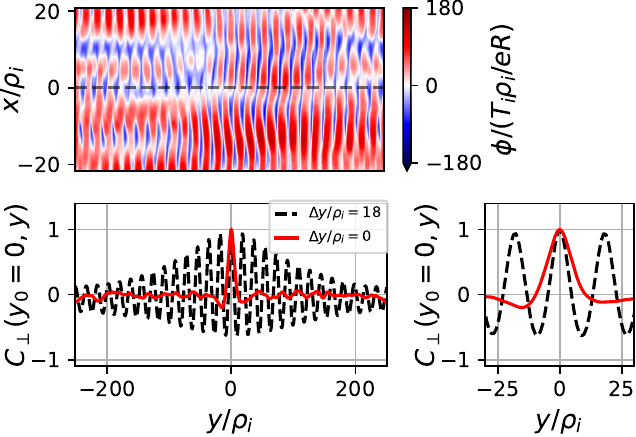}
    \caption{}
\end{subfigure}
\begin{subfigure}{0.65\textwidth}
    \centering
    \includegraphics[width=1\textwidth]{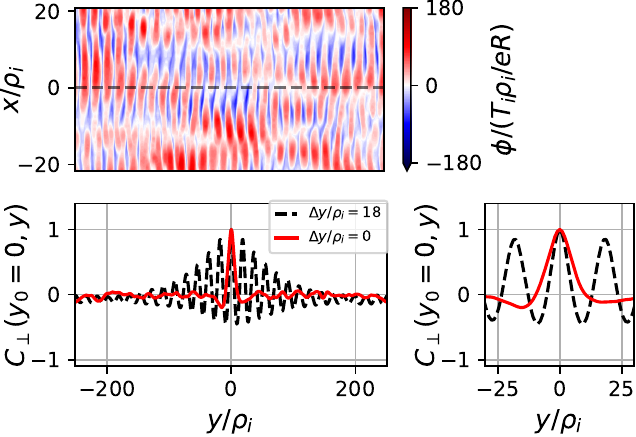}
    \caption{}
\end{subfigure}
\caption{Snapshots of electrostatic potential at $t \simeq 200 R/c_{s}$ (top subplot) and binormal correlation $C_{\perp}(y_{0}=0, y) = \langle C(y_{0}=0, y)\rangle_{x, t}$ at $z=0$ (bottom subplot) for pITG simulations with artificially heavy electrons and (a) $\Delta y / \rho_{i} = 18$, $N_{pol}=1$ or (b) $\Delta y / \rho_{i} = 18$, $N_{pol}=2$. In each case, the bottom right subplot provides a zoomed-in view of the correlation over the $y\in[-30,30] \rho_{i}$ range. Note that the snapshots of electrostatic potential are shown only for the fraction $x \in [-20,20] \rho_{i}$ of the radial simulation domain ($L_{x} \simeq 110 \rho_{i}$), while the full binormal domain ($L_{y} \simeq 500 \rho_{i}$) is shown. These simulations were performed using the parameters shown in Table \ref{tab:parameters_nonlinear_kyScan_KE_squeezing_Study} of Appendix \ref{appendix:simulation_parameters}.}
\label{fig:s0_pITG_etaScan_corrcontour_squeezingStudy}
\end{figure}

\begin{figure}[H]
    \centering
    \includegraphics[width=0.5\textwidth]{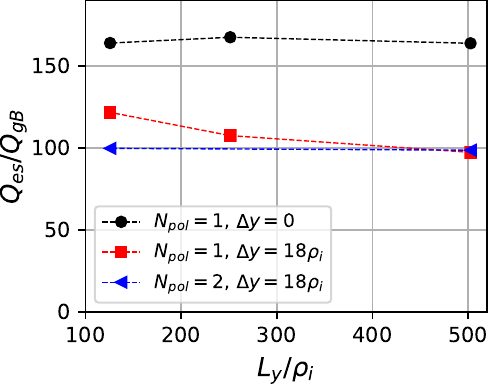}
    \caption{The dependence of the total electrostatic heat flux on $L_{y}$ for pITG simulations with artificially heavy electrons. These simulations were performed using the parameters shown in Table \ref{tab:parameters_nonlinear_kyScan_KE_squeezing_Study} in Appendix \ref{appendix:simulation_parameters}.}
    \label{fig:ITG_mex10_kyScan_Dy18_resIssueResolved}
\end{figure}

Our final discussion concerns how eddy squeezing relates to larger devices. The $\Delta y$ scan in Fig. \ref{fig:IoP_CBCpITGetaScan_full_221122} used $k_{y,min} \rho_{i}=0.05$, which is equivalent to $L_{y} = 126 \rho_{i}$. If this is taken to be the full flux surface, it corresponds to a relatively small tokamak with $1/\rho^{*} \sim O(L_{y}/\rho_{i})\simeq 100$. This is similar to the value in the core of TCV \cite{Hofmann1994}. Here, we aim to investigate whether eddy squeezing can still be achieved in larger devices with much smaller $\rho^{*}$. This is important because power plant-scale devices may be too large for a single eddy to fully cover a flux surface. To this end, we simulated pITG scenarios where the flux surface is large enough such that a single eddy can not cover the entire surface. To accomplish this we increased the domain size to $L_{y} \simeq 500 \rho_{i}$. However, for computational efficiency, we also used $10$ times heavier electrons to shorten the eddy length $l_{\parallel, \text{turb}}$ by a factor of $\sqrt{10} \sim 3$. Fig. \ref{fig:s0_pITG_etaScan_corrcontour_squeezingStudy} shows the correlation profiles, which demonstrates that the eddies are indeed unable to entirely span the whole binormal domain. It can be seen from Figs. \ref{fig:s0_pITG_etaScan_corrcontour_squeezingStudy}(a) and \ref{fig:s0_pITG_etaScan_corrcontour_squeezingStudy}(b) that, as the eddies pass through the parallel boundary with a binormal shift $\Delta y$, they are partially tiling the binormal domain. However, due to their finite parallel extent, they are unable to completely cover the whole surface while staying perfectly correlated and the binormal correlation therefore gradually decreases. This differs from the cases in Figs. \ref{fig:s0_pITG_etaScan_corrcontour_test}(b) or \ref{fig:s0_pITG_etaScan_corrcontour_test}(c) where the eddies remain perfectly correlated as they tile the surface. Fig. \ref{fig:s0_pITG_etaScan_corrcontour_squeezingStudy}(b) shows the same case as Fig. \ref{fig:s0_pITG_etaScan_corrcontour_squeezingStudy}(a), but with $N_{pol}=2$.  The binormal correlation decreases faster than with $N_{pol} = 1$ since the eddy must cover twice the parallel distance before coming back close enough to squeeze itself. However, in all of these cases the eddies are still squeezed similarly, even when the eddy does not occupy the full flux surface. Fig. \ref{fig:ITG_mex10_kyScan_Dy18_resIssueResolved} compares the change in the average electrostatic heat flux when varying $k_{y,min}=2\pi / L_{y}$ between simulations with and without eddy squeezing. As expected, for simulations with $\Delta y = 0$ changing $L_{y}$ has no significant effect. Importantly, simulations with $\Delta y = 18 \rho_{i}$ display a $40 \%$ reduction compared to the $\Delta y=0$ cases due to eddy squeezing. Moreover, in the $\Delta y = 18 \rho_{i}$ cases the heat flux \textit{does not grow} as $L_{y}$ is increased. This all suggests that eddy will still be squeezed and the heat flux would be reduced, even in larger devices. We believe this motivates an experimental investigation of turbulent eddy squeezing using accurate and detailed turbulence diagnostics (e.g. Beam Emission Spectroscopy (BES) \cite{McKee1999} or Tangential Phase Contrast Imaging (TPCI) \cite{Coda2024}).

\subsection{Turbulence self-interaction at low but finite magnetic shear } \label{section:NonlinearNonzeroShear}

So far, we have focused on the impact of magnetic topology on plasma turbulence at zero magnetic shear. For $\hat{s}=0$, the turbulence is radially homogeneous, as all field lines have the same topology. However, at low but finite shear different flux surfaces have different topologies due to the twist-and-shift parallel boundary condition. This leads to a significant radial variation in turbulence properties and flow drive across the simulation domain \cite{Parra2015a}. Thus, it is also important to investigate low but finite shear to understand which trends and effects from the zero magnetic shear scenario carry over to finite shear. In this section, we will exclusively consider CBC-like parameters while varying magnetic shear, since they are more realistic than the ones of the pITG case.

\begin{figure}[H]
\centering
\centering
\includegraphics[width=0.8\textwidth]{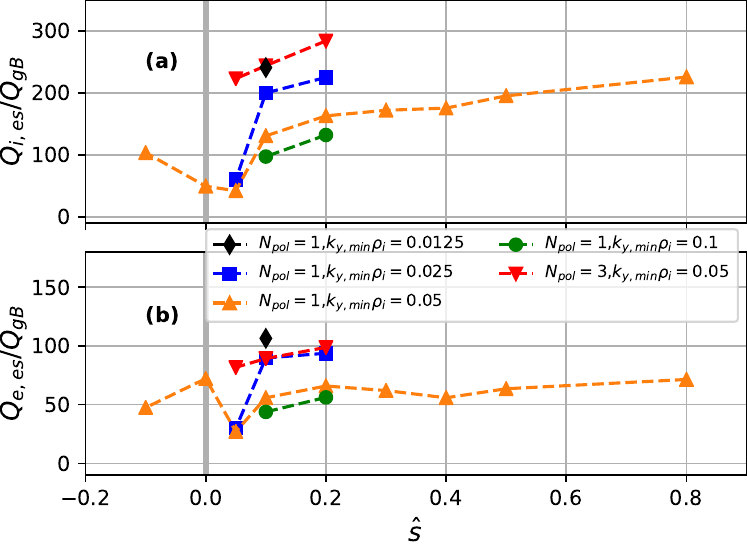}
\caption{Dependence of the (a) ion and (b) electron electrostatic heat flux on magnetic shear. These simulations were performed using kinetic electrons and CBC parameters (except for $\hat{s}$). The different curves show the heat flux for different $N_{pol}$ and $k_{y,min} \rho_{i}$ values. These simulations were performed using the parameters shown in Table \ref{tab:parameters_nonlinear_sScan_KE} of Appendix \ref{appendix:simulation_parameters}.}
\label{fig:sscanNL_Qes_ion_230821}
\end{figure}

Fig. \ref{fig:sscanNL_Qes_ion_230821} shows how the nonlinear heat flux depends on the magnetic shear and the parallel domain length. As we will show, the considerable heat flux variation with $\hat{s}$ observed in this figure can be attributed to self-interaction at multiple low order rational surfaces when magnetic shear is low. It is worth noting that due to the domain quantization condition (i.e. Eq. \eqref{eq:domain_quantization}) low magnetic shear simulations require very large radial domains, resulting in more expensive simulations compared to zero or standard magnetic shear ($\hat{s} \sim 1$). It is important to highlight that the simulations with low but finite magnetic shear considered in Fig. \ref{fig:sscanNL_Qes_ion_230821} follow the same trends due to self-interaction as identified in reference \cite{Ball2020} at standard magnetic shear. These trends are:
\begin{itemize}
    \item Increasing $N_{pol}$ increases heat flux. In finite shear simulations, parallel self-interaction appears to stabilize turbulence. If the parallel domain becomes longer, parallel self-interaction becomes weaker (analogously to the cases with $\hat{s}=0$) resulting in weaker stabilization of turbulence.
    \item Reducing $k_{y,min}$ (equivalent to increasing $L_{y}$) increases heat flux. Due to the domain quantization condition of Eq. \eqref{eq:domain_quantization}, as $k_{y,min}$ is decreased the radial domain size will expand. However, if the magnetic shear is held constant, the number of rational surfaces within the domain will remain unchanged. Consequently, the effects of self-interaction at low order rational surfaces are gradually "diluted," as the radial density of these surfaces decreases.
\end{itemize}
Notably, the magnitude of these effects is much larger at low $\hat{s}$. Regardless, the agreement with previously published findings for varying magnetic shear further stresses the important role of self-interaction and the necessity to accurately account for it in numerical studies of plasma turbulence. 

Furthermore, simulations with low but finite magnetic shear ($\hat{s} \sim 0.1$) display steady radial corrugations in plasma density and temperature profiles around low order rational surfaces. Such corrugations have been also reported in previous studies on self-interaction \cite{Ball2020, Waltz2006, Dominski2015, Ajay2020} at standard magnetic shear ($\hat{s} \sim 1$). At low $\hat{s}$, we find the corrugations to be much stronger in terms of amplitude and width compared to their counterparts at standard magnetic shear. Moreover, they occur even at higher order rational surfaces. 

\begin{figure}[H]
\centering
\includegraphics[width=0.8\textwidth]{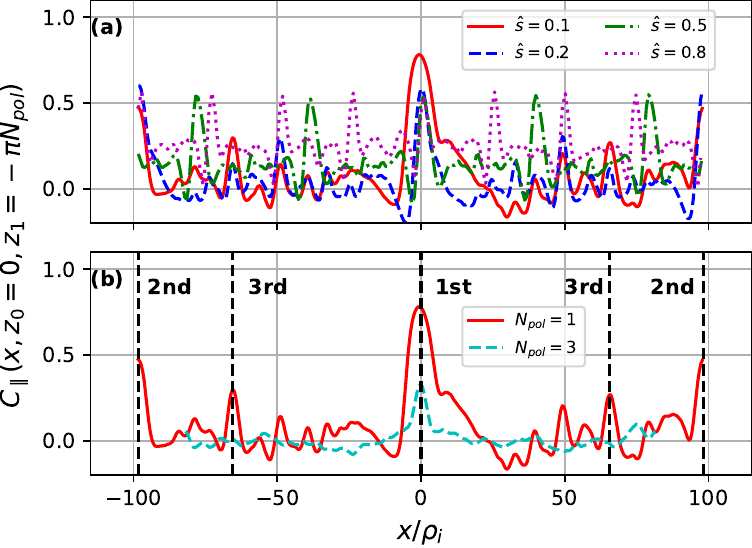}
 \caption{A comparison of the radial profile of the parallel correlation measured along the magnetic field lines for (a) different values of magnetic shear with $N_{pol}=1$ and (b) different $N_{pol}$ with $\hat{s}=0.1$. Plot (a) shows $\hat{s}=0.8$ (magenta dotted), $\hat{s}=0.5$ (green dash dotted), $\hat{s}=0.2$ (blue dashed) and $\hat{s}=0.1$ (red solid) with $L_{x} \simeq 200 \rho_{i}$, $k_{y, min} \rho_{i} = 0.05$, and $M/\hat{s}$ kept constant. The peaks in correlation indicate integer surfaces. Plot (b) shows $\hat{s}=0.1$ with $N_{pol}=1$ (red solid) and $N_{pol}=3$ (cyan dashed). The $N_{pol}=3$ simulation was performed with a lower $k_{y,min}$ value ($k_{y, min} \rho_{i} = 0.02$) therefore it covers a narrower range in $x$ ($L_{x} \simeq 160 \rho_{i}$). The vertical black dashed lines show 1st, 2nd, and 3rd order rational surfaces for the case with $\hat{s}=0.1$ and $N_{pol}=1$. These simulations were performed using the parameters shown in Table \ref{tab:parameters_nonlinear_sScan_KE} of Appendix \ref{appendix:simulation_parameters}.}
\label{fig:maximumcorrelationVSshear_kymin005_230821}
\end{figure}

A key indicator of self-interaction is the parallel correlation between the inboard and outboard midplanes $C_{\parallel}(x, z_{1}=0, z_{2}= - \pi N_{pol})$. As shown in Fig. \ref{fig:maximumcorrelationVSshear_kymin005_230821}(a), when shear is reduced, $C_{\parallel}(x,z_{1}=0, z_{2}= - \pi N_{pol})$ remains fairly constant at integer surfaces (located at $x=p/(k_{y,min}\hat{s})$, $p \in \mathbb{Z}$) and small everywhere else as long as $\hat{s} \gtrsim 0.1$. However, at $\hat{s}=0.1$ the correlation at integer and half-integer surfaces notably increases. An increase in the parallel correlation at a low order rational surface indicates that eddies are longer on this surface. While these eddies stretch further in the parallel direction compared to the $\hat{s} = 0.8$ case, they are nonetheless much shorter than those seen in zero magnetic shear simulations. For instance, as displayed in Fig. \ref{fig:maximumcorrelationVSshear_kymin005_230821}(b), a simulation with $N_{pol}=3$ and $\hat{s}=0.1$ has only a small peak in the correlation at the lowest order rational surface (i.e. 3rd order). This shows that turbulent eddies for $\hat{s}=0.1$ are approximately $3$ poloidal turns long. Note that, the height of the correlation peak at the lowest order rational surface (located at $x=0$) in the $\hat{s}=0.1$, $N_{pol}=3$ simulation is essentially the same as at the third order rational surface for the $\hat{s}=0.1$, $N_{pol}=1$ simulation, showing consistency. As discussed in Sec. \ref{section:NonlinearZeroShear}, the eddy drift due to particle orbits is negligible at low order rational surfaces for simulations with kinetic electrons. However, it is stronger away from low order rational surfaces where electron behavior is more adiabatic causing the asymmetry of $C_{\parallel}$ about $x = 0$ in Fig. \ref{fig:maximumcorrelationVSshear_kymin005_230821}. Overall, these results suggest that strong self-interaction still occurs in tokamaks when magnetic shear is low but finite, indicating that it is not necessary to have strictly zero magnetic shear.

\begin{figure}[H]
\centering
\includegraphics[width=0.8\textwidth]{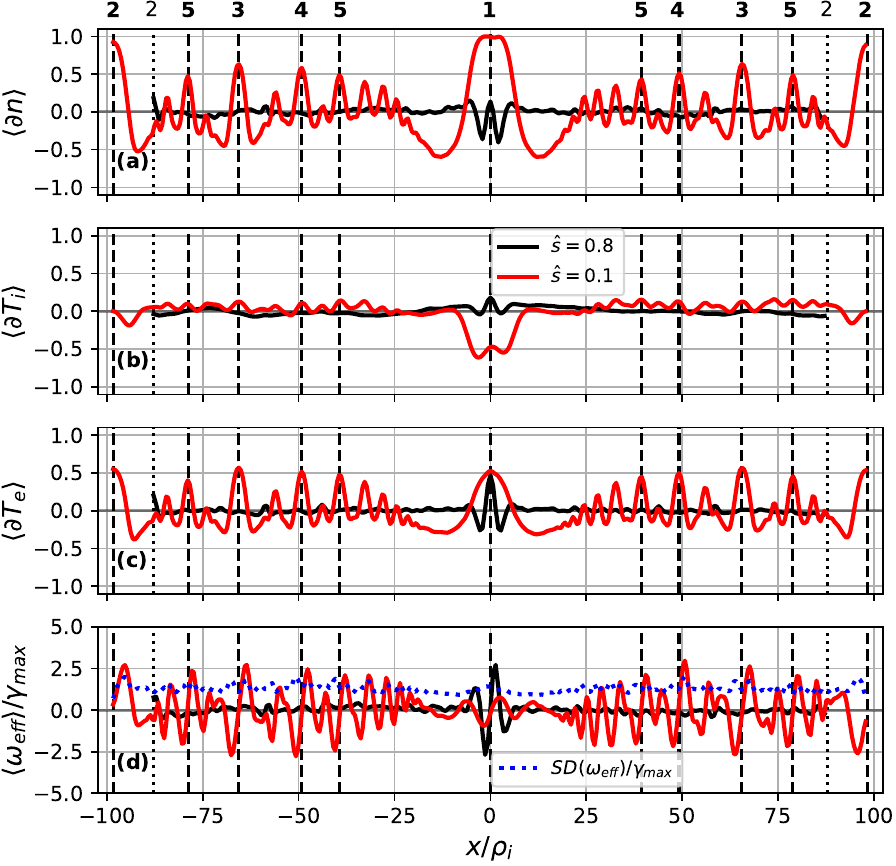}
\caption{The radial profile of $\langle \partial A \rangle = - \frac{d \langle \delta A \rangle}{dx}/ \frac{d A}{d x} $ where $A$ is the (a) density, (b) ion temperature or (c) electron temperature for $\hat{s}=0.8$ (black solid) and $\hat{s}=0.1$ (red solid). Also shown is the (d) time-averaged $E \times B$ shearing rate $\omega_{eff} = \langle \omega_{E\times B} \rangle_{\tau}$ normalized to $\gamma_{max}$ together with the standard deviation of the fluctuating component for the $\hat{s}=0.1$ case (blue dotted). The black vertical dotted lines indicate rational surfaces up to 2nd order for the $\hat{s}=0.8$ case and dashed lines indicate rational surfaces up to 5th order for the $\hat{s}=0.1$ case. The $\hat{s}=0.1$ simulation was performed with $M=1$, $k_{y,min} \rho_{i} = 0.05$ and $N_{pol} = 1$ (same as in Fig. \ref{fig:maximumcorrelationVSshear_kymin005_230821} (a)), and using other parameters shown in Table \ref{tab:parameters_nonlinear_sScan_KE} of Appendix \ref{appendix:simulation_parameters}. The $\hat{s}=0.8$ simulation was performed with $M=1$ and $k_{y,min} \rho_{i} = 0.007$, and using other parameters shown in Table \ref{tab:parameters_nonlinear_s08_KE} of Appendix \ref{appendix:simulation_parameters}.}
\label{fig:IoP_s01_ky005_MomCorrugations_220723}
\end{figure}

Due to self-interaction, turbulence at low-order rational surfaces behaves differently than at other locations, modifying the flux-gradient relationship. Since the fluxes must be constant across the flux tube in quasi-steady state, this leads to stationary zonal corrugations in the profiles of density, temperature, as well as $\mathbf{E} \times \mathbf{B}$ and parallel flow. This modifies the imposed background gradients that drive turbulence. For instance, the radial profile of a physical quantity $A$ (e.g. density $n$, temperature $T$, flow $u_{\parallel}$) modulated by the fluctuation is
\begin{equation}
    \Bar{A}=A_{0}+\frac{d A}{d x}x +\langle \delta A \rangle,
\end{equation}
where $A_{0}$ is the imposed background value, $d A / d x$ is the imposed background gradient, and $\langle \delta A \rangle$ is the contribution from the fluctuations. The effective gradient is then
\begin{equation}
    \frac{d\Bar{A}}{dx} = \frac{d A}{d x} + \frac{d \langle \delta A \rangle}{d x}.
\end{equation}
Here $\langle \delta A \rangle$ stands for the flux surface and time average of the fluctuating quantity $\delta A$
\begin{equation}
    \langle \delta A \rangle = \langle A(x,y,z,t) \rangle_{y,z,t},
\end{equation}
where $\langle.\rangle_{t},\langle.\rangle_{y},\langle.\rangle_{z}$ represent temporal, $y$ and $z$ averages respectively. Additionally, we define the ratio of the corrugation gradient to the background gradient as
\begin{equation}
    \left<\partial A \right> = - \frac{\frac{d}{dx}\langle \delta A \rangle}{\frac{d A}{d x}}.
\end{equation}
If $\left<\partial A \right>$ is positive, the plasma profile is locally flattened, and if it is negative, the profile is steepened. The value $\langle \partial A \rangle = 1$ represents complete flattening. Fig. \ref{fig:IoP_s01_ky005_MomCorrugations_220723} shows this ratio for density $\left<\partial n \right> $ as well as ion $\left<\partial T_{i} \right> $ and electron temperature $\left<\partial T_{e} \right> $ gradients for the $\hat{s}=0.1$ simulation. In this simulation, we observe corrugations in the density and temperature profiles that are comparable to the imposed plasma profiles. As can be seen in Fig. \ref{fig:IoP_s01_ky005_MomCorrugations_220723} (a), the density corrugations around integer and half-integer surfaces are large enough to completely flatten the imposed density profile. The electron temperature corrugations around integer and half-integer surfaces also tend to flatten the profile, albeit not completely. On the other hand, the ion temperature corrugation around the integer surface leads to a \textit{steepening} of the profile, thus forming a modest energy transport barrier. Such a steepening in the ion temperature profile was not found in simulations with $\hat{s} \geq 0.2$, which is consistent with results in \cite{Dominski2015}. A weak steepening (an order of magnitude smaller than in Fig. \ref{fig:IoP_s01_ky005_MomCorrugations_220723}) in the ion temperature profile was also found in simulations using adiabatic electrons at the extremely low shear of $\hat{s}=0.05$. Changing the minimum binormal wavenumbers from $k_{y, min} \rho_{i} = 0.05$ to $k_{y, min} \rho_{i} = 0.025$ shows that the modest transport barrier in the ion temperature channel is maintained in effectively larger devices. As $k_{y, min}$ is decreased, the amount of steepening and the width of the steepened regions remain fairly unchanged, while the distance between the rational surfaces increases. This indicates that the effect is connected to the size of turbulent eddies (of order $\rho_{i}$), as opposed to the distance between rational surfaces (of order $a$).

In addition to the plasma profile corrugations, we observe stationary corrugations in the zonal component of the electrostatic potential $\langle \phi \rangle_{yz}$ that leads to stationary $E \times B$ shearing flows. Fig. \ref{fig:IoP_s01_ky005_MomCorrugations_220723}(d) shows the time average of the effective zonal shearing rate \cite{ChandrarajanJayalekshmi2020Thesis}
\begin{equation}
    \omega_{eff}(x,t) = \langle \frac{1}{B_{0}} \frac{\partial^{2} \langle \phi \rangle_{y,z}}{\partial x^{2}} \rangle_{\tau},
\end{equation}
where $\langle . \rangle_{\tau}$ is a moving time average over the short turbulent time scale $\tau \sim 1/\gamma_{max}$ ($\gamma_{max}$ is the largest growth rate from the ion-scale $k_{y}$ spectra with $k_{x}=0$), and the Standard Deviation (SD) of the effective zonal shearing rate:
\begin{equation}
    SD(\omega_{eff})(x) = [ \langle (\omega_{eff} - \langle \omega_{eff} \rangle_{t})^{2} \rangle_{t}  ]^{1/2}.
\end{equation}
While $\langle \omega_{eff} \rangle_{t}(x)$ corresponds to the shearing rate contribution from the stationary component, $SD(\omega_{eff})(x)$ provides an estimate of the contribution from the fluctuating component. 

In Fig. \ref{fig:IoP_s01_ky005_MomCorrugations_220723} we see that the strongest shearing rate is measured in the vicinity of rational surfaces, owing to the radial variation of turbulent characteristics \cite{Parra2015a}. The radial variation in turbulence characteristics is a direct consequence of the magnetic field topology - at rational surfaces magnetic field lines reconnect after a few passes through the parallel boundary, leading to strong parallel turbulent self-interaction and different turbulence characteristics at these radial locations. Details of how turbulence drives $E \times B$ shear layers around rational surfaces are given in Ref. \cite{Ajay2020}. The shearing rate is known to play an important role in turbulence self-organization. At radial locations where $\langle \omega_{eff}(x)\rangle_{t} \gtrsim \gamma_{max}$, ITG turbulence stabilization due to shear flow is expected \cite{Waltz1998}. However, from Fig. \ref{fig:IoP_s01_ky005_MomCorrugations_220723} we see that around the integer surface at $x=0$, where the corrugation in the radial plasma profiles is most prominent, the stationary component of the shearing rate has the lowest amplitude and the shearing rate profile does not correspond well to density or temperature corrugations. This suggests that an additional mechanism related to strong parallel self-interaction must be at play at this location in agreement with Ref. \cite{Ajay2020}. Furthermore, the fluctuating component of the shearing rate is effectively constant across the domain and therefore cannot account for radial profile variations in Fig. \ref{fig:IoP_s01_ky005_MomCorrugations_220723}. Around higher-order rational surfaces we see that the locations where $|\langle \omega_{eff}(x)\rangle_{t}|$ is maximum corresponds to steeper effective plasma profiles and in the immediate vicinity where $\langle \omega_{eff}(x)\rangle_{t} = 0$ the profiles flatten. Around these surfaces stationary and fluctuating $E \times B$ shearing rate components have similar magnitude. While the stabilizing effect of shear flow around these surfaces cannot be entirely dismissed, the observed plasma profile characteristics near the lowest-order rational surfaces suggest that it is not the main mechanism responsible for the plasma profile corrugations and diminished transport. Rather, these phenomena are primarily driven by intense self-interaction at the rational surfaces.

\begin{figure}[H]
\centering
\includegraphics[width=0.85\textwidth]{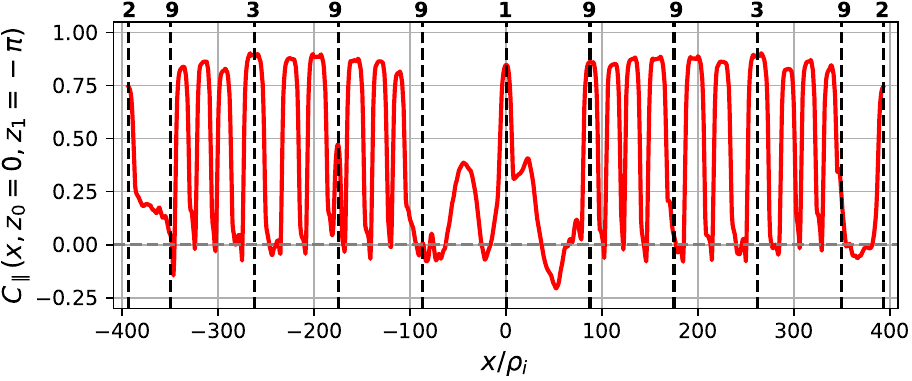}
\caption{The radial profile of the parallel correlation for $\hat{s}=0.05$, $k_{y,min} \rho_{i} = 0.025$ and $N_{pol}=1$ with 1st, 2nd, 3rd, and 9th order rational surfaces indicated by vertical black dashed lines. These simulations were performed using the parameters shown in Table \ref{tab:parameters_nonlinear_s005_KE} of Appendix \ref{appendix:simulation_parameters}.}
\label{fig:s005_ky0025_correlation_place_holder}
\end{figure}

As the $\hat{s}=0.1$ simulation shows strong stationary profile corrugations around lowest and second order surfaces, we also performed a $\hat{s}=0.05$ simulation (with $N_{pol}=1$), which is shown in Fig. \ref{fig:s005_ky0025_correlation_place_holder} and exhibits a far more intense behavior. Even 9th-order rational surfaces can sustain long turbulent eddies, as indicated by the radial parallel correlation profile at $x \simeq - 175 \rho_{i}$ and $ 90 \rho_{i}$. This happens because the magnetic shear is not strong enough to decorrelate eddies as they traverse through the domain many times and can "bite their own tails" even on very high order rational surfaces. Notably, we observe eddy squeezing effects in $\hat{s}=0.05$ simulation around several low order rational surfaces, which may play a role in the observed reduction in the heat flux. Interestingly, this appears to be a result of electromagnetic fluctuations modifying the background safety factor profile due to small but finite plasma $\beta$ ($\beta=10^{-4}$ was used to increase the simulation time step). This mechanism is outside the scope of this work and will be explored in greater detail in the future.

Simulations with $\hat{s} \lesssim 0.05$ are extremely computationally demanding as they require large radial domains due to the magnetic shear dependence in the domain quantization condition of Eq. \eqref{eq:domain_quantization}. This computational cost is further increased as it takes an extremely long time for the radial profiles to adjust and develop the steady-state corrugations. The turbulent modifications of the background profiles build up on a timescale $t_{transp}$, which is much longer than turbulence decorrelation time. This timescale is essentially the transport time scale over the simulation domain, which can be estimated based on a simple random walk argument to be $t_{transp} \sim \tau_{turb}(L_{x}/l_{corr,x})^{2} \sim 10^{4} R/c_{i}$ for $\hat{s}=0.05$ (where $\tau_{turb} \sim 2.5 R/c_{i}$ is the turbulence decorrelation timescale calculated from simulation results). Simulations this costly are impractical for use in parameter scans. However, some of these issues can be circumvented using more advanced flux tube methods \cite{Ball2022}, making even low shear cases computationally manageable. This will be the focus of future work.

Finally, given the significant impact of magnetic shear on turbulent eddy length, we investigated the scaling between the two. This analysis involved using data from simulations conducted at low magnetic shear, where changes in the parallel eddy length were identifiable, in conjunction with an analytical calculation.

In the gyrokinetic model, magnetic shear appears in the computation of geometric coefficients and enters into the equations in two places: the finite Larmor radius (FLR) stabilization and the magnetic drift frequency \cite{Plunk2014}. Separately investigating these two contributions, we show in Appendix \ref{appendix:eddy_scaling_law} that both FLR stabilization and magnetic drifts lead to the eddy length scaling as $l_{\parallel} \sim 1/\hat{s}$.

\begin{figure}[H]
\centering
\includegraphics[width=0.45\textwidth]{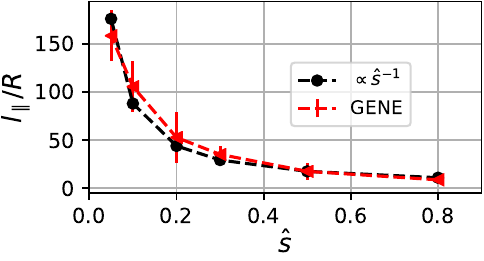}
\caption{The parallel eddy length as a function of magnetic shear for CBC-like simulations using kinetic electrons (same simulations as in Fig. \ref{fig:maximumcorrelationVSshear_kymin005_230821}). Red triangles indicate the minimum eddy length based on the radial corrugations in shearing rate profile. Black circles show the best fit of Eq. \eqref{eq:eddy_lenght_scaling_final}.}
\label{fig:ParallelCorrelationScalingWShear_28032023}
\end{figure}

The scaling $l_{\parallel} \sim 1/\hat{s}$ suggests that eddy length diverges as magnetic shear decreases to zero. However, from the nonlinear $\hat{s}=0$ simulation results with varying $N_{pol}$ presented in Fig. \ref{fig:IoP_pITG_s0_NpolScan_wCollisions_011122} and \ref{fig:IoP_CBC_s0_NpolScan_wCollisions_011122}, we know that there actually is an upper limit to the eddy length $l_{\parallel, max} \sim v_{th, e} t_{\text{turb}}$. This arises from set by critical balance, i.e. how far kinetic electrons can communicate information along the field line during a turbulent correlation time. Therefore, we estimate the eddy length as
\begin{equation}
    \label{eq:eddy_lenght_scaling_final}
    l_{\parallel} = \mathrm{min}[\frac{\alpha}{\hat{s}}, l_{\parallel, max}],
\end{equation}
where $\alpha$ is a constant of proportionality. We anticipate that $\alpha$ and $l_{\parallel, max}$ will depend on geometric parameters including the safety factor and profile gradients. In Fig. \ref{fig:ParallelCorrelationScalingWShear_28032023} we show the best fit of Eq. \eqref{eq:eddy_lenght_scaling_final} against estimates of our parallel eddy length from the simulations, where the fit parameter $\alpha = 2 \pi q_{0} R$ is obtained from finite shear simulations and $l_{\parallel, max} \simeq 100 \pi q_{0}  R$  is obtained from $\hat{s}=0$ simulations discussed in Sec.  \ref{section:NonlinearZeroShear}. The parallel lengths of the eddy at different magnetic shear values were estimated by identifying the highest order rational surfaces where the corrugations in the shearing rate profile were still above the standard deviation of the fluctuating shearing rate. For instance, from the shearing rate profile at $\hat{s}=0.1$ and $N_{pol}=3$ (not shown), we deduced that the eddy length for $\hat{s}=0.1$ is at most $l_{\parallel}/(2 \pi q_{0} R)=12$ poloidal turns since the 4th order rational surface was the highest order surface on which we could still identify a clear corrugation. In conclusion, this relationship further illustrates that we can anticipate unique physics to occur at very low magnetic shear as it enables strong turbulent self-interaction.

\section{Discussion and conclusions} \label{section:discussion}

In this work we have presented an in-depth gyrokinetic study of turbulent self-interaction in tokamak plasmas under conditions favorable for ITB formation. One of the key insights made was that at low magnetic shear $|\hat{s}| \lesssim 0.1$ turbulent eddies can span tens of poloidal turns and when $\hat{s} \simeq0$ they can extend for hundreds of poloidal turns. Given these ultra-long eddies, it is essential to correctly handle the magnetic field line topology by including the physically accurate parallel boundary condition and domain length. One must ensure that the flux tube uses the magnetic topology that corresponds to a given value of safety factor.

At zero magnetic shear, we explored different magnetic topologies (integer, rational, close to rational, and irrational magnetic surfaces) by employing a binormal shift $\Delta y$ at the parallel boundary. This investigation demonstrated that, depending on the turbulent regime, a slight deviation from an integer surface can lead to complete turbulence suppression (as seen in the pITG case) or, a two-fold increase in the heat flux (for the CBC-like case). In the extreme case of complete turbulence suppression in the pITG case, three times larger temperature gradient was needed to recover the same heat flux value as in the nominal case. This phenomenon is likely playing a significant role in the formation of ITBs, which, as experimental observations show, tend to form preferentially at low-order rational surfaces. Moreover, we observed that ultra-long eddies are long enough that, for specific $q$ values, a single eddy can cover the entire magnetic surface. This leads to an eddy squeezing effect due to \textit{perpendicular} self-interaction. This effect can significantly stabilize turbulence and potentially provide experimentally testable predictions for a new improved confinement regime.

In finite magnetic shear simulations, long turbulent eddies and pronounced parallel self-interaction lead to strong corrugations in the radial profiles of plasma parameters at low order rational surfaces. Building on earlier work using $\hat{s} \simeq 1$, our study showed that as magnetic shear decreases, the plasma profile corrugations become increasingly pronounced. At low (but non-zero) $\hat{s}$ values, these corrugations can entirely flatten the imposed gradients or steepen them by up to $50\%$. When $\hat{s} = 0.05$, even very high-order rational surfaces (such as the $9$th order) can exhibit strong self-interaction, resulting in rapidly varying turbulent characteristics and a strong $E \times B$ shear flow drive. We believe that self-interaction is essential to the experimental observation that ITBs preferentially form when magnetic shear is low and when $q$ is \textit{near} low order rational values.

Lastly, several significant considerations remain for future research. As standard flux tube simulations become exceedingly computationally expensive as magnetic shear is lowered, we plan to use a novel flux tube model with non-uniform magnetic shear \cite{Ball2022}. This model is less costly to run and can be used to space out rational surfaces as well as include the curvature of the safety factor profile. Moreover, we anticipate that even weak electromagnetic effects may play a critical role in the physics of turbulence at extremely small magnetic shear due to their ability to modify the safety factor profile. Finally, the robustness of the features discussed in this work (e.g. ultra-long eddies) must be verified by considering plasma shaping and collisions to ensure that they still hold true in realistic experimental conditions.

\section*{Acknowledgements}
The authors would like to thank Antoine Hoffmann, Moahan Murugappan, Oleg Krutkin, Alessandro Geraldini, Ben McMillan, and Garud Snoep for useful discussions pertaining to this work. \textit{
This work has been carried out within the framework of the EUROfusion Consortium, via the Euratom Research and Training Programme (Grant Agreement No 101052200 — EUROfusion) and funded by the Swiss State Secretariat for Education, Research and Innovation (SERI). Views and opinions expressed are however those of the author(s) only and do not necessarily reflect those of the European Union, the European Commission, or SERI. Neither the European Union nor the European Commission nor SERI can be held responsible for them. We acknowledge the CINECA award under the ISCRA initiative, for the availability of high performance computing resources and support. This work was supported by a grant from the Swiss National Supercomputing Centre (CSCS) under project ID s1050 and s1097. 
This work was supported in part by the Swiss National Science Foundation.}

\appendix
\section{Phase factor benchmark}
\label{appendix:eta_benchmark}

To validate the correctness of the parallel boundary condition phase factor implementation, we compared numerical and analytical linear mode frequencies. To enable analytically tractable solutions to the gyrokinetic system we took the cold ion limit and adiabatic electrons in a slab geometry with $\hat{s}=0$. We examined two instabilities: (i) the Parallel Velocity Gradient (PVG) and (ii) the Ion Temperature Gradient (ITG).

We begin with the PVG case, where the starting equations are the two lowest order velocity moments of the Fourier space ion gyrokinetic equation (Eqs. (20) and (21) in Ref. \cite{Ball2019}), which before applying the Fourier transform in the $z$ direction read:

\begin{align}
\label{eq:Ball2019_eq_20}
    & \left. \frac{\partial}{\partial t} \right\vert_{k_{x}}[(1+k_{\bot}^{2}\rho_{s}^{2})\phi]+\frac{T_{e}}{e} \hat{\mathbf{b}} \cdot \nabla z \frac{\partial \delta u_{\parallel}}{\partial z} +\frac{1}{2B} \sum_{\mathbf{k}'=\mathbf{k}-\mathbf{k}''} (\mathbf{k}' \times \mathbf{k}'') \cdot \hat{\mathbf{b}}(k_{\bot}^{\prime 2}-k_{\bot}^{\prime \prime 2}) \rho_{s}^{2} \phi' \phi '' \nonumber \\
    & +i(k_{x}\rho_{s}\omega_{Mx}+k_{y} \rho_{s} \omega_{My}) \phi = 0,
\end{align}
\begin{align}
\label{eq:Ball2019_eq_21}
     \left. \frac{\partial}{\partial t}\right\vert_{k_{x}} \delta u_{\parallel} +\frac{1}{2B} \sum_{\mathbf{k}'=\mathbf{k}-\mathbf{k}''} (\mathbf{k}' \times \mathbf{k}'') \cdot \hat{\mathbf{b}}(\delta u^{\prime}_{\parallel} \phi^{\prime \prime}-\delta u^{\prime \prime}_{\parallel} \phi^{\prime}) =-\frac{Z_{i}e}{m_{i}}\hat{\mathbf{b}} \cdot \nabla z \frac{\partial \phi}{ \partial z} + ik_{y}\frac{\omega_{V_{\parallel}}}{B} \phi, 
\end{align}
where $\rho_{s}=\sqrt{Z_{i}T_{e}/m_{i}}/\Omega_{i}$ is the sound gyroradius, $T_{e}$ is the electron temperature, $\omega_{Mx}$ and $\omega_{My}$ contain the effects of the magnetic drifts and density gradient, and $\omega_{V_{\parallel}}$ is the shear in the parallel flow. $\phi$ and $\delta u_{\parallel}$ are Fourier coefficients of electrostatic potential and parallel velocity respectively. \par

Linearising Eqs. \eqref{eq:Ball2019_eq_20} and \eqref{eq:Ball2019_eq_21} by considering only small perturbations (i.e. neglecting the nonlinear terms), assuming no density gradient and slab geometry $\omega_{My}=\omega_{Mx}=0$, and Fourier analyzing in time (e.g. $\phi(t) = \hat{\phi} \exp{[-i\omega t]}$) we can rewrite the above equations as
\begin{align}
    -i \omega (1+k_{\bot}^{2}\rho_{s}^{2})\hat{\phi}+\frac{T_{e}}{e} \hat{\mathbf{b}} \cdot \nabla z \frac{\partial \hat{\delta u_{\parallel}}}{\partial z}= 0,
\end{align}
\begin{align}
     -i \omega \hat{\delta u_{\parallel}} =-\frac{Z_{i}e}{m_{i}} \hat{\mathbf{b}} \cdot \nabla z \frac{\partial \hat{\phi}}{ \partial z} + ik_{y}\frac{\omega_{V_{\parallel}}}{B} \hat{\phi}.
\end{align}
The only linear instability drive in this system is the PVG term $\omega_{V_{\parallel}}$. For simplicity of notation, the hat symbol will be dropped from the Fourier coeficients. Combining the system of equations into a single differential equation for $\phi$ gives
\begin{align}
     \frac{Z_{i}e}{m_{i}} \hat{\mathbf{b}} \cdot \nabla z \frac{\partial^{2} \phi}{\partial z^{2}} - i \frac{k_{y} \omega_{V_{\parallel}}}{B} \frac{\partial \phi}{ \partial z} +\frac{e \omega^{2}}{T_{e} \hat{\mathbf{b}} \cdot \nabla z}(1+k_{\bot}^{2}\rho^{2}_{S}) \phi = 0,
\end{align}
which can be rewritten as
\begin{align}
    \frac{\partial^{2} \phi}{\partial z^{2}} - i A \frac{\partial \phi}{ \partial z} +\omega^{2} D \phi = 0,
\end{align}
where $A=k_{y} \rho_{s} \omega_{v_{\parallel}} / (c_{s} \hat{\mathbf{b}} \cdot \nabla z)$ and $D=(1+k^{2}_{\bot}\rho_{s}^{2})/(c_{s} \hat{\mathbf{b}} \cdot \nabla z)^{2}$.
Solving the above equation with the Ansatz $\phi \propto \exp{(ik_{z}z)}$ to find $k_{z}$ and applying periodic parallel boundary conditions with the phase factor
\begin{equation}
    \phi(z+L_{z}) = \phi(z) e^{i2\pi \eta}
\end{equation}
allows us to express the mode frequency as
\begin{equation}
\label{eq:PVG_frequency}
   \omega^{2} = \frac{k_{z}(k_{z}-A)}{D},
\end{equation}
where $L_{z}$ is parallel domain length and $k_{z}=2\pi(N+\eta)/L_{z}$ is the parallel wavenumber ($N \in \mathbb{N}$). From this, it can be shown that the mode is unstable when
\begin{equation}
    \label{eq:unstable_mode_limits}
    0 < \frac{k_{z}}{A}<1,
\end{equation}
since $D$ is always positive.

\begin{figure}[H]
\centering
\begin{subfigure}{0.45\textwidth}
    \centering
    \includegraphics[width=1\textwidth]{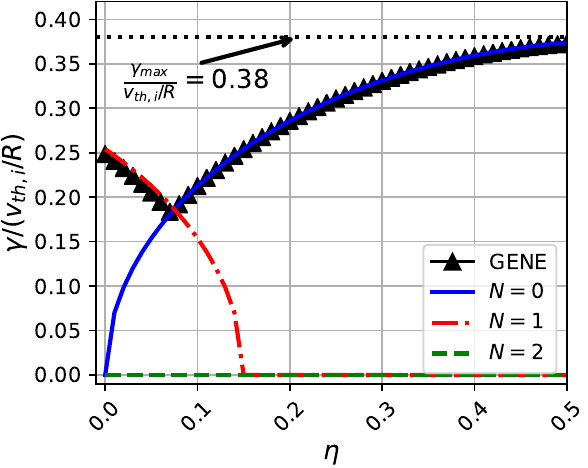}
    \caption{}
\end{subfigure}
\begin{subfigure}{0.45\textwidth}
    \centering
    \includegraphics[width=1\textwidth]{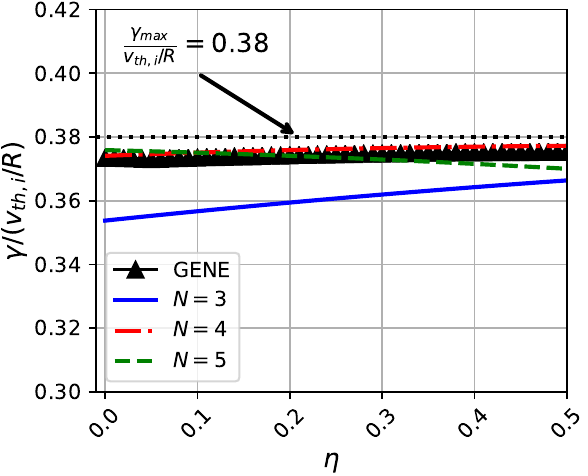}
    \caption{}
\end{subfigure} 
\caption{A comparison between linear GENE simulation growth rates (black triangles) and analytical results from Eq. \eqref{eq:PVG_frequency} for the PVG instability in the cold ion limit with varying parallel boundary shift $\eta$ and (a) $L_{z}=R$ or (b) $L_{z}=8R$. The different style lines in the legend indicate different $N$ values. Note that the two plots do not have the same $y$-axis scale. For GENE simulations $T_{i}=10^{-3} T_{e}$, $\omega_{v \parallel}=R/c_{i}$, $k_{x}\rho_{i}=0$ and $k_{y} \rho_{i}=1.15$ (i.e. $k_{\perp} = k_{y}$).}
\label{fig:etaScan_TP4studentLike_Lz_T10m3_170222}
\end{figure}

Analyzing Eq. \eqref{eq:unstable_mode_limits}, we see that, for a fixed $N+\eta$, the mode gets destabilized for sufficiently strong drive $\omega_{v_{\parallel}}$, large $k_{y} \rho_{s}$ values and long parallel length domains $L_{z} \propto (\mathbf{b} \cdot \nabla z)^{-1}$. Moreover, as the domain length $L_{z}$ increases, larger $|k_{z}|$ values, i.e. larger $N \in \mathbb{N}$ values, become necessary for the instability to stay close to the maximum growth rate $\gamma_{max}$ for $k_{z, max} = A/2$, subsequently diminishing the role of $\eta$ as the limit $N \gg \eta$ is reached. This shows that the impact of the phase factor is particularly significant in shorter domains. These theoretical predictions and analytical results can be compared with GENE results in the appropriate cold ion limit $T_{i} \ll T_{e}$ by fixing $\omega_{v_{\parallel}}$, testing two different $L_{z}$ values, and changing $\eta$, as shown in Fig. \ref{fig:etaScan_TP4studentLike_Lz_T10m3_170222}. This figure also underscores a crucial aspect mentioned in Section \ref{section:LinearZeroShear}: when a mode is permitted sufficient extension in the parallel direction (the $L_{z}=8$ case), the growth rate becomes unaffected by the phase factor and matches the maximum growth rate $\gamma_{max}$ derived from the phase factor scan in the restricted domain ($L_{z}=1$). This suggests that, given a sufficiently long parallel domain, the mode can freely explore all phase factors and select the one that maximizes the growth rate. Conversely, in a short domain, imposing a fixed phase factor can artificially alter the mode growth rate.

Following a similar approach to the one used for the PVG instability, we can formulate a cold ion ITG dispersion relation under the assumption that $\omega_{T,i} T_{i} / T_{e} \sim O(1)$, where $\omega_{T,i}$ is the normalized temperature gradient. Note that even though the ion temperature is small $T_{i}/T_{e} \ll 1$, the ITG drive can be retained by ordering the normalized ion temperature gradient to be large $\omega_{T, i} \gg 1$. This case is particularly useful to our study as we primarily focus on turbulence driven by the ITG mode. In the context of shearless slab geometry, the dispersion relation can be shown to be:
\begin{equation}
\label{eq:cold_ion_limit_ITG}
    \omega^{3}-\omega \frac{c_{i}^{2} k_{z}^{2}}{(1+k_{\bot}^{2}\rho_{i}^{2})}+\frac{k_{z}^{2} k_{y} \rho_{i}}{(1+k_{\bot}^{2}\rho_{i}^{2})} \frac{c_{i}}{m_{i}}\omega_{T_{i}}T_{i} = 0,
\end{equation} 
where the phase factor again comes in through the parallel boundary condition and appears in the parallel wavenumber $k_{z}=2 \pi(N+\eta)/L_{z}$.

\begin{figure}[H]
\centering
\begin{subfigure}{0.45\textwidth}
    \centering
    \includegraphics[width=1\textwidth]{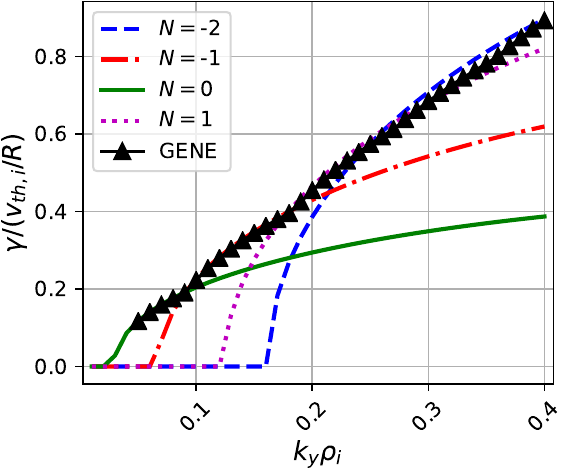}
    \caption{}
\end{subfigure}
\begin{subfigure}{0.465\textwidth}
    \centering
    \includegraphics[width=1\textwidth]{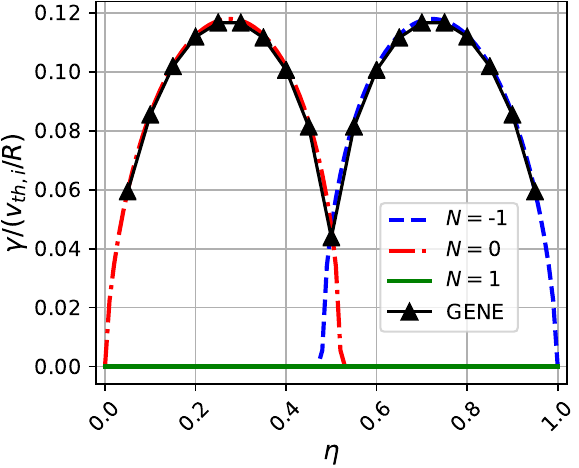}
    \caption{}
\end{subfigure}
\caption{A comparison between linear GENE simulation growth rates (black triangles) and analytical results from Eq. \eqref{eq:cold_ion_limit_ITG} for the ITG instability in the cold ion limit with varying (a) binormal wavenumber $k_{y} \rho_{i}$ at $\eta=0.3$ and (b) parallel boundary phase factor $\eta$ at $k_{y} \rho_{i} = 0.05$. The $L_{z} = R$ and the different style lines in the legend indicate different $N$ values.}
\label{fig:coldIonLimit_eta_kyScan_gamma_130422}
\end{figure}

In Fig. \ref{fig:coldIonLimit_eta_kyScan_gamma_130422} we compare the dispersion relation in Eq.  \eqref{eq:cold_ion_limit_ITG} with numerical results from GENE in the appropriate limit and we find a good match. GENE simulations were performed with adiabatic electrons in slab geometry, using $\omega_{T,i}=4 \times 10^{4}$ and $T_{i}/T_{e}=10^{-4}$. When examining the $k_{y} \rho_{i}$ scan, the best agreement is seen at low $k_{y} \rho_{i} \ll 1$ values, for which approximating the Finite Larmor Radius (FLR) effects to the lowest order, introduced through the denominator $1+k_{y}^{2}\rho_{i}^{2}$, is valid. From the $\eta$ scan, we again observe that the fastest growth rate occurs for $\eta \neq 0$. On the other hand, $\eta=0$ corresponds to a standard periodic boundary condition, which results in a stable linear mode.

The benchmarks shown in Figs. \ref{fig:etaScan_TP4studentLike_Lz_T10m3_170222} and \ref{fig:coldIonLimit_eta_kyScan_gamma_130422} verify the implementation of the parallel boundary condition phase factor in the GENE code and provides insight into the impact of the phase factor on the physics of linear modes.

\section{Floquet-Bloch theory for linear gyrokinetics}
\label{appendix:Floquet_Bloch}

When $\hat{s}=0$, we can use the parallel periodicity of the gyrokinetic model to find a general form for a linear eigenmode by invoking Bloch's theorem. This allows us to relate the growth rates and real frequencies of modes from systems with different parallel phase factors $C=\exp{(i2 \pi \eta)}$, $\eta \in ( -0.5, 0.5]$ and parallel lengths $2 \pi N_{pol}$, where $N_{pol} \in \mathbb{N}^{*}$. 

After having solved for the self-consistent electromagnetic fields in terms of the distributions and substituted the corresponding expressions into the linearized gyrokinetic equation, the equations of the fluctuating part of the distributions $f=[\delta f_{1,s}]$ (which is a vector when considering multiple kinetic species) are
\begin{equation}
    \frac{\partial f}{\partial t} + \hat{G}f=0,
\end{equation}
where $\hat{G}$ is a linear integro-differential operator with respect to $x$, $y$, $z$, $v_{\parallel}$ and $\mu$. In particular, note that for $\hat{s}=0$, $\hat{G}$ is invariant with respect to a translation in $z$ by $2 \pi$.

We can then Fourier-analyze the above expression in time to get \par
\begin{equation}
\label{eq:gk_operator_fourier}
     \hat{G}f = - i \omega f,
\end{equation}
which appears as an eigenvalue equation for $f$ with eigenvalue $\lambda=-i\omega$. From here our derivation will closely follow the standard derivation of Bloch's theorem in condensed matter physics \cite{ashcroftmermin1976}. \par

First, we define the translation operator $\hat{T}_{z}$ on function $f(z)$ according to
\begin{equation}
\label{eq:translation_op}
    \hat{T}_{l_{z}} f(z) = f(z+l_{z}),
\end{equation} which translates the function by $l_{z}$ in the parallel direction. We will consider $l_{z}$ to be equal to a single poloidal turn, i.e. $l_{z}=2\pi$.

In the particular case of $\hat{s}=0$, the operator $\hat{G}$ is invariant with respect to translation by $l_{z} = 2 \pi$. This is not the case for $\hat{s} \neq 0$ as certain geometric coefficients (e.g. the metric elements $g^{xy}$ and $g^{yy}$) are not periodic in $z$ due to a secular dependence on $\hat{s}$ \cite{Lapillonne2009}. Therefore, if $\hat{s}=0$ the gyrokinetic operator will be periodic $\hat{G}[z]f(z)=\hat{G}[z+l_{z}]f(z)$. Taking this periodicity into account one obtains 
\begin{equation}
    \hat{T}_{l_{z}} \hat{G}[z]f(z) = \hat{G}[z+l_{z}]f(z+l_{z}) = \hat{G}[z] f(z+l_{z}) = \hat{G}[z] \hat{T}_{l_{z}}f(z).
\end{equation}
Since the above relationship holds for any $f(z)$, it follows that the two operators $\hat{G}$ and $\hat{T}_{l_{z}}$ commute
\begin{equation}
    \hat{G}\hat{T}_{l_{z}}=\hat{T}_{l_{z}}\hat{G}.
\end{equation}
As a result, $\hat{T}_{l_{z}}$ leaves the eigenspace of $\hat{G}$ invariant, and one can thus always choose the eigenfunctions of $\hat{G}$ to also be eigenfunctions of $\hat{T}_{l_{z}}$. This can be written as
\begin{align}
\label{eq:G_eigenstates}
    & \hat{G} f = \lambda f, \\
\label{eq:T_eigenstates}
    & \hat{T}_{l_{z}} f = \nu f,
\end{align}
where $\lambda$ and $\nu$ are eigenvalues of $\hat{G}f$ and $\hat{T}_{l_{z}}f$ respectively. 

We can show that translation operator $\hat{T}_{l_{z}}$ is unitary and that its eigenvalues lie on the unit circle in the complex plane according to
\begin{equation}
\label{eq:proof_unitary}
    |\nu|^{2} \int |f(z)|^{2} dz = \int |T_{l_{z}}f(z)|^{2} dz = \int |f(z+l_{z})| dz = \int |f(z)|^{2} dz.
\end{equation}
Since $f(z) \neq 0$ this equation leads to $|\nu|=1$ and $\nu=\exp{(i\theta)}$, $\theta \in \mathbb{R}$.

We make an Ansatz that Eq. \eqref{eq:T_eigenstates} has a solution of the form 
\begin{equation}
\label{eq:f_Fexp}
f(z)=F(z)e^{i K_{z} z}  ,  
\end{equation}
where $F(z)=F(z+l_{z})$ is periodic in $l_{z}$. Substituting this and $\nu=\exp{(i\theta)}$ into Eq. \eqref{eq:T_eigenstates} gives
\begin{equation}
    F(z+l_{z}) e^{i K_{z} z + i  K_{z} l_{z}} = e^{-i \theta} F(z) e^{i K_{z} z} 
\end{equation}
\begin{equation}
    F(z+l_{z}) e^{i K_{z} l_{z} } = e^{-i \theta} F(z). 
\end{equation}
Hence, the anstaz is correct if $\theta = - K_{z} l_{z} + 2 \pi p$, where $p \in \mathbb{Z}$.

We finally substitute Eq. \eqref{eq:f_Fexp} into the parallel boundary condition of Eq. \eqref{eq:zero_shear_parallel_boundry_phase} to find
\begin{equation}
    1 = e^{i 2 \pi K_{z} N_{pol}} e^{-i 2\pi j \eta}.
\end{equation}
To satisfy this equation, we require that
\begin{equation}
\label{eq:kz_Bloch}
     K_{z} = (p + j \eta)/N_{pol},
\end{equation}
where $j = k_{y} / k_{y, min} \in \mathbb{Z}$ and $p \in \mathbb{Z}$. Given that we are considering linear calculations that only include one $k_{y}$ values, we take $j=1$. \par

Combining Eqs. \eqref{eq:f_Fexp} and \eqref{eq:kz_Bloch} gives
\begin{equation}
\label{eq:f_in_bloch_form}
    f(z)=F(z)e^{i(p+\eta)z/N_{pol}},
\end{equation}
where $f(z)$ is the particle distribution function and $F(z)$ is a $2 \pi$ periodic function. Eq. \eqref{eq:f_in_bloch_form} is the general form for a linear eigenmode in the gyrokinetic model with $\hat{s}=0$. Now let's consider that  $f_{multi}(z)$ is a solution to the gyrokinetic model in a domain that is $N_{pol}$ turns long, i.e. $L_{z} = 2 \pi N_{pol}$. If we denote the phase shift after $N_{pol}$ poloidal turns by $\eta_{multi}$ (which in simulations is set as an input parameter), we can evaluate that the phase shift $\eta_{single}$ after a single poloidal turn has to satisfy
\begin{equation}
\label{eq:f_etasingle}
    f_{multi}(z+2 \pi) = f_{multi}(z) e^{i 2 \pi \eta_{single} }.
\end{equation}
In other words, we are calculating the phase factor $\eta_{single}$ that leads to an equivalent solution in a single poloidal turn domain as in multiple poloidal turn domain. Substituting Eq. \eqref{eq:f_in_bloch_form} into Eq. \eqref{eq:f_etasingle} and using the $2 \pi$ periodicity of $F(z)$ yields
\begin{equation}
\label{eq:eta_relationship_appendix}
    \eta_{single} =\frac{ \eta_{multi} + p }{N_{pol}},
\end{equation}
where $p \in \mathbb{Z}$ is a free parameter. In other words, Eq. \eqref{eq:eta_relationship_appendix} tells us that the phase shift after each $2 \pi$ segment $\eta_{single}$ must add up to the imposed total phase shift $\eta_{multi}$ after $N_{pol}$ segments. This constrains what value of $\eta_{single}$ are possible as the only free parameter is $p$, which must be an integer. The system will adjust $p$ such that it selects the allowed value of $\eta_{single}$ with the largest growth rate. \par

\section{Scaling of eddy length with magnetic shear}
\label{appendix:eddy_scaling_law}

The magnetic shear enters into the gyrokinetic model in two ways -- the FLR factor and the magnetic drift term. A simple analysis of these can provide insights into how they influence the parallel eddy length.

Beginning with the FLR effects, we propose that the parallel eddy length is determined by the eddy becoming too sheared as it extends along the field line. FLR effects enter into the gyrokinetic model through the zeroth order Bessel function of the first kind $J_{0}(k_{\perp} \rho_{s})$. We postulate that an eddy will extend along the magnetic field line until it is sheared such that $|J_{0}(k_{\perp} \rho_{i})| < J_{crit}$. Given that the Bessel function decreases continuously to $0$ as $k_{\perp} \rho_{i}$ is increased from small values, our critical value $J_{crit}$ corresponds to a critical value of the perpendicular wavenumber $(k_{\perp} \rho_{s})_{crit}$. To lowest order in the inverse aspect ratio, we can approximate the perpendicular wavenumber as $k_{\bot} \rho_{s} \simeq (1+(\hat{s} z)^{2})^{1/2} k_{y} \rho_{s}$ where we have assumed circular flux surfaces \cite{Ball_2016} and considered a representative mode with $k_{x}=0$ and $k_{y}$. Using $l_{\parallel} \simeq Rq\Delta z$, where $\Delta z$ is the extent of the eddy in the poloidal angle $z$, this relation provides a simple scaling of the parallel eddy length
\begin{equation}
    \left(1+\left(\hat{s} \frac{l_{\parallel}}{qR}\right)^{2}\right)^{\frac{1}{2}} k_{y} \rho_{s} = (k_{y} \rho_{s})_{crit}.
\end{equation}
Assuming that $k_{y} \rho_{s}$ and $(k_{y} \rho_{s})_{crit}$ are constant with $\hat{s}$ this implies
\begin{equation}
\label{eq:l_par_shat_FLR_estimate}
    l_{\parallel} \propto \frac{1}{\hat{s}}.
\end{equation}

Next, let us examine the scaling of magnetic drifts with magnetic shear in a circular equilibrium geometry at a large aspect ratio, as outlined in \cite{Ball_2016}. Again we will consider a mode with $k_{x} = 0$, so our interest lies solely in the binormal drift
\begin{equation}
\label{eq:twidle_binormal_drifts}
    v_{y,d} = (\mathbf{v}_{C}+\mathbf{v}_{\nabla B}) \cdot \nabla y \propto \cos(z)+\hat{s} z \sin(z),
\end{equation}
which also exhibits a secular dependence on $z$ resulting from finite magnetic shear.

A basic analysis of the toroidal ITG instability shows that it requires a resonance between the diamagnetic frequency $\omega_{*}$ and the drift frequency $\omega_{d}$ to remain unstable \cite{Beer1995thesis}. As a mode extends along the magnetic field to larger values of $z$, the diamagnetic frequency $\omega_{*}$ remains constant, while the drift frequency $\omega_{d}$ increases due to its secular dependence on $z$. In the limit of large $z$, the drift frequency $\omega_{d}$ of a given mode with a fixed $k_{y}$ thus increases according to
\begin{equation}
    \omega_{d} \simeq \mathbf{v}_{d} \cdot \mathbf{k}_{\bot} \propto k_{y} \hat{s} z.
\end{equation}
We will postulate that a mode will be stable if $\omega_{d} > \omega_{d,crit}$, which produces a simple scaling for the parallel eddy length
\begin{equation}
\label{eq:eddy_lenght_scaling}
    l_{\parallel} \propto \frac{\omega_{d, crit}}{k_{y} \hat{s}} \propto \frac{1}{\hat{s}}.
\end{equation}
This again assumes that $k_{y}$ and $\omega_{d,crit}$ remain approximately constant as $\hat{s}$ varies.

From the above analyses of the FLR effects and magnetic drifts, we find that they both motivate the same scaling of turbulent eddy length with magnetic shear
\begin{equation}
    l_{\parallel} = \frac{\alpha}{\hat{s}},
\end{equation}
where the fitting parameter $\alpha$ may depend on many other properties of the equilibrium, but not on $\hat{s}$.

\section{Simulation parameters}
\label{appendix:simulation_parameters}

In this appendix we present the parameters used for numerical simulations using GENE v2. All simulations use Miller local equilibrium geometry \cite{Miller1998}. In the tables below, $\epsilon$ is the inverse aspect ratio, $T_{i}/T_{e}$ is the ion to electron temperature ratio, $m_{i}/m_{e}$ is the ion to electron mass ratio, $L_{v}$ and $L_{w}$ are the extensions of the simulation domain in $v_{\parallel}$ and $\mu$ respectively. Lastly, $N_{x}$, $N_{y}$, $N_{z}$, $N_{v_{\parallel}}$, $N_{\mu}$ are number of grid points in the $x$, $y$, $z$, $v_{\parallel}$ and $\mu$ directions respectively, and $N_{species}$ is the number of species. Finally, $M \in \mathbb{N}$ determines the domain size according to the quantization condition in Eq. \eqref{eq:domain_quantization}.

\begin{table}[H]
\caption{Key parameters for the linear $\hat{s}$ scan simulations. Note $N_{x}=1$ was used for $\hat{s}=0$ simulations. The corresponding results are shown Figs. \ref{fig:IoP_LinModeFreqVSshear}, \ref{fig:IoP_LinModeBallooningAngleVSShear_220223} and \ref{fig:IoP_LinModeEnergyVSShear}.}
\label{tab:parameters_linear_miller}
\begin{tabular}{|lllll|}
\hline
\multicolumn{1}{|l|}{$\epsilon = 0.18$} & \multicolumn{1}{l|}{$q_{0}=1.4$} & \multicolumn{1}{l|}{$\hat{s} \in [-0.8, 0.8]$} & \multicolumn{1}{l|}{$\beta = 0$} & $m_{i}/m_{e}=3670$ \\ \hline
\multicolumn{1}{|l|}{$T_{i}/T_{e}=1$} & \multicolumn{1}{l|}{$R/L_{N}\in \{0,1.11,2.22\}$} & \multicolumn{1}{l|}{$R/L_{T_{i}}=6.96$} & \multicolumn{1}{l|}{$R/L_{T_{e}} \in \{0,3.48,6.96\}$} & $N_{pol}=1$ \\ \hline
\multicolumn{1}{|l|}{$L_{z}=2\pi$} & \multicolumn{1}{l|}{$L_{v}$=3 $(2T_{s}/m_{s})^{1/2}$} & \multicolumn{1}{l|}{$L_{w}$=12 $T_{s}/B_{ref}$} & \multicolumn{1}{l|}{$k_{y} \rho_{i} = 0.45$} & $k_{x} \rho_{i} = 0$ \\ \hline
\multicolumn{5}{|l|}{$N_{x} \times N_{y} \times N_{z} \times N_{v_{\parallel}} \times N_{\mu} \times N_{species} = 256 \times 1 \times 32 \times 32 \times 12 \times 2$.} \\ \hline
\end{tabular}
\end{table}

\begin{table}[H]
\caption{Key parameters for the linear $N_{pol}$ and $\Delta y$ scan simulations. The corresponding results are shown in Figs. \ref{fig:IoP_LinModeEnergandFreqVSNpol}, \ref{fig:IoP_LinModeFreqVSEta}, \ref{fig:IoP_LinModeFreqVSEta_Npol10_030323} and \ref{fig:IoP_gamma_scanSafetyFactor_030323}.}
\label{tab:parameters_linear_npolScan_miller}
\begin{tabular}{|lllll|}
\hline
\multicolumn{1}{|l|}{$\epsilon = 0.18$} & \multicolumn{1}{l|}{$q_{0}=1.4$} & \multicolumn{1}{l|}{$\hat{s} = 0$} & \multicolumn{1}{l|}{$\beta = 0$} & $m_{i}/m_{e}=3670$ \\ \hline
\multicolumn{1}{|l|}{$T_{i}/T_{e}=1$} & \multicolumn{1}{l|}{$R/L_{N}=\{0,1.11,2.22\}$} & \multicolumn{1}{l|}{$R/L_{T_{i}}=6.96$} & \multicolumn{1}{l|}{$R/L_{T_{e}}=\{0,3.48,6.96\}$} & $N_{pol} \in [1, 12]$ \\ \hline
\multicolumn{1}{|l|}{$L_{z}=2\pi N_{pol}$} & \multicolumn{1}{l|}{$L_{v}$=3 $(2T_{s}/m_{s})^{1/2}$} & \multicolumn{1}{l|}{$L_{w}$=12 $T_{s}/B_{ref}$} & \multicolumn{1}{l|}{$k_{y} \rho_{i} = 0.45$} & $k_{x} \rho_{i} = 0$ \\ \hline
\multicolumn{5}{|l|}{$N_{x} \times N_{k_{y}} \times N_{z} \times N_{v_{\parallel}} \times N_{\mu} \times N_{species} = 1\times 1 \times 32N_{pol} \times 64 \times 12 \times 2$} \\ \hline
\end{tabular}
\end{table}

\begin{table}[H]
\caption{Key parameters for the nonlinear $N_{pol}$ scan simulations using adiabatic electrons. Note that simulations with $N_{pol}=21$ used $N_{x}=192$ and $N_{z}=20$ to decrease numerical costs. The corresponding simulations are shown in Figs. \ref{fig:IoP_AE_s0_NpolScan} and \ref{fig:IoP_AE_CorrEnvelopeNpol10_180324}.}
\label{tab:parameters_nonlinear_adiabatic_electron}
\begin{tabular}{|lllll|}
\hline
\multicolumn{1}{|l|}{$\epsilon = 0.18$} & \multicolumn{1}{l|}{$q_{0}=1.4$} & \multicolumn{1}{l|}{$\hat{s} =0$} & \multicolumn{1}{l|}{$\beta = 0$} & $m_{i}/m_{e} \rightarrow \infty$ \\ \hline
\multicolumn{1}{|l|}{$T_{i}/T_{e}=1$} & \multicolumn{1}{l|}{$R/L_{N}=2.22$} & \multicolumn{1}{l|}{$R/L_{T_{i}}=6.96$} & \multicolumn{1}{l|}{Adiabatic electrons} & $N_{pol} \in [1, 21]$ \\ \hline
\multicolumn{1}{|l|}{$L_{z}=2\pi N_{pol}$} & \multicolumn{1}{l|}{$L_{v}$=3 $(2T_{s}/m_{s})^{1/2}$} & \multicolumn{1}{l|}{$L_{w}$=9 $T_{s}/B_{ref}$} & \multicolumn{1}{l|}{$k_{y,min} \rho_{i} = 0.05$} & $L_{x}=350$ $ \rho_{i}$ \\ \hline
\multicolumn{5}{|l|}{$N_{x} \times N_{k_{y}} \times N_{z} \times N_{v_{\parallel}} \times N_{\mu} \times N_{species} = 384 \times 64 \times 24 \times 32 \times 9 \times 1$} \\ \hline
\end{tabular}
\end{table}

\begin{table}[H]
\caption{Key parameters for the nonlinear $N_{pol}$ scan simulations using kinetic electrons. The table indicates the minimal resolution used. The results at this resolution were checked against higher resolution simulations for a set of different $N_{pol}$ values and no significant changes were noted. The corresponding simulations are shown in Figs. \ref{fig:IoP_CorrEnvelopeNpol171_091122}, \ref{fig:IoP_pITG_s0_NpolScan_wCollisions_011122} and \ref{fig:IoP_CBC_s0_NpolScan_wCollisions_011122}.}
\label{tab:parameters_nonlinear_NpolScan_kinetic_electron}
\begin{tabular}{|lllll|}
\hline
\multicolumn{1}{|l|}{$\epsilon = 0.18$} & \multicolumn{1}{l|}{$q_{0}=1.4$} & \multicolumn{1}{l|}{$\hat{s} =0$} & \multicolumn{1}{l|}{$\beta = 10^{-5}$} & $m_{i}/m_{e}=$3670 \\ \hline
\multicolumn{1}{|l|}{$T_{i}/T_{e}=1$} & \multicolumn{1}{l|}{$R/L_{N}= \{0, 2.22\}$} & \multicolumn{1}{l|}{$R/L_{T_{i}}=6.96$} & \multicolumn{1}{l|}{$R/L_{T_{e}}=\{0, 6.96\}$} & $N_{pol} \in [1, 171]$ \\ \hline
\multicolumn{1}{|l|}{$L_{z}=2\pi N_{pol}$} & \multicolumn{1}{l|}{$L_{v}$=3 $(2T_{s}/m_{s})^{1/2}$} & \multicolumn{1}{l|}{$L_{w}$=12 $T_{s}/B_{ref}$} & \multicolumn{1}{l|}{$k_{y,min} \rho_{i} = 0.1$} & $L_{x}= 90 \rho_{i}$ \\ \hline
\multicolumn{5}{|l|}{$N_{x} \times N_{k_{y}} \times N_{z} \times N_{v_{\parallel}} \times N_{\mu} \times N_{species} = 96 \times 32 \times 16 \times 32 \times 12 \times 2$} \\ \hline
\end{tabular}
\end{table}

\begin{table}[H]
\caption{Key parameters for the nonlinear $\Delta y$ scan simulations using adiabatic electrons. The corresponding simulations are shown in Figs. \ref{fig:IoP_AE_s0_etaScan_ky005_Npol1andNpol11} and \ref{fig:IoP_AE_s0_etaScan_ky005_strechingIlustration}.}
\label{tab:parameters_nonlinear_adiabatic_electron_eta_scan}
\begin{tabular}{|lllll|}
\hline
\multicolumn{1}{|l|}{$\epsilon = 0.18$} & \multicolumn{1}{l|}{$q_{0}=1.4$} & \multicolumn{1}{l|}{$\hat{s} =0$} & \multicolumn{1}{l|}{$\beta = 0$} & $m_{i}/m_{e} \rightarrow \infty$ \\ \hline
\multicolumn{1}{|l|}{$T_{i}/T_{e}=1$} & \multicolumn{1}{l|}{$R/L_{N}=2.22$} & \multicolumn{1}{l|}{$R/L_{T_{i}}=6.96$} & \multicolumn{1}{l|}{Adiabatic electrons} & $N_{pol} \in \{1, 11\}$ \\ \hline
\multicolumn{1}{|l|}{$L_{z}=2\pi N_{pol}$} & \multicolumn{1}{l|}{$L_{v}$=3 $(2T_{s}/m_{s})^{1/2}$} & \multicolumn{1}{l|}{$L_{w}$=12 $T_{s}/B_{ref}$} & \multicolumn{1}{l|}{$k_{y,min} \rho_{i} = 0.05$} & $L_{x}=100\rho_{i}$ \\ \hline
\multicolumn{5}{|l|}{$N_{x} \times N_{k_{y}} \times N_{z} \times N_{v_{\parallel}} \times N_{\mu} \times N_{species} = 128 \times 64 \times 16 \times 32 \times 12 \times 1$} \\ \hline
\end{tabular}
\end{table}

\begin{table}[H]
\caption{Key parameters for the nonlinear $\Delta y$ scan simulations using kinetic electrons. The table indicates the minimal resolution used. The corresponding simulations are shown in Figs. \ref{fig:IoP_CBCpITGetaScan_full_221122}, \ref{fig:s0_pITG_eta006_nrg_Q_es_ions_190522} and \ref{fig:s0_pITG_etaScan_corrcontour_test}.}
\label{tab:parameters_nonlinear_etaScan_kinetic_electron}
\begin{tabular}{|lllll|}
\hline
\multicolumn{1}{|l|}{$\epsilon = 0.18$} & \multicolumn{1}{l|}{$q_{0}=1.4$} & \multicolumn{1}{l|}{$\hat{s} =0$} & \multicolumn{1}{l|}{$\beta = 0.00001$} & $m_{i}/m_{e}=$3670 \\ \hline
\multicolumn{1}{|l|}{$T_{i}/T_{e}=1$} & \multicolumn{1}{l|}{$R/L_{N}=\{0,2.22\}$} & \multicolumn{1}{l|}{$R/L_{T_{i}}=6.96$} & \multicolumn{1}{l|}{$R/L_{T_{e}}=\{0,6.96\}$} & $N_{pol} = 1$ \\ \hline
\multicolumn{1}{|l|}{$L_{z}=2\pi N_{pol}$} & \multicolumn{1}{l|}{$L_{v}$=3 $(2T_{s}/m_{s})^{1/2}$} & \multicolumn{1}{l|}{$L_{w}$=12 $T_{s}/B_{ref}$} & \multicolumn{1}{l|}{$k_{y,min} \rho_{i} = 0.05$} & $L_{x}= 150 \rho_{i}$ \\ \hline
\multicolumn{5}{|l|}{$N_{x} \times N_{k_{y}} \times N_{z} \times N_{v_{\parallel}} \times N_{\mu} \times N_{species} = 196 \times 64 \times 16 \times 32 \times 12 \times 2$} \\ \hline
\end{tabular}
\end{table}

\begin{table}[H]
\caption{Key parameters for the nonlinear $k_{y,min}$ scan simulations using heavy kinetic electrons. The table indicates the minimal resolution used. The corresponding simulations are shown in Figs. \ref{fig:s0_pITG_etaScan_corrcontour_squeezingStudy} and \ref{fig:ITG_mex10_kyScan_Dy18_resIssueResolved}.}
\label{tab:parameters_nonlinear_kyScan_KE_squeezing_Study}
\begin{tabular}{|lllll|}
\hline
\multicolumn{1}{|l|}{$\epsilon = 0.18$} & \multicolumn{1}{l|}{$q_{0}=1.4$} & \multicolumn{1}{l|}{$\hat{s} =0$} & \multicolumn{1}{l|}{$\beta = 10^{-5}$} & $m_{i}/m_{e}=$367 \\ \hline
\multicolumn{1}{|l|}{$T_{i}/T_{e}=1$} & \multicolumn{1}{l|}{$R/L_{N}=0$} & \multicolumn{1}{l|}{$R/L_{T_{i}}=6.96$} & \multicolumn{1}{l|}{$R/L_{T_{e}}=0$} & $N_{pol} = 1$ \\ \hline
\multicolumn{1}{|l|}{$L_{z}=2\pi N_{pol}$} & \multicolumn{1}{l|}{$L_{v}$=3 $(2T_{s}/m_{s})^{1/2}$} & \multicolumn{1}{l|}{$L_{w}$=12 $T_{s}/B_{ref}$} & \multicolumn{1}{l|}{$k_{y,min} \rho_{i} = [0.0125, 0.025, 0.05]$} & $L_{x}= 110 \rho_{i}$ \\ \hline
\multicolumn{5}{|l|}{$N_{x} \times N_{k_{y}} \times N_{z} \times N_{v_{\parallel}} \times N_{\mu} \times N_{species} = 128  \times 256 \times 32 \times 32 \times 12 \times 2$} \\ \hline
\end{tabular}
\end{table}

\begin{table}[H]
\caption{Key parameters for the nonlinear $\hat{s}$ scan simulations using kinetic electrons. The table indicates the most common numerical resolution used, however some simulations were performed with more (at lower magnetic shear) or less (at higher magnetic shear) grid points. Note that $M=1$ was used for $\hat{s}= \pm 1$ and $M=|\hat{s}|/0.1$ was used for all other values of $\hat{s}$ in the $N_{pol}=1$ case. The corresponding simulations are shown in Figs. \ref{fig:sscanNL_Qes_ion_230821},  \ref{fig:maximumcorrelationVSshear_kymin005_230821} and \ref{fig:IoP_s01_ky005_MomCorrugations_220723}}
\label{tab:parameters_nonlinear_sScan_KE}
\begin{tabular}{|lllll|}
\hline
\multicolumn{1}{|l|}{$\epsilon = 0.18$} & \multicolumn{1}{l|}{$q_{0}=1.4$} & \multicolumn{1}{l|}{$\hat{s} = \in [-0.1, 0.8]$} & \multicolumn{1}{l|}{$\beta = 10^{-5}$} & $m_{i}/m_{e}=$3670 \\ \hline
\multicolumn{1}{|l|}{$T_{i}/T_{e}=1$} & \multicolumn{1}{l|}{$R/L_{N}=2.22$} & \multicolumn{1}{l|}{$R/L_{T_{i}}=6.96$} & \multicolumn{1}{l|}{$R/L_{T_{e}}=6.96$} & $N_{pol} = \{1,3\}$ \\ \hline
\multicolumn{1}{|l|}{$L_{z}=2\pi N_{pol}$} & \multicolumn{1}{l|}{$L_{v}$=3 $(2T_{s}/m_{s})^{1/2}$} & \multicolumn{1}{l|}{$L_{w}$=9 $T_{s}/B_{ref}$} & \multicolumn{1}{l|}{$k_{y,min} \rho_{i} \in \{0.0125, 0.025, 0.05, 0.1 \}$} & $L_{x}$ is given by Eq. \eqref{eq:domain_quantization} \\ \hline
\multicolumn{5}{|l|}{$N_{x} \times N_{k_{y}} \times N_{z} \times N_{v_{\parallel}} \times N_{\mu} \times N_{species} = 384  \times 96 \times 16 \times 64 \times 9 \times 2$} \\ \hline
\end{tabular}
\end{table}

\begin{table}[H]
\caption{Key parameters for the nonlinear $\hat{s}=0.8$ simulation using kinetic electrons shown in Fig. \ref{fig:IoP_s01_ky005_MomCorrugations_220723}.}
\label{tab:parameters_nonlinear_s08_KE}
\begin{tabular}{|lllll|}
\hline
\multicolumn{1}{|l|}{$\epsilon = 0.18$} & \multicolumn{1}{l|}{$q_{0}=1.4$} & \multicolumn{1}{l|}{$\hat{s} = 0.8$} & \multicolumn{1}{l|}{$\beta = 10^{-3}$} & $m_{i}/m_{e}=$3670 \\ \hline
\multicolumn{1}{|l|}{$T_{i}/T_{e}=1$} & \multicolumn{1}{l|}{$R/L_{N}=2.22$} & \multicolumn{1}{l|}{$R/L_{T_{i}}=6.96$} & \multicolumn{1}{l|}{$R/L_{T_{e}}=6.96$} & $N_{pol} = \{1,3\}$ \\ \hline
\multicolumn{1}{|l|}{$L_{z}=2\pi$} & \multicolumn{1}{l|}{$L_{v}$=3 $(2T_{s}/m_{s})^{1/2}$} & \multicolumn{1}{l|}{$L_{w}$=9 $T_{s}/B_{ref}$} & \multicolumn{1}{l|}{$k_{y,min} \rho_{i} = 0.007$} & $L_{x}=175.6 \rho_{i}$ \\ \hline
\multicolumn{5}{|l|}{$N_{x} \times N_{k_{y}} \times N_{z} \times N_{v_{\parallel}} \times N_{\mu} \times N_{species} = 256 \times 512 \times 16 \times 32 \times 9 \times 2$} \\ \hline
\end{tabular}
\end{table}

\begin{table}[H]
\caption{Key parameters for the nonlinear $\hat{s}=0.05$ simulation using kinetic electrons shown in Fig. \ref{fig:s005_ky0025_correlation_place_holder}.}
\label{tab:parameters_nonlinear_s005_KE}
\begin{tabular}{|lllll|}
\hline
\multicolumn{1}{|l|}{$\epsilon = 0.18$} & \multicolumn{1}{l|}{$q_{0}=1.4$} & \multicolumn{1}{l|}{$\hat{s} = 0.05$} & \multicolumn{1}{l|}{$\beta = 10^{-4}$} & $m_{i}/m_{e}=$3670 \\ \hline
\multicolumn{1}{|l|}{$T_{i}/T_{e}=1$} & \multicolumn{1}{l|}{$R/L_{N}=2.22$} & \multicolumn{1}{l|}{$R/L_{T_{i}}=6.96$} & \multicolumn{1}{l|}{$R/L_{T_{e}}=6.96$} & $N_{pol} = 1$ \\ \hline
\multicolumn{1}{|l|}{$L_{z}=2\pi$} & \multicolumn{1}{l|}{$L_{v}$=3 $(2T_{s}/m_{s})^{1/2}$} & \multicolumn{1}{l|}{$L_{w}$=9 $T_{s}/B_{ref}$} & \multicolumn{1}{l|}{$k_{y,min} \rho_{i} = 0.025$} & $L_{x}=786.933 \rho_{i}$ \\ \hline
\multicolumn{5}{|l|}{$N_{x} \times N_{k_{y}} \times N_{z} \times N_{v_{\parallel}} \times N_{\mu} \times N_{species} = 1024 \times 128 \times 16 \times 64 \times 9 \times 2$} \\ \hline
\end{tabular}
\end{table}

\bibliographystyle{unsrt}
\bibliography{bibliography.bib}

\end{document}